\documentclass[fleqn,twocolumn]{openjournal}
\usepackage[T1]{fontenc}
\usepackage{ae,aecompl}
\usepackage{graphicx}   % Including figure files

\usepackage{ulem}
\usepackage{hyperref}
\usepackage{amsmath}    % Advanced maths commands
\usepackage{amssymb}    % Extra maths symbols
\usepackage{natbib}
\usepackage{color}      
\definecolor{dgreen}{rgb}{0,0.5,0}
\definecolor{dp}{rgb}{0.5,0,0.5}
\definecolor{magen}{rgb}{0.79,0.08,0.48}
\definecolor{darkred}{rgb}{0.65,0.06,0.37}
\definecolor{red}{rgb}{.65,0,0}
\usepackage{booktabs}
\usepackage{multirow}
\usepackage{makecell}

\newcommand{\grale}{\textsc{grale}}
\newcommand{\simgt}{\hbox{\,\rlap{\raise0.425ex\hbox{$>$}}\lower0.65ex\hbox{$\sim$}\,}}
\newcommand{\simlt}{\hbox{\,\rlap{\raise0.425ex\hbox{$<$}}\lower0.65ex\hbox{$\sim$}\,}}
\newcommand{\llrw}{\textcolor{black}}

\newcommand{\lasko}{\textcolor{black}}

\newcommand{\final}{\textcolor{black}}

\hypersetup{
    colorlinks=true,  
    urlcolor=blue
}

\shorttitle{Multiple image and cluster properties}
\shortauthors{Lasko et al.}

\begin{document}
\nocite{*}

\title[]{What multiple images say about the large-scale mass maps of galaxy clusters}
% Force line breaks with \\

\author{Kekoa Lasko}
\email{lasko062@umn.edu}
%\affiliation{ 
%Authors' institution and/or address%\\This line break forced with \textbackslash\textbackslash
%}%
\author{Liliya L.R. Williams}
\email{llrw@umn.edu}
\author{Agniva Ghosh}
\email{ghosh116@umn.edu}
\affil{School of Physics and Astronomy, University of Minnesota, 116 Church St SE, Minneapolis MN 55455, USA}

%\date{\today}//% It is always \today, today,
             %  but any date may be explicitly specified

\begin{abstract}
All lens modeling methods, simply-parametrized, hybrid, and free-form, use assumptions to reconstruct galaxy clusters with multiply imaged sources, though the nature of these assumptions (priors) can differ considerably between methods. This raises an important question in strong lens modeling: how much information about the mass model comes from the lensed images themselves, and how much is a consequence of model priors. One way to assess the relative contributions of the lensing data vs. model priors is to estimate global lens properties through images alone, without any prior assumptions about the mass distribution. This is our approach.
We use 200 mock cluster lenses, half of which have substructures which vary from clumpy and compact to smooth and extended; a simulated cluster Ares; and real clusters Abell 1689 and RXJ1347.5-1145 to show that the center, ellipticity, and position angle can be estimated quite well, and nearly perfectly for weakly substructured clusters, implying that the recovery of these properties is largely driven by the images, not priors. However, the correlation between the true and image-estimated amount of substructure has a lot of scatter, suggesting that multiple images do not uniquely constrain substructure. Therefore in general, lens model priors have a stronger effect on smaller scales. Our analysis partly explains why reconstructions using different methodologies can produce qualitatively different mass maps on substructure scales. Our analysis is not meant to aide or replace lens inversion methods, but only to investigate what cluster properties are constrained with multiple images.
%% ~244 words
\\\\
\textbf{Key words:} gravitational lensing: strong - galaxies: clusters: general
\end{abstract}

\maketitle

\section{Introduction}\label{sec:intro}

Mass distribution in galaxy clusters is important for constraining the properties of dark matter \citep[e.g.,][]{and22,veg21,har15},  and for using clusters as natural telescopes to observe the very high redshift Universe \citep[e.g.,][]{bou22,sal20,liv17}. Most of the existing work on galaxy cluster lensing reconstructs their sky-projected mass distributions, using one or more lens inversion methods. The methods range from light-traces-mass, where the distribution of cluster stellar light is the basis for reconstructing their total mass \citep{zit15,bro05}, to parametric, where simple functional forms describe individual galaxy members and a few cluster-scale dark matter halos \citep{lap21,nie20,gri15,ogu10,jul09}, to free-form, which do not have a strict relation between galaxies' light and mass so lensed images have a greater say in how cluster-scale mass is distributed \citep{tor22,cha22,lam19,bra08,lie06,lie07}, to hybrid, which combine free-form and parametric features \citep{die18}. All types of methods have their strengths as well as weaknesses. 

With the increasing amount and precision of the lensing data, from {\it Hubble Space Telescope's} Frontier Fields \citep{lot17}, and BUFFALO \citep{ste20}, {\it James Webb Space Telescope} \citep{mah22,gol22}, as well as spectroscopic data from the {\it Multi-Unit Spectroscopic Explorer} \citep[MUSE, e.g.,][]{ric21,jau21,mah18,lag17}, it has become apparent that clusters have complex mass distributions. Parametric methods have adapted by including more flexible variables in their models \citep[e.g.,][]{bea21}, and some free-form methods have increased their resolution on small scales \citep{lie20}.

Despite the quantity and quality of the data, the clusters are still underconstrained, and all lens inversion methods need to make modeling assumptions: the main difference is in the type of assumptions. As a result of differing model priors, the details of reconstructed mass distributions differ between models. This is the case not just for parametric vs. free-form methods, but even within parametric models created using the same software \citep{pri17,lim16}. It follows that the lensing data by itself does not \llrw{completely} determine the lens mass distribution. In other words, lens mass reconstructions, when based on the same high quality data suffer from degeneracies: a range of mass models can reproduce observed images, even when 100-200 images are used as input. The degeneracies are reduced with 1000 images \citep{gho20}, but observations are not there yet, though the early spectacular results from JWST may change that.

In view of this situation, it is important to determine how much information about the mass models comes from the lensed images themselves, and how much is a consequence of model priors. This is the question we try to address in this paper. One way to address this question is by finding out how much information can be estimated from the lensing data alone, without resorting to any mass modeling. This is the approach we take. It is model-free in the sense that we \llrw{do not carry out lens inversion to recover mass models of our mock clusters. Our analysis to estimate cluster properties is based solely on the multiple images.}

Our approach is supported by the analysis of \cite{men07}, who concludes that ellipticity, asymmetry and substructure in clusters are important in determining clusters' strong lensing properties. We turn that around, and ask how well can strong lensing properties---without a lens model---constrain clusters' center, ellipticity and substructure. Our definition of substructure is very broad: they range from compact clumps to smoothly varying density perturbations; \llrw{in other words, if a cluster's mass distribution deviates from elliptical, we call it substructured.}

Several papers in the literature aim to recover lens structure using multiple images, under certain specific conditions. For example, model-independent analysis can extract maximum information about cluster properties, without assuming a mass model, but only locally, in small lens plane regions covered by extended lensed images \citep[e.g.,][]{wag22,gri21,wag19,wag17}. Other works seek to find deviations from pure ellipticity using single point-like quads \citep{wit96} in galaxy lenses, or constrain isolated substructures using extend images near critical curves \citep{ala08}. \cite{cla16} explores how a systematic decomposition of large extended images using certain basis sets can potentially constrain lens properties.

The aim of this paper is to determine how well one can estimate {\it global} cluster properties, namely, the center, the amplitude and position angle of ellipticity, and the amount of substructure \llrw{or deviations from ellipticity}, using several multiply-imaged point sources. In this regard our analysis is different from other published work.  It is based on the work by \cite{wil08,wol12,wol15}, and \cite{gom18} on galaxy-scale lenses, which used observed properties of quad images. 

\begin{figure}
    \centering
    \includegraphics[trim={2cm 17.5cm 2cm 4.5cm},clip,width=0.65\textwidth]{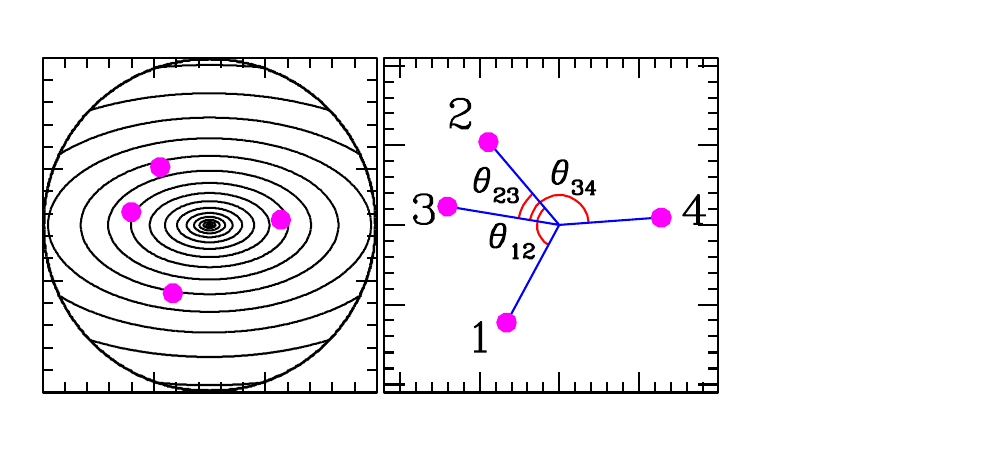}
    \caption{Definition of relative image angles in a quad lens system, illustrated using a smooth elliptical lens. {\it Left:} Black contours represent the mass distribution, and magenta dots are the 4 images of a quad.  {\it Right:} The same images, numbered by arrival order, with the relative angles labelled.}
    \label{fig:angles}
    \vskip0.15cm
\end{figure}

\begin{figure}
    \centering
    \includegraphics[trim={0cm 0cm 0cm 0cm},clip,width=0.45\textwidth]{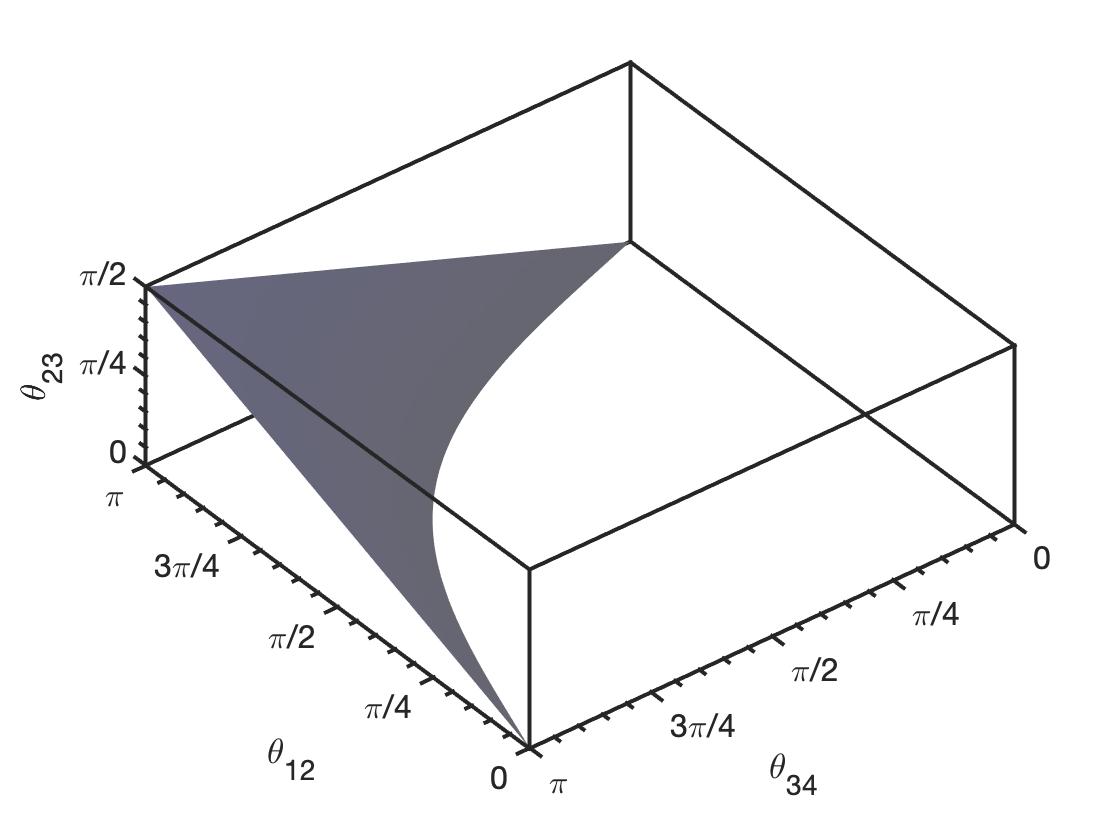}
    \caption{Two-dimensional slightly curved surface in 3D space of relative image angles (in radians) of quad lenses, called the Fundamental Surface of Quads (FSQ) \citep{wol12}. All quads of elliptical mass distributions of a wide range of density profile lie either on or very close on this surface. Deviations of quad distribution form the FSQ indicate deviations from ellipticity in the lens. \llrw{(The units on the axes are in radians.)}}
    \label{fig:FSQ}
    \vskip0.15cm
\end{figure}

Their analysis method is based on a curious property of lensing mass distributions: smooth, purely elliptical lenses of arbitrary ellipticity which is at least approximately constant with radius and density profile shape which is within the range of astrophysically plausible ones generate quads whose polar angles around the lens center obey a certain well defined relation. These angles are shown in Figure~\ref{fig:angles}: they are the relative image angles $\theta_{12},\theta_{23}$, and $\theta_{34}$. The subscripts indicate the arrival sequence of images, $1\rightarrow 4$, which for nearly all quads can be determined by quad morphology alone, without time delay data \citep{sah03}.
The relation between these 3 angles---the Fundamental Surface of Quads (FSQ)---is shown in Figure~\ref{fig:FSQ}.  \llrw{The FSQ is defined by a specific analytical lensing potential \citep{wol12}. Quads from other elliptical potentials and mass distributions follow FSQ closely, but not exactly.}\footnote{For example, the average separation of quads from a Singular Isothermal Elliptical mass distribution with ellipticity $\epsilon=0.3$ from the FSQ is $\sim\!0.6^\circ$, which for a cluster with an Einstein radius of $30"$ translates into image positions on the sky that differ from those predicted by the FSQ by $\sim 0.3"$. For a circular de Vaucouleurs profile with external shear $\gamma=0.4$ the corresponding value is $\sim 0.024"$.}
Each quad is a single point on the FSQ. Since all quads from any elliptical lens lie very close to the FSQ, it follows that deviations from the FSQ signal deviations from pure ellipticity, i.e., presence of substructure.

The FSQ-based technique to characterize substructure has already been used on galaxies. Using a population of 40 quads, \citep{gom18} concluded that $\Lambda$CDM substructure alone cannot account for the image properties, and other deviations from elliptical symmetry must be present in lensing galaxies, either in the lens plane or along the line of sight.

In addition to the difference in angular resolution,
the difference between galaxies and clusters in this context is that the former have simpler mass distributions because they are dynamically older, and their multiple image regions are quite small compared to their virial radii---only a few percent---therefore galaxies are closer to being relaxed. The clusters, on the other hand, are dynamically younger, and their multiple image regions sample a much larger central portion the cluster, which is expected to be more abundant in substructure. 

In this paper we generalize the quad-based technique, and apply it to smooth and substructured clusters to characterize global cluster properties---center, ellipticity,  position angle, and the amount of substructure. Unlike nearly all published papers on cluster lensing we do not reconstruct mass distributions in clusters; our estimation is independent of any lens modeling technique. 

All our estimators, including those that do not use FSQ directly, rely only on the relative image angles of quads. Not using image distances from the cluster lens center is a limitation of our method, because we are not using all the positional information provided by the quads. However, we are not aware of any model-free, global (i.e., cluster-wide) method that makes use of all positional information of multiple images. 
%To compensate for lensing data we cannot utilize, we use more quads, and therefore more total number of images, than is currently available for clusters.

Because we use only quads, not all clusters are amenable to our analysis.  Merging clusters, which can be highly elongated, are often dominated by triply-imaged systems---naked cusp systems---which we cannot use. 

The paper is organized as follows.
The construction of our mock clusters is described in Section~\ref{sec:mock}. Our quad-based estimators of cluster properties are presented in Section~\ref{sec:estimated}. How we measure the corresponding true properties, i.e., those measured directly from the known mass distribution in the mock clusters, is described in Section~\ref{sec:measured}. In Sections~\ref{sec:smooth} and \ref{sec:substruc} we examine how our quad-based estimators correlate with the true cluster properties, for 100 smooth and 100 substructured clusters, respectively.
In Section~\ref{sec:examples} we apply our method to one simulated cluster, Ares \citep{Meneghetti_2017}, generated using a semi-analytic code \citep{gio12}, and two observed clusters, Abell 1689, and RX J1347.5-1145. The discussion and conclusions are presented in Section~\ref{sec:last}.

\final{The projected surface mass densities of our clusters are quoted in dimensionless units of convergence $\kappa$, which is the surface mass density normalized by the critical surface mass density for lensing, $\Sigma_{\rm crit}=\frac{c^2}{4\pi G}\frac{D_s}{D_l\,D_{ls}}$, where $D$'s are angular diameter distances between the observer, lens, and source. For the standard cosmology with $\Omega_m=0.3$, $\Omega_\Lambda=0.7$, $h=0.7$, with a cluster at redshift $z=0.3$, and a source at $z=2.5$, $\Sigma_{\rm crit}=0.476$ g~cm$^{-2}$. }

Before we proceed we stress that there are 3 distinct pieces in our analysis: (i) making our mock clusters (Section~\ref{sec:mock}), (ii) measuring their true properties (Section~\ref{sec:measured}), and (iii) estimating their properties from multiple images (Section~\ref{sec:estimated}). Mass models are used in (i) only to construct the mocks; no mass modeling is done in (ii) and (iii), making our estimation analysis model-free.

\section{Mock galaxy clusters}\label{sec:mock}

We apply our method of estimating cluster properties to a set of $100$ mock smooth, purely elliptical clusters, and $100$ mock substructured clusters. \llrw{Smooth clusters without any small-scale or cluster-scale substructure are not realistic; we use them as a control sample for substructures ones.} The latter were obtained by adding mass clumps (see below) to the corresponding purely elliptical clusters. We construct these to approximate real clusters, but with a wider range of substructure properties, to stress test our method.

Smooth mock clusters have just the elliptical lensing potential, while substructured ones also have superimposed mass clumps. The smooth part is described by the {\tt alphapot} potential, $\psi = b\big(s^2 + x^2 + \frac{y^2}{q^2} + K^2xy\big)^{\frac{\alpha}{2}}$ \citep{kee01}, with slope $\alpha=1.1$ (or projected density slope of $-0.9$), and core radius ranging from $0.25\,r_{\rm Ein}$ to $0.70\,r_{\rm Ein}$, where $r_{\rm Ein}$ is the Einstein radius of the cluster lens. This range is consistent with the recent findings of \cite{lim22}.  (We estimate the Einstein radius\footnote{For circularly symmetric lenses Einstein radius is a model-independent quantity, and established the mass scale of the lens. In this paper we focus on morphological features of clusters, not scalings.}, $r_{\rm Ein}$, of each cluster as the average distance from the true cluster center to the 4 images of all its quads, which come from sources at a range of redshifts; see below.)

The slope of $\alpha=1.1$ is chosen to be somewhat shallower than isothermal ($\alpha=1$) to approximate the typical slope of cluster in the region where the images usually form. We give our clusters a range of ellipticities of the lensing potential, $\psi$, with axes ratios between $0.77$ and $0.95$, which correspond to mass axis ratio between $0.48$ and $0.87$.\footnote{An approximate relation between the axis ratio of the lensing potential iso-contours, $(b/a)_\psi$, and mass density iso-contours, $(b/a)_\kappa$, is $(b/a)_\kappa\approx {(b/a)_\psi}^{2.75}$.} We avoid circular clusters as these do not generate any quads. The ellipticity position angles were chosen randomly.

\llrw{Our substructures have a wide range of properties: compactness, normalization, and number density. Depending on their specific properties they can describe massive galaxies, or large-scale deviations from ellipticity in the dark matter, that are the result of past mergers.}

The amount of substructure varies considerably in observed clusters: equilibrium clusters can be described by an overall elliptical distribution with some modest amount of superimposed mass clumps, whereas some merging clusters can be so distorted in shape that describing them as elliptical is an oversimplification.

The substructure clumps are modeled by circularly symmetric Einasto profiles \citep{dha21}, with a range of parameters, picked randomly from within the following ranges: the Einasto shape parameter, $\alpha_E=2-5$, and the Einasto scale radius, $r_s=0.005\,r_{\rm Ein}-0.6\,r_{\rm Ein}$. The number of substructures in each cluster varies between 20 and 80. \llrw{The distribution of Einasto substructure centers is random in polar angle with respect to the cluster center, and random in distance, $r$, from center. That means their projected number density decreases as $1/r$.} Because of the wide range of parameters and because they are combined randomly in any given cluster, the range of substructure types we generate---scale, concentration, abundance, etc.---is wider than real clusters probably have. It is important to keep in mind that many of our substructures are not subhalos, but instead add to the cluster-scale mass perturbations and deviations from pure ellipticity.

Sources are placed at a range of redshifts, such that the critical surface mass density for lensing spans a factor of $2$. For a typical lens redshift, $z_l\sim 0.4$ this covers $z_s$ between $\sim 0.7$ and $\sim 4$, typical of observed sources. Sources at different redshifts are valued for cluster reconstruction because they usually \citep[but not always; see][]{pri17} break the mass sheet degeneracy \citep[MSD,][]{gor88,sah00}. In this paper we are not concerned with MSD; for us, sources at different redshifts result in image distribution that span a wider range of radii within clusters, which helps in extracting properties we are interested in. \llrw{All lensing mass is assumed to be at the same redshift, in a thin lens plane.}

We then scatter sources randomly in the source plane, and collect only those that produce quads around the cluster center. (Quads that are formed by individual substructures are not used.)  We generate 300 such quads for each of the smooth and substructured clusters.
%(500 in mock clusters without substructure clumps), but in some cases we use only a subset of these.

\section{Estimation of cluster properties using quads}\label{sec:estimated}

\final{The center of mass of a cluster is an important property. If it does not coincide with the brightest cluster galaxy (BCG) that would imply that the BCG is wobbling in the potential of the cluster \citep[e.g.,][]{zit12,lau14,sep23}. This wobbling may be consistent with the expectation of the standard $\Lambda$CDM, or, if large offsets are observed then self-interacting dark matter is a possibility \citep{har19,fis23}. The ellipticity position angle is also important for the understanding of the relation between clusters and their environment, for example, connection to cosmic filaments that feed clusters with infalling mass, or nearby merging clusters 
\citep[e.g.,][]{tam20,cho22,fur23}.}

We first describe how to find global cluster properties based solely on the relative image angles of its quadruply imaged background sources.

\subsection{Cluster center}\label{sec:est_cen}

\begin{figure*}
    \centering
    \includegraphics[width=.5\textwidth]{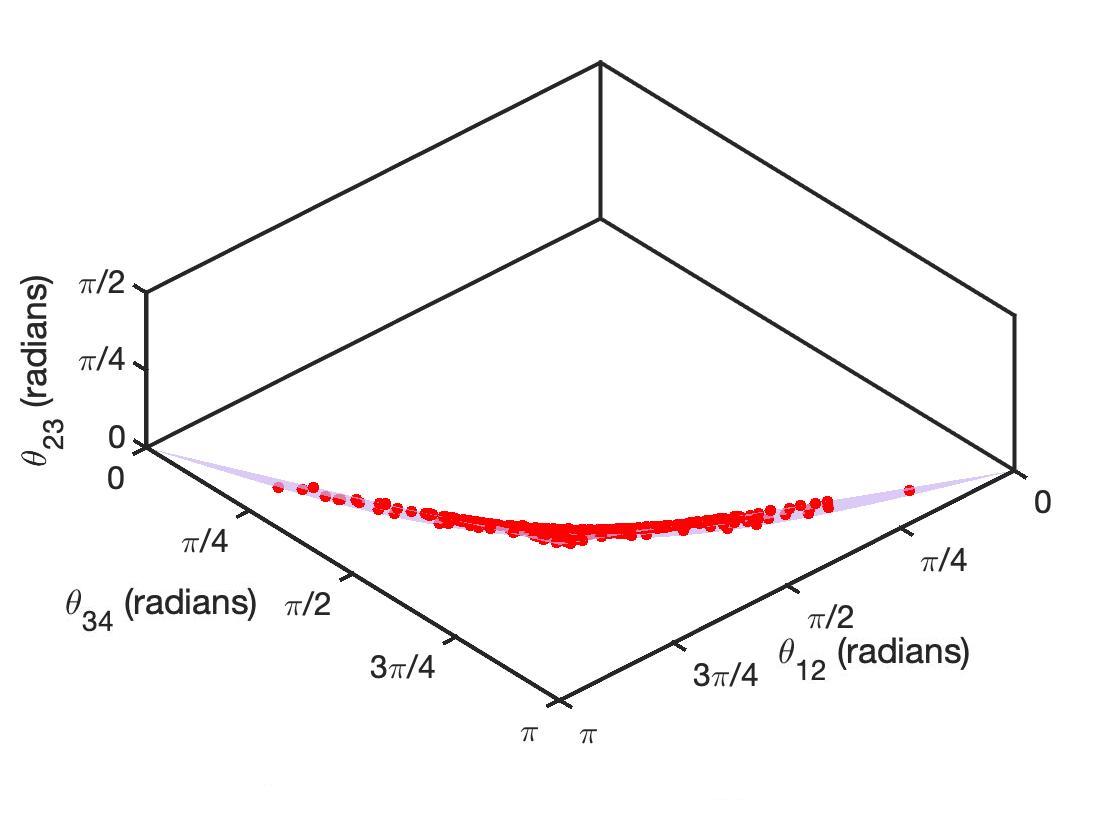}%
    \includegraphics[width=.5\textwidth]{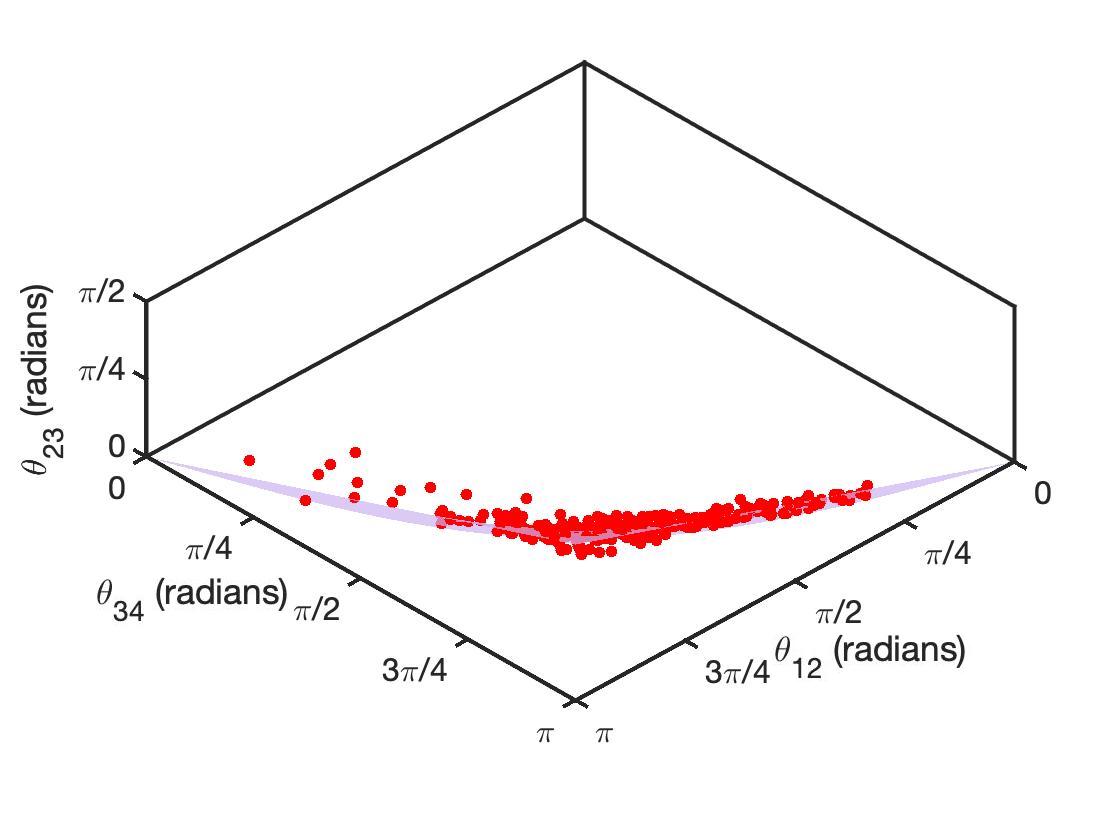}%
    \caption{Hundred quads of the clusters shown in the third rows of Figure \ref{fig:smooth} ({\it left}; purely elliptical clusters) and \ref{fig:structured} ({\it right}; substructured clusters). (Our calculations use a total of 300 quads per cluster.) The FSQ is the magenta surface. The 3D space of relative image angles has been rotated such that it appears as edge-on as possible, to show the quads' dispersion around the FSQ. Quads from the smooth, purely elliptical cluster have visibly less deviation from the FSQ, while the quads from the substructured cluster deviate from the FSQ.}
    \label{fig:FSQDeviations}
\end{figure*}

The 3D space of quad image angles, FSQ, is defined by 3 polar relative image angles, $\theta_{12},\,\theta_{23}$, and $\theta_{34}$, with respect to the lens center (Figure~\ref{fig:angles}).  Figure~\ref{fig:FSQDeviations} shows two examples (discussed later in the paper) each with $100$ quads (red points), alongside the FSQ (pink shaded surface). The left panel shows a purely elliptical cluster, while the right panel shows a substructured cluster. As expected, the quads in the former case are located very close to the FSQ, while in the latter case they are dispersed around it. We use such distribution of quads to estimate the cluster center. 

The principle behind the method is that for a purely elliptical lens all its quads will lie on, or very near the FSQ. Therefore to find the center of a purely elliptical lens, one needs to find the point in the lens plane that, when used as the center of the polar image distribution, results in the smallest dispersion of quads around the FSQ. Because \llrw{to first order} clusters are elliptical, the center of substructured cluster can also be found as the point that minimizes the deviations of its quads from the FSQ. \llrw{The corresponding rms dispersion is called $\delta_{\rm FSQ}$, and will be used later, in Section~\ref{sec:est_sub}.}

Observed galaxy clusters can have up to $\sim 15$ quads per cluster; the three clusters we discuss in Section~\ref{sec:examples} have 13 quads each. Aiming at more average clusters, we use 10 quads per mock cluster to carry out the procedure described in the previous paragraph. \llrw{We stress again that our cluster property estimation is done not for its own sake, but to figure out what global properties are already determined by the quads, even without modeling.} For each mock cluster we repeat the procedure for 30 independent sets of 10 quads each. Using each of these determinations we calculate mean and standard deviation. \final{We refer to the center generated using the mean of all 30 subset centers, with coordinates $(x_Q, y_Q)$, as the quad-estimated center of the cluster, and allow the standard deviation to become our estimate of the uncertainty.} \llrw{In other words, we use bootstrapping without replacement to estimate uncertainty.}

\final{We considered instead obtaining uncertainties from the same optimization algorithm used to minimize deviations from the FSQ. Ultimately, we chose to use bootstrapping to define our uncertainties because this method provided us estimates which were larger by orders of magnitude and which better predicted separation of the cluster centers between subsets of quads.}

\subsection{Ellipticity and position angle}\label{sec:est_ell}

Once the center has been located using the method described above, quads can be used to find the position angle of ellipticity of a cluster. \cite{wil08} have shown that the bisector of the angle between the 1st and 2nd arriving images is well aligned with the ellipticity position angle of the lens' mass distribution. The average of all bisectors from the 300 quads provides us with our quad-estimated position angle, which we call $\theta_Q$. We use the 30 independent samples of 10 quads per mock cluster to calculate the rms, which provides us with our uncertainty. 

To measure the amplitude of ellipticity we use another property of quads. The ellipticity is reflected in the distribution of quad images around the center. For high ellipticity clusters \llrw{(mass axis ratio $\lesssim 0.75$)}, most of the 1st and 2nd arriving images are along the minor cluster axis, and so tend to be across from each other around the cluster center. Very elliptical clusters tend to have $\theta_{12}\sim 180^\circ$, while less elliptical ones will have smaller $\theta_{12}$. We use $\theta_{12}$ as an indicator of cluster ellipticity. We calculate the average $\theta_{12}$ for 30 independent sets of 10 quads each, per mock cluster. The average and dispersion of these 30 measurements give us a quantity, $\langle\theta_{12}\rangle=\epsilon_Q$, which we call elongation, and the associated uncertainty. Note that this measure of elongation is expressed in radians.

\final{We again tested our choices of uncertainty estimation used above by comparing them against propagation of the uncertainty on the cluster center. We found that the rms deviations from subsets of quads were the larger estimate of the two methods, and that, for position angle, they better matched the difference between the quad-estimated value $\theta_Q$ and the true postition angle of the cluster, $\theta_T$, as described in Section \ref{sec:true_ell}. To better represent the distribution of quad-estimated values in this method, we chose to use bootstrapping without replacement as our method of determining uncertainties.}

\subsection{Substructure, \llrw{ or deviations from ellipticity}}\label{sec:est_sub}

We define substructure as deviations from a purely elliptical mass distribution.  Because quads of purely elliptical lenses lie \llrw{very near the FSQ (indistinguishable from it given the astrometric errors)} , the amount of substructure can be estimated as the rms dispersion of cluster quads (in the 3D space of angles) around the FSQ, $\delta_{\rm FSQ}$, which we defined in Section~\ref{sec:est_cen}. The value of $\delta_{\rm FSQ}$ generated by all 300 quads (using the quad-estimated center) is our quad-based estimate of substructure, $s_Q$. We also calculate the average $\delta_{\rm FSQ}$ for 30 independent samples of 10 quads each, per mock cluster, and the rms dispersion of these gives us the uncertainty. $s_Q$ will be very near $0$ for smooth elliptical lenses, and increases with increasing prevalence of substructure. 

\section{Measuring true cluster properties}\label{sec:measured}

Here, we describe how we quantify true global properties of clusters, using their known mass distribution. As we describe in Section~\ref{sec:mock}, we use 100 purely elliptical, smooth clusters, and 100 substructured clusters, generated by superimposing mass clumps onto purely elliptical clusters.

\subsection{Cluster center}\label{sec:true_cen}

The true cluster center, $(x_T,y_T)$, is taken to be the center of the mass distribution of the purely elliptical part of the cluster. \llrw{We checked that for substructured clusters this center is nearly identical (median difference is $0.01\,r_{\rm Ein}$)} to the center of mass of the cluster, determined using the mass distribution within the Einstein radius.

\subsection{Ellipticity and position angle}\label{sec:true_ell}

Because many of our clusters are substructured, and in some cases the added substructure changes the appearance of clusters considerably, ellipticity and its position angle (PA) are not just those of the purely elliptical part of the cluster, but are due to the total projected mass distribution. The latter is quantified by dimensionless convergence $\kappa(x,y)$.
%, the surface mass density normalized by the critical surface mass density for lensing, $\Sigma_{\rm crit}=\frac{c^2}{4\pi G}\frac{D_s}{D_l\,D_{ls}}$, where $D$'s are angular diameter distances between the observer, lens, and source.

We measure the ellipticity and PA using the first moment of the distribution of $\kappa(x,y)$ around different trial axes, assumed to go through the true cluster center, $(x_T,y_T)$. For each trial PA, we calculate the average distance from the center, along the trial PA axis, $\langle r_{||}\rangle$:
\begin{equation}
   \langle r_{||}\rangle=\epsilon_T=\frac{\iint\, r_{||}\,\kappa(x,y)\,dx\,dy}{\iint\,\kappa(x,y)\,dx\,dy},
\end{equation} \label{eq:ellipticity}
where each location in the lens plane is weighted by its surface mass density, $\kappa$. The integrals are evaluated as discrete summations over small mass pixels, $\sim 0.005$ of Einstein radius. The subscript of $r_{||}$ means that only the distance along the trial PA axis is used; the perpendicular component of the distance from the center is disregarded. The axis with the largest value of $\langle r_{||}\rangle$ is the true PA, which we call $\theta_T$, and the corresponding value of $\langle r_{||}\rangle$ is a measure of ellipticity, $\epsilon_T$.\footnote{Our method of finding the PA and ellipticity is very similar to the one used in SExtractor; see \\ {\tt https://sextractor.readthedocs.io/en/latest/Position.html\\ \#basic-shape-parameters-a-b-theta}}, but we use $r_E$ as a length measure, instead of a semi-major axis. Note that because this measure is the first moment of a distribution of distances, it is not straightforwardly related to the usual definition of ellipticity, and is not dimensionless. It is expressed in units of the Einstein radius, $r_{\rm Ein},$ for each cluster lens. Since we are interested only in how well the true and quad-estimated properties correlate, the units of the measures are not important.

\subsection{Substructure, \llrw{or deviations from ellipticity}}\label{sec:true_sub}

There is no unique way to separate the cluster mass distribution into the smooth component and substructure without making assumptions \citep[e.g.,][]{wag18}. Therefore there is no unique definition of substructure in a cluster. Various definitions have been used in the literature. In parametric models substructures are mostly the cluster's member galaxies and their associated dark matter halos. These are represented by mass clumps at or very near the locations of the galaxies, and can be described by a subhalo mass function \citep[e.g.,][]{nat17}. However, some cluster-scale dark matter halos that are not centered on the cluster center can also be considered substructure. In free-form methods, substructure is less tightly associated with member galaxies, and is better represented by a power spectrum of cluster's projected mass distribution \citep{moh16}. 

Because ours is the first attempt to quantify globally distributed substructure using point-like lensed images only, without mass modeling, we seek a simple, one-parameter measure of the amount of substructure. \final{Due to the differences in galaxies and clusters described in Section~\ref{sec:intro}, we use a different measure of substructure from the one used for galaxies.}\footnote{\final{The subhalo mass fraction, which is commonly used to quantify the amount of substructure in galaxies is well suited when substructures are compact, like $\Lambda$CDM substructure} \citep[\final{e.g.,}][]{des17,ori23}.
\final{In the case of clusters, which are not as relaxed as centers of galaxies, compact substructure is only one type of deviation from ellipticity, therefore the metric we use is different from subhalo mass fraction.}} We also want this measure to quantify deviations from pure ellipticity, similar to what $\delta_{\rm FSQ}$ does; see Section~\ref{sec:est_sub}. 

\begin{figure}
    \centering
    \vskip-2cm
    \includegraphics[width=0.5\textwidth]{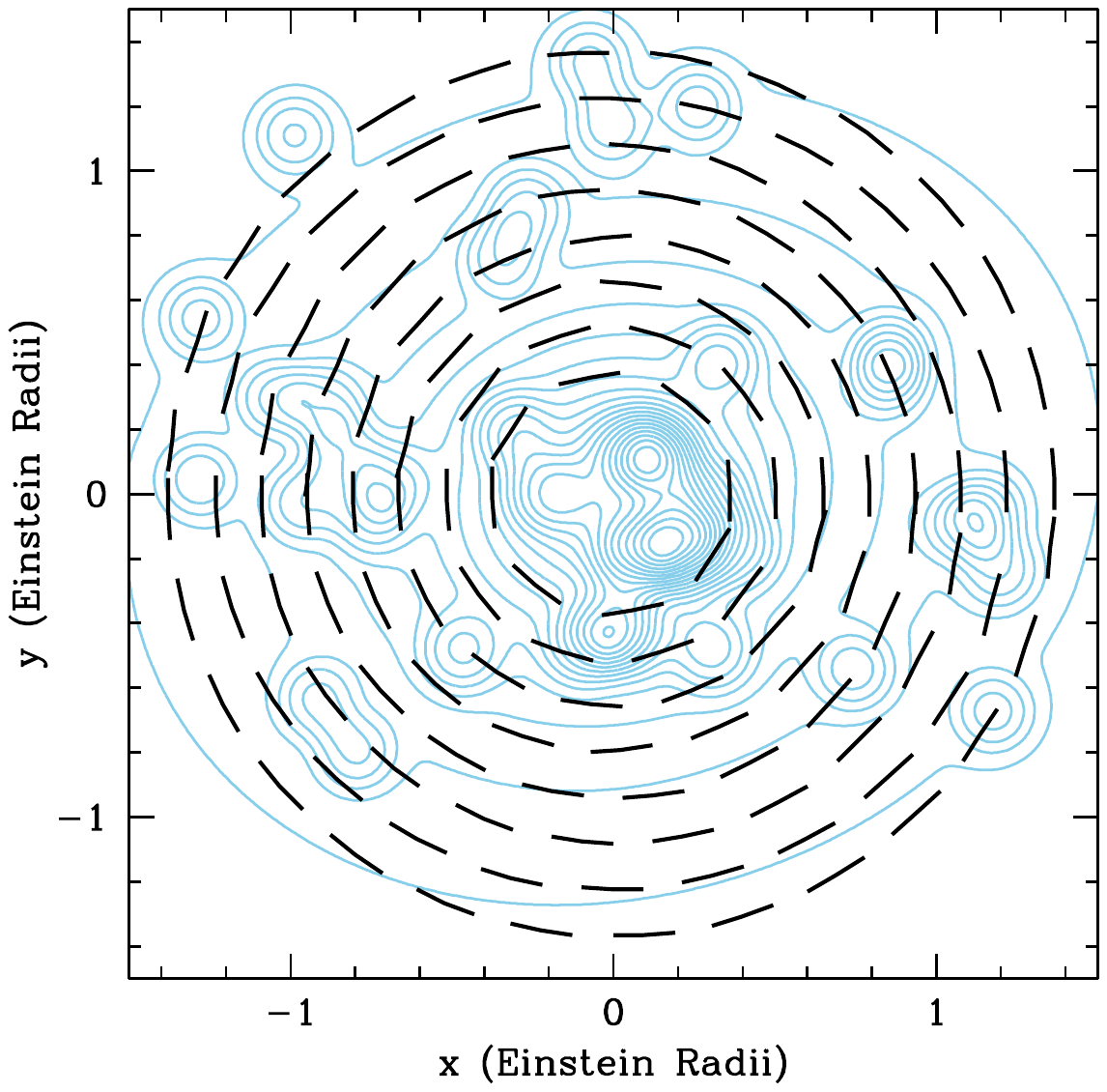}
    \vskip-2cm
    \caption{Measuring true substructure. The background light blue contours represent the mass density of one of our substructured clusters. We quantify the amount of substructure by measuring the absolute value of the density differences between the two ends of each of the black line segments, $|\Delta\kappa(r,\theta)|$. Here, $(r,\theta)$ refer to the center of the black line segment. The difference between $|\Delta\kappa$|'s on the opposite sides of the cluster (i.e., across the cluster center from each other) are subtracted to eliminate inversion symmetric (i.e., not substructure) mass features, such as ellipticity. See Section~\ref{sec:true_sub} and eq.~\ref{eq:sT} for details.}
    \label{fig:sticks}
    \vskip0.5cm
\end{figure}

We use the fact that \llrw{all types of} substructure produce local deviations from smooth elliptical mass distribution.  Such local deviations will introduce local density gradients, beyond those due to the smooth ellipticity. To quantify these we cover clusters with short line segments, oriented tangentially with respect to the center, $(x_T,y_T)$; see Figure~\ref{fig:sticks}. (In our analysis, we use many more segments than shown in the Figure.) We measure the absolute value of the density difference, $|\Delta\kappa(r,\theta)|$ between the two ends of each of these segments. Each line segment is labelled by the location of its center, $(r,\theta)$, where $r$ and $\theta$ are with respect to the center of the cluster. To take out the global cluster ellipticity or any other inversion symmetric component of the mass distribution, we use the difference between $\Delta\kappa$'s on the opposite sides of the cluster, $|\Delta\kappa(r,\theta)|-|\Delta\kappa(r,\theta+\pi)|$. Our measure of substructure is the average over radii and angles,
\begin{equation}
  s_T=\Big\langle\Big|\Big[ |\Delta\kappa(r,\theta)|-|\Delta\kappa(r,\theta+\pi)|\Big]\Big|\Big\rangle_{r>0,\theta=[0,\pi]},\label{eq:sT}
\end{equation}
which is dimensionless.
We tried a range of different line segment lengths, from $0.05$ to $0.5$ of the Einstein radius. While the values of $s_T$ depended on the choice of line segment length, the important metric for us is the nature of the correlation between true and quad-estimated amount of substructure. That relation, discussed in Section~\ref{sec:substruc}, did not depend on the length of the line segment used. We use $0.2\,r_{\rm Ein}$.

%The properties of clusters predicted from quads alone are designated with a lower subscript $T$: cluster center locations are $(x_T,y_T)$, ellipticity is $\epsilon_T$, its position angle $\theta_T$, and the degree of substructure is $s_T$. We want to see if these correlate with the corresponding true measures of these cluster, as determined from their true mass distributions. We use subscripts $M$ to denote these. The true center $(x_T, y_T)$ is found as the center of mass within XX\% of the Einstein radius. Ellipticity parameters, $\epsilon_T$ and $\theta_T$ are found as the first moment of the mass distribution around trial PA axes. Finally, $s_T$ is quantified as the clumpiness of the cluster. We describe the quantitative measurements of these in Sections XX, etc. 

In the following sections we will show how well global cluster parameters can be estimated, by comparing the quad-estimated properties (Section~\ref{sec:estimated}) vs. true properties (Section~\ref{sec:measured}).

\begin{figure*}
    \centering
    \includegraphics[height=0.31\textwidth]{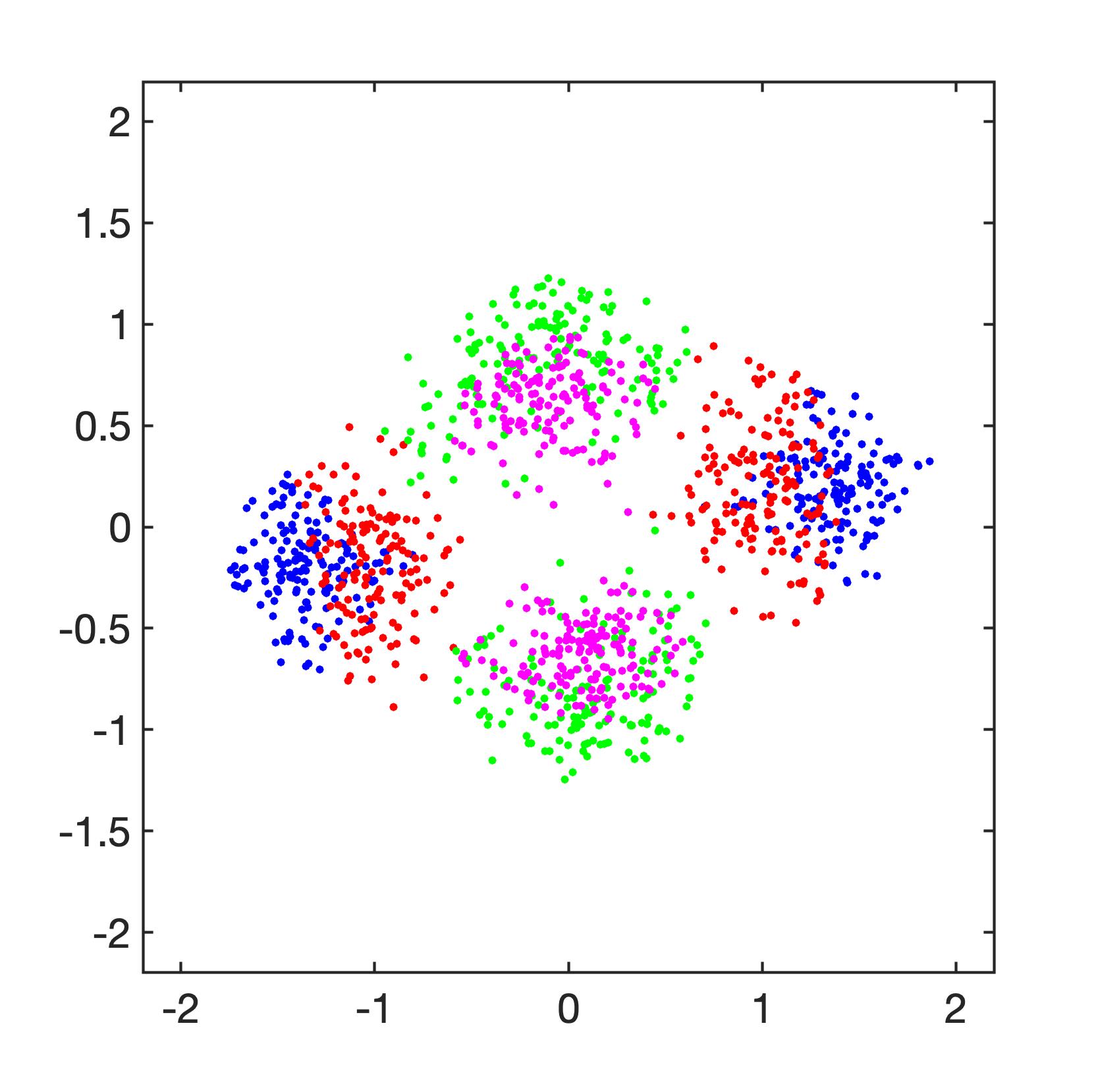}%
    \includegraphics[height=0.31\textwidth]{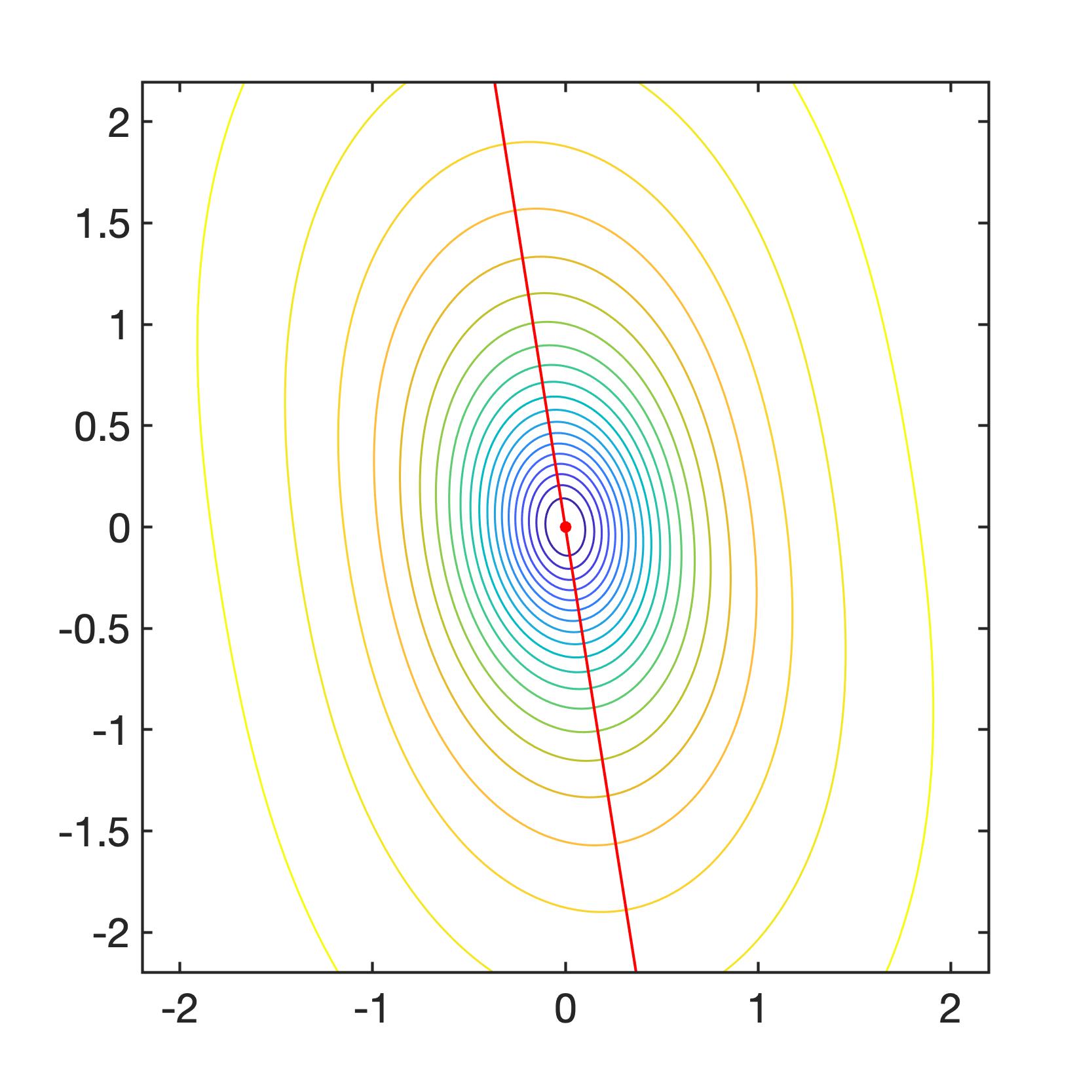}%
    \includegraphics[height=0.31\textwidth]{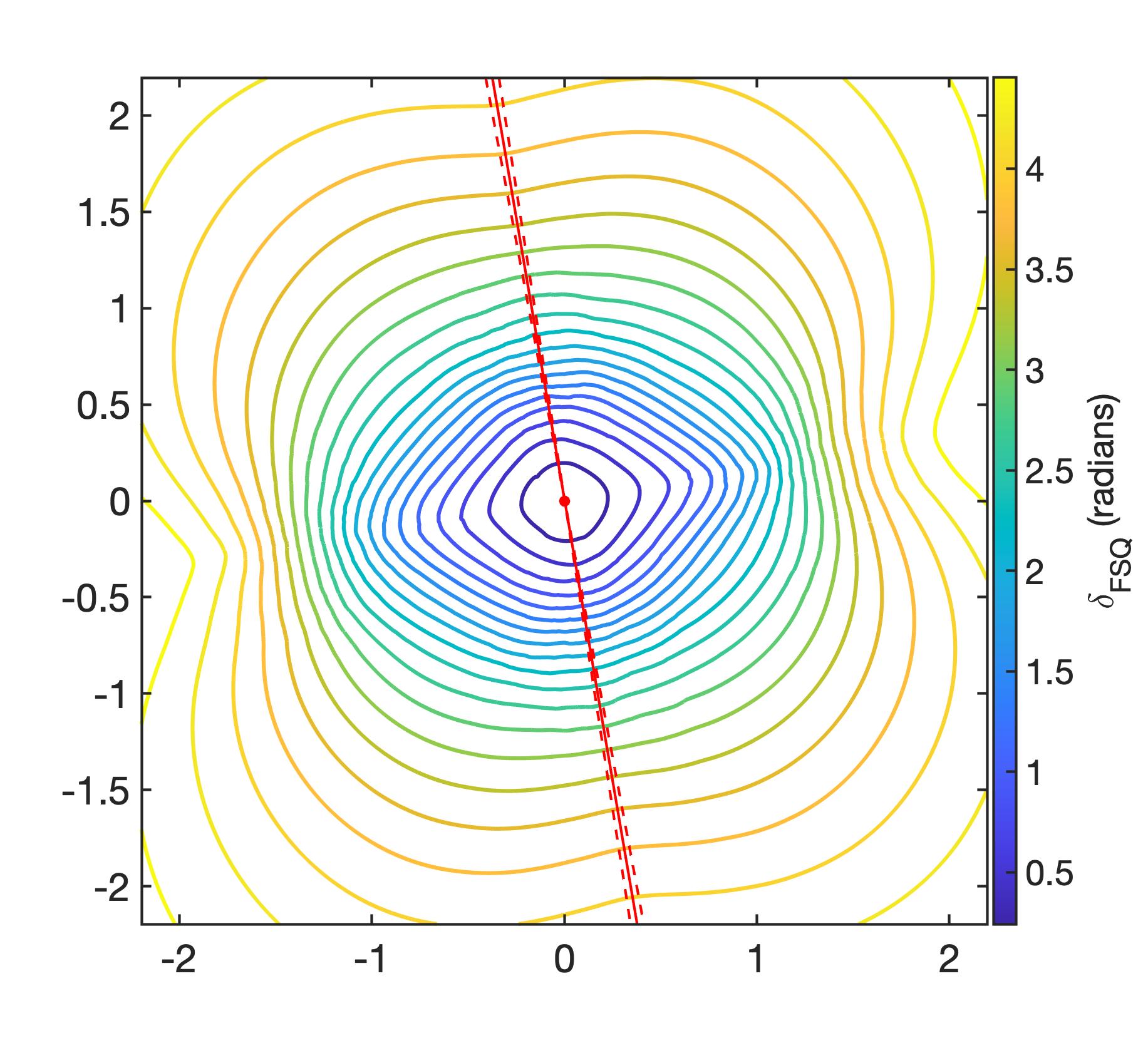}\\
    \includegraphics[height=0.31\textwidth]{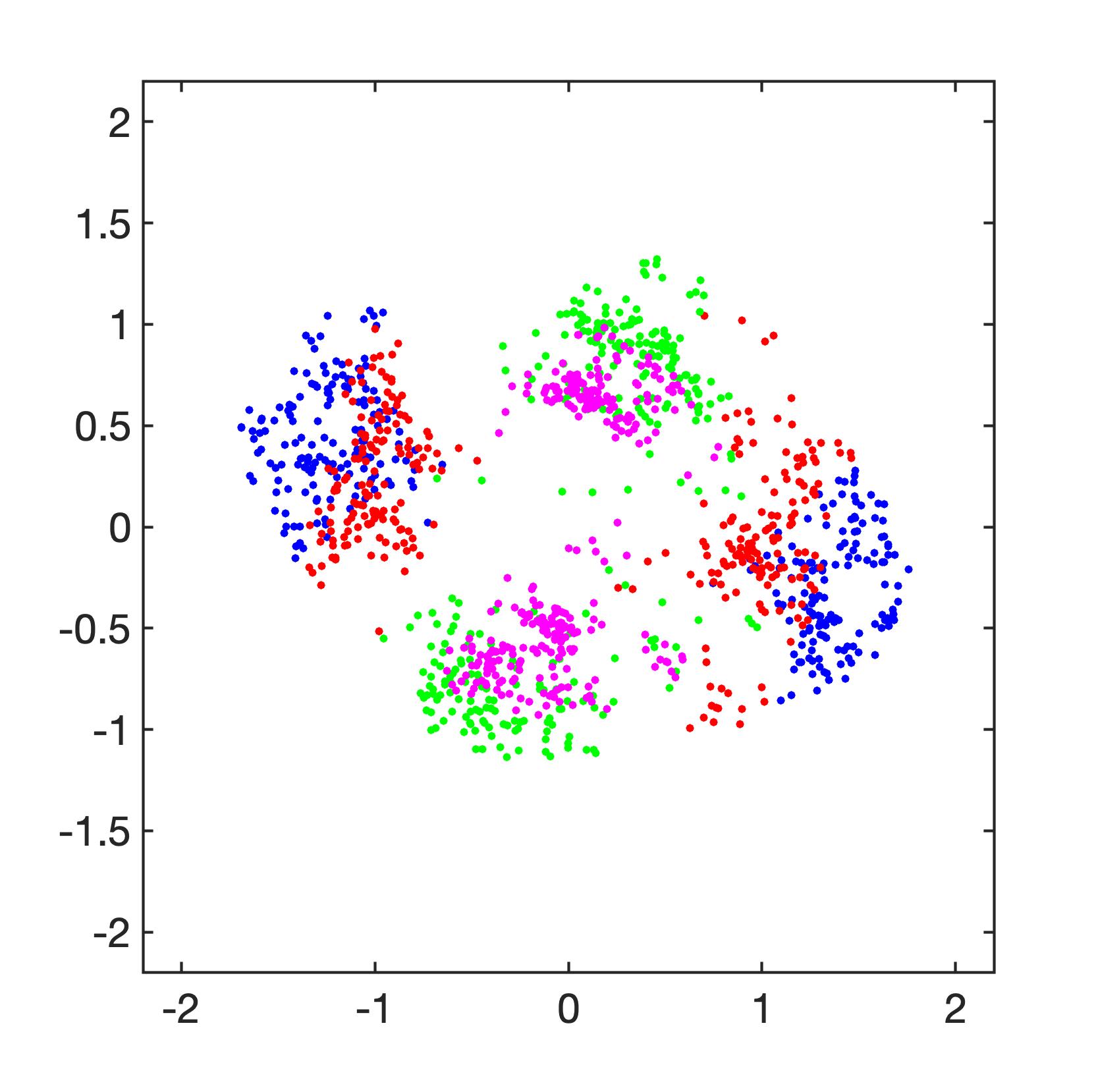}%
    \includegraphics[height=0.31\textwidth]{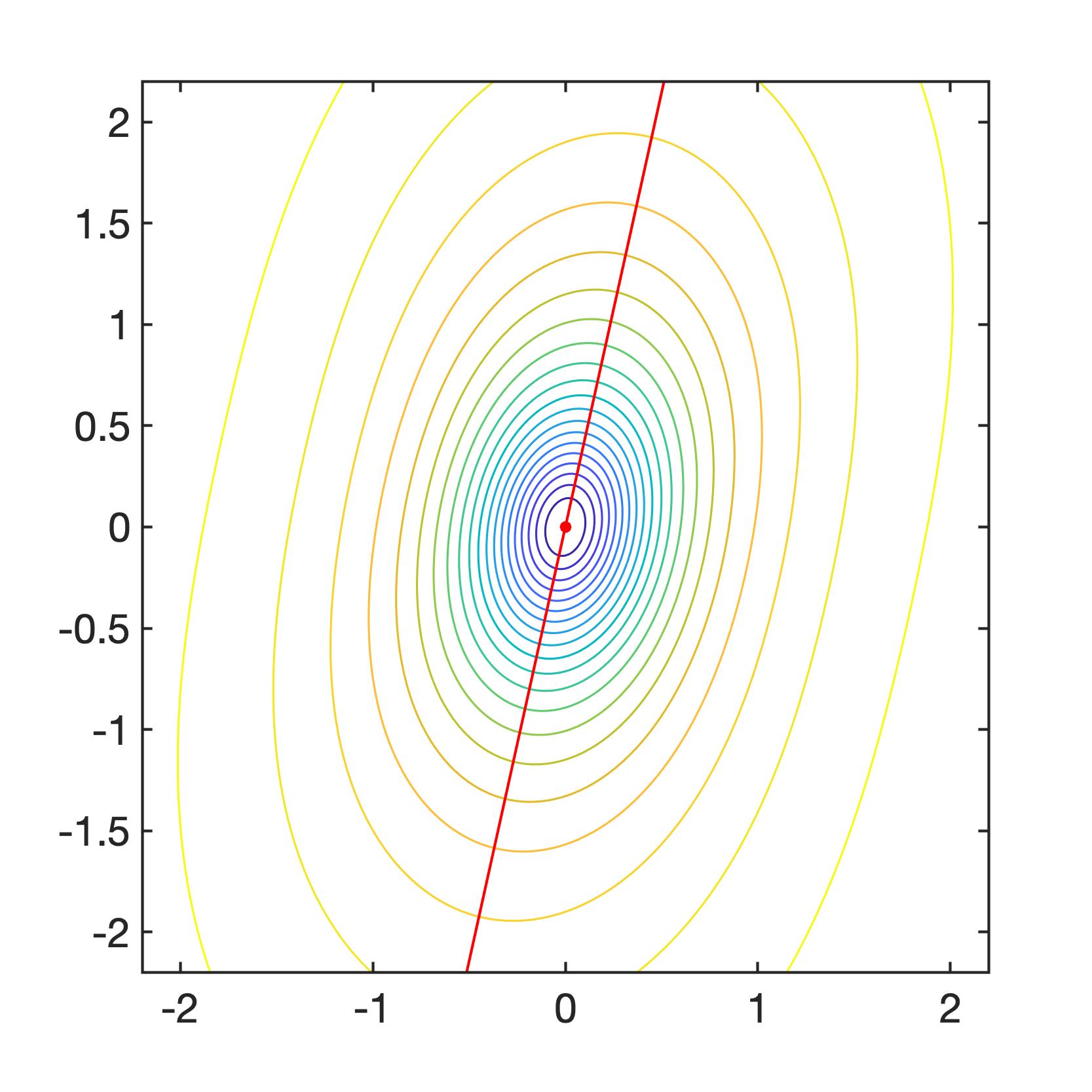}%
    \includegraphics[height=0.31\textwidth]{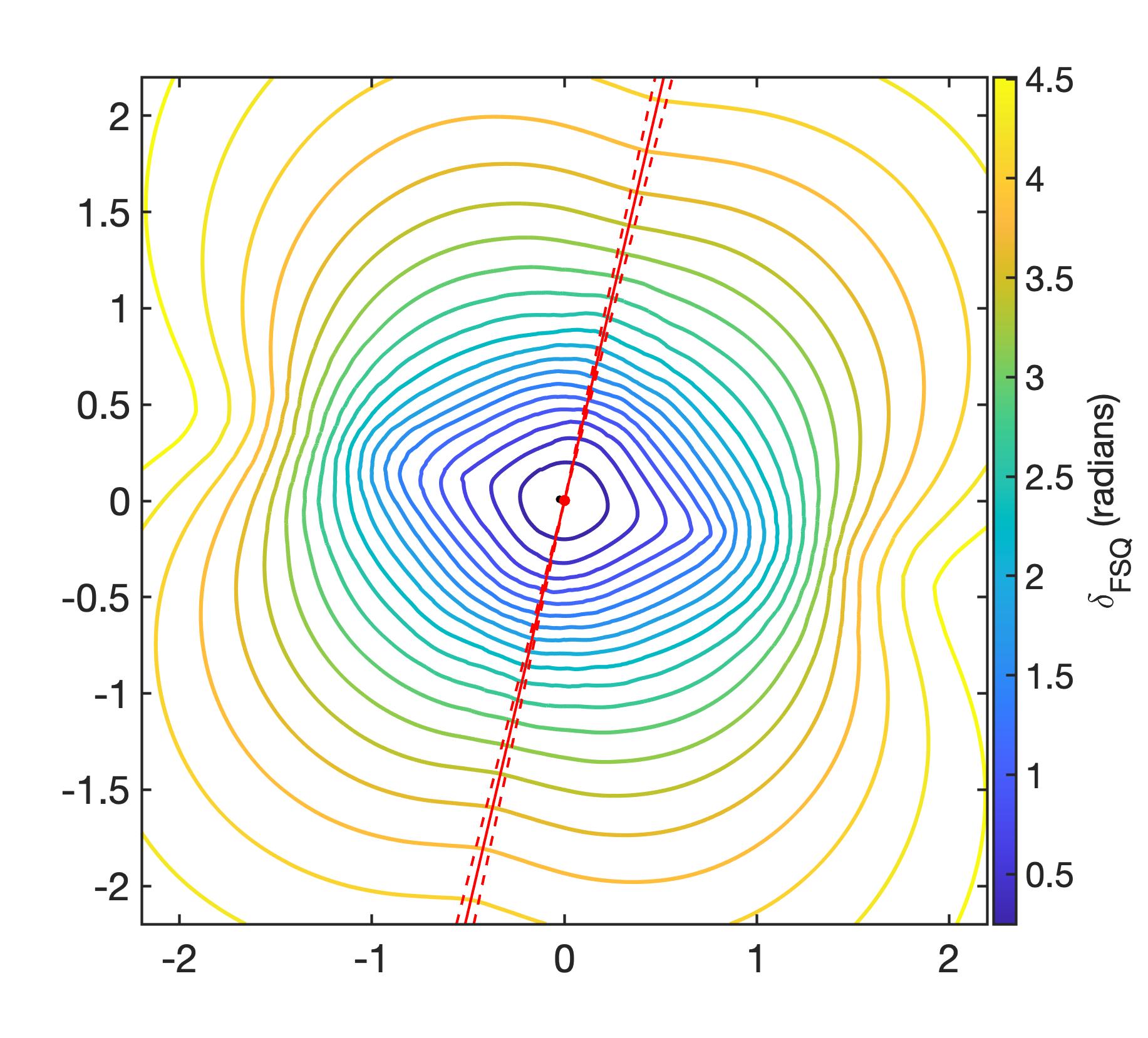}\\
    \includegraphics[height=0.31\textwidth]{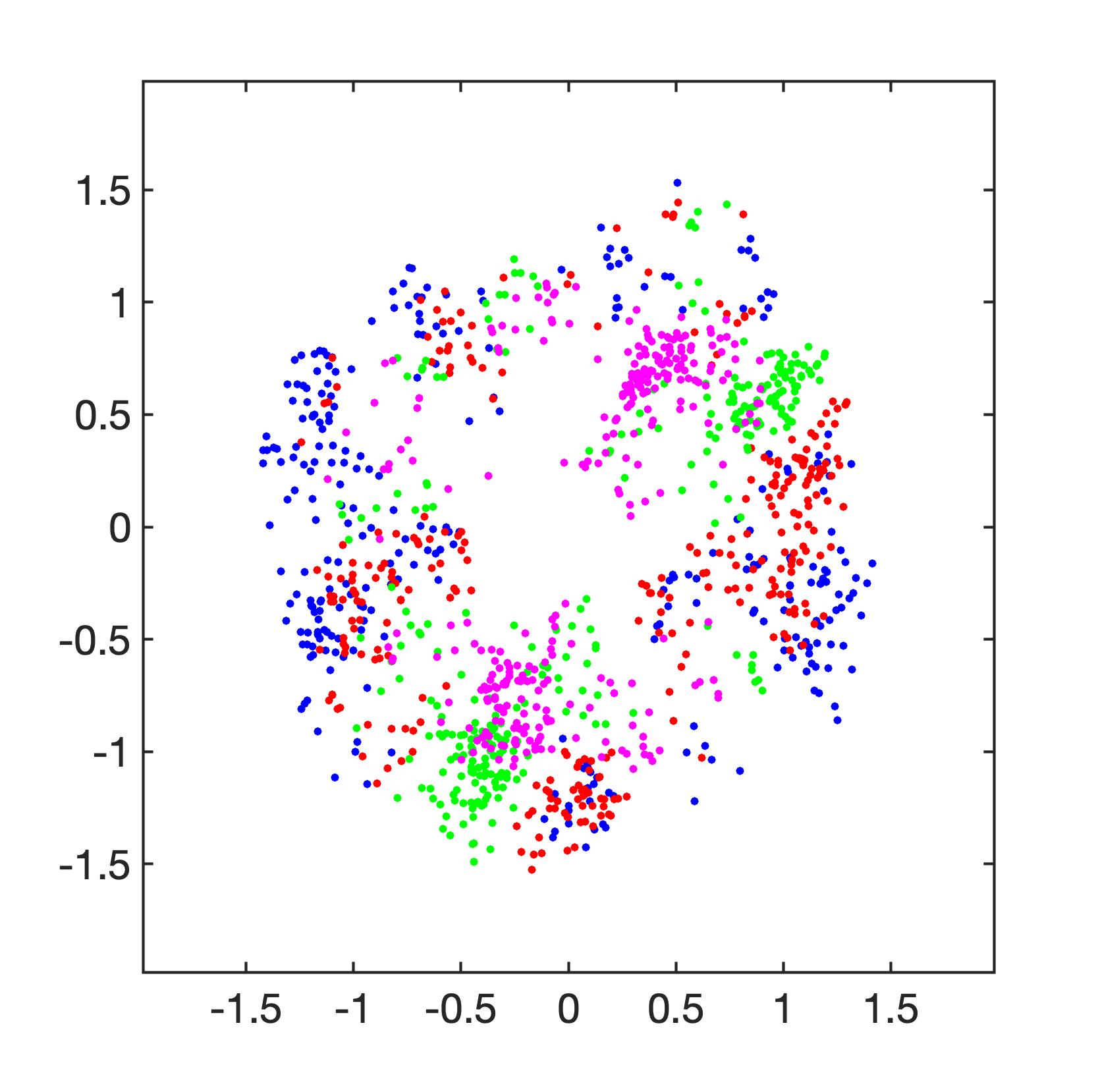}%
    \includegraphics[height=0.31\textwidth]{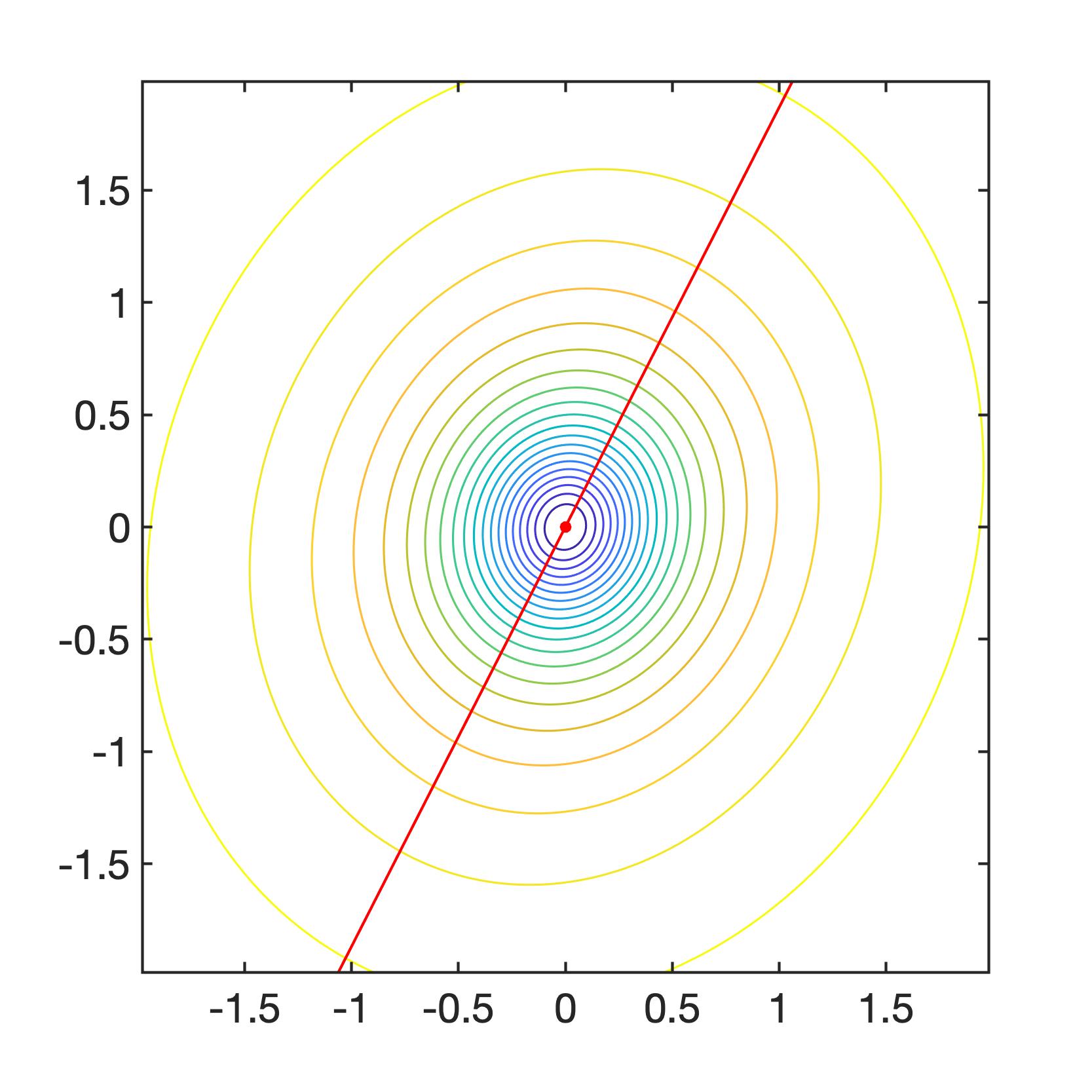}%
    \includegraphics[height=0.31\textwidth]{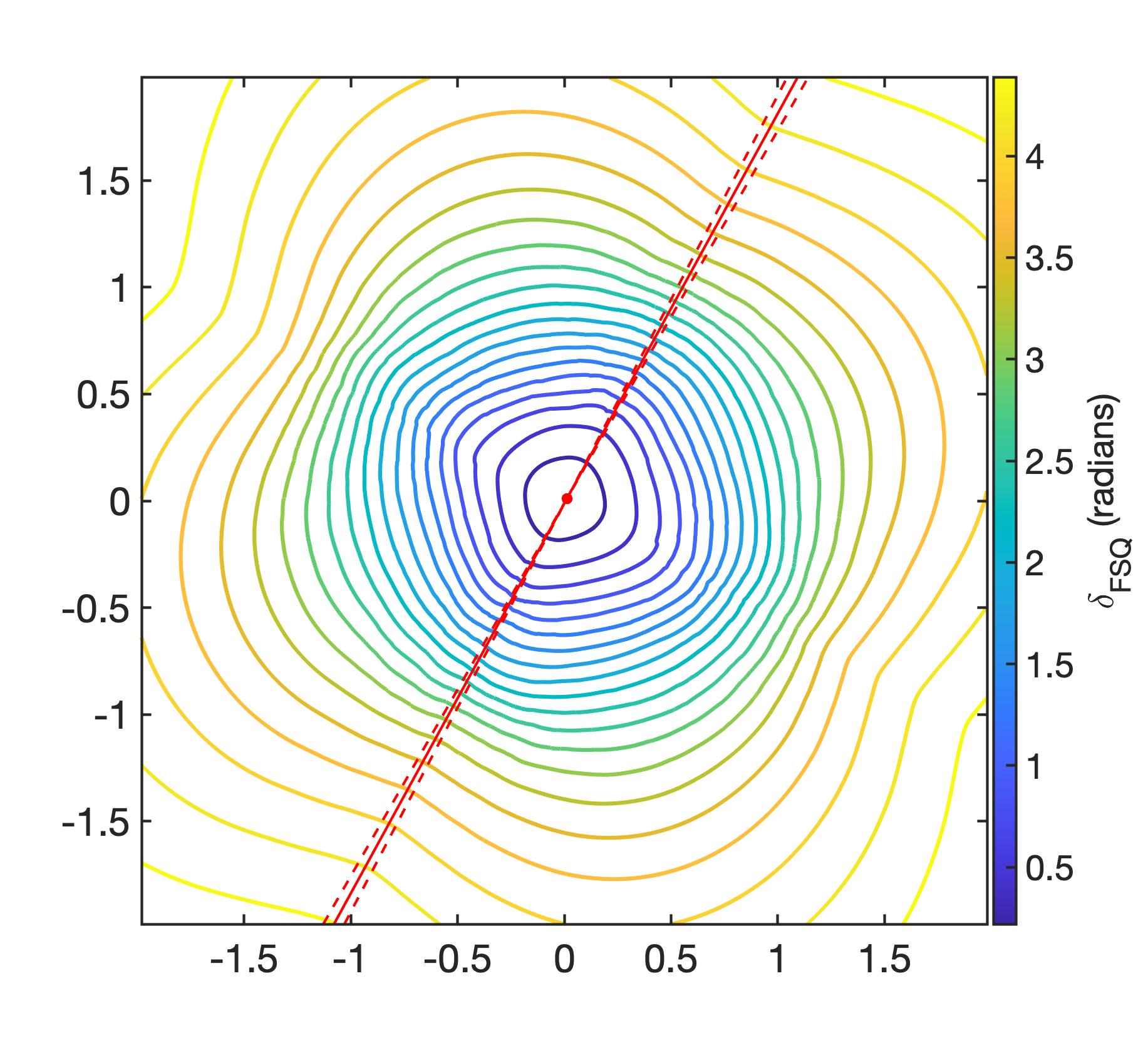}\\
    \includegraphics[height=0.31\textwidth]{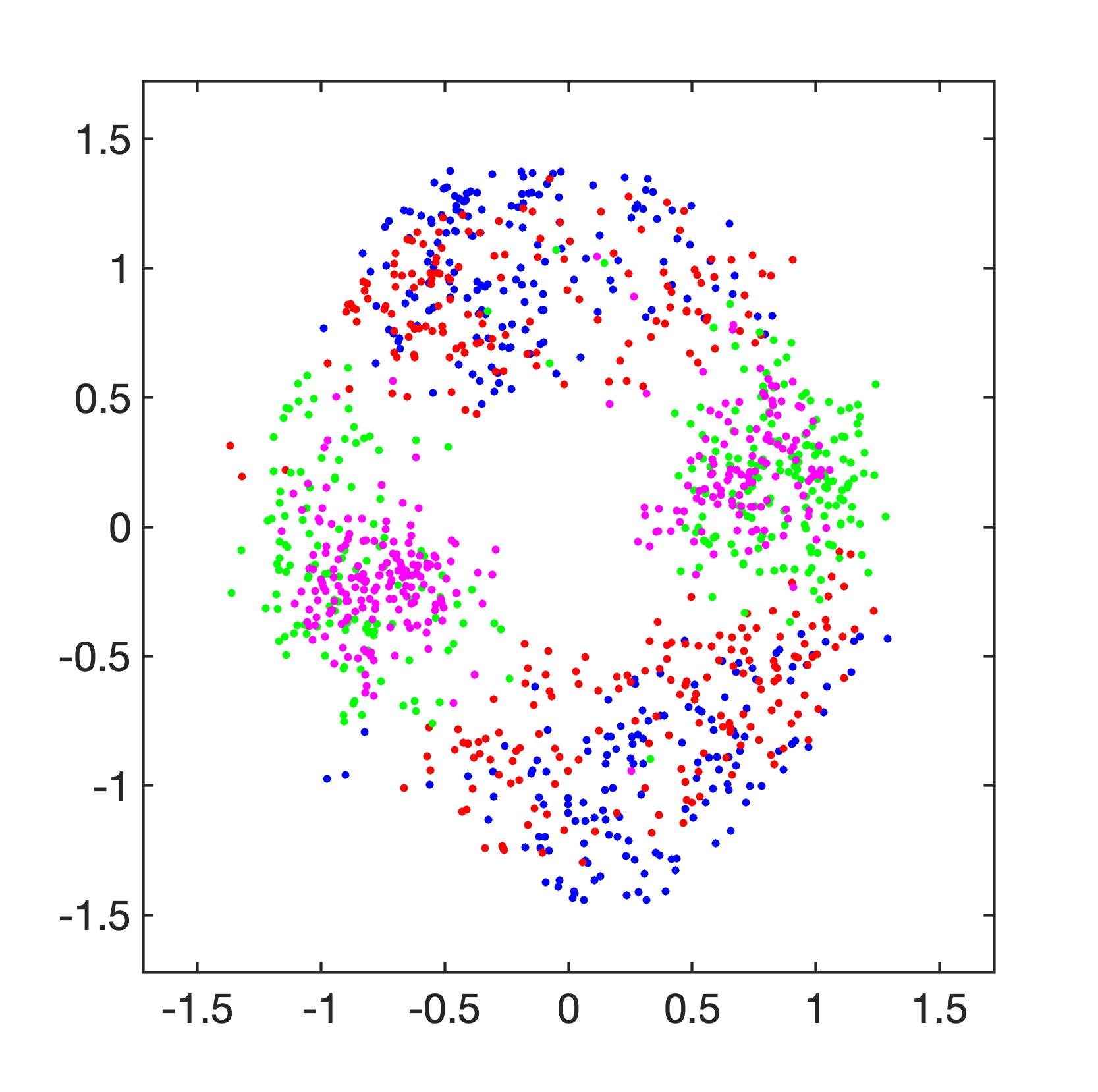}%
    \includegraphics[height=0.31\textwidth]{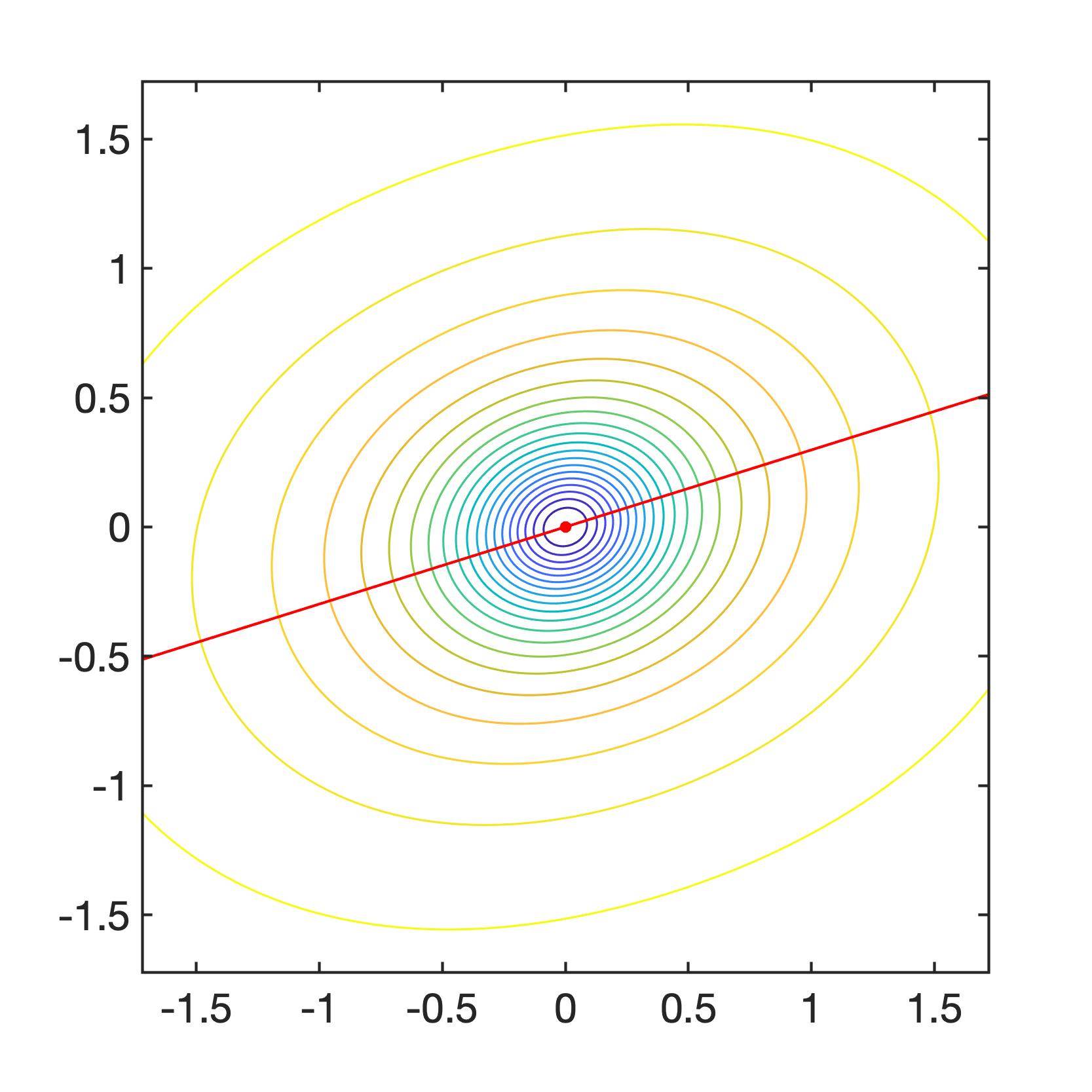}%
    \includegraphics[height=0.31\textwidth]{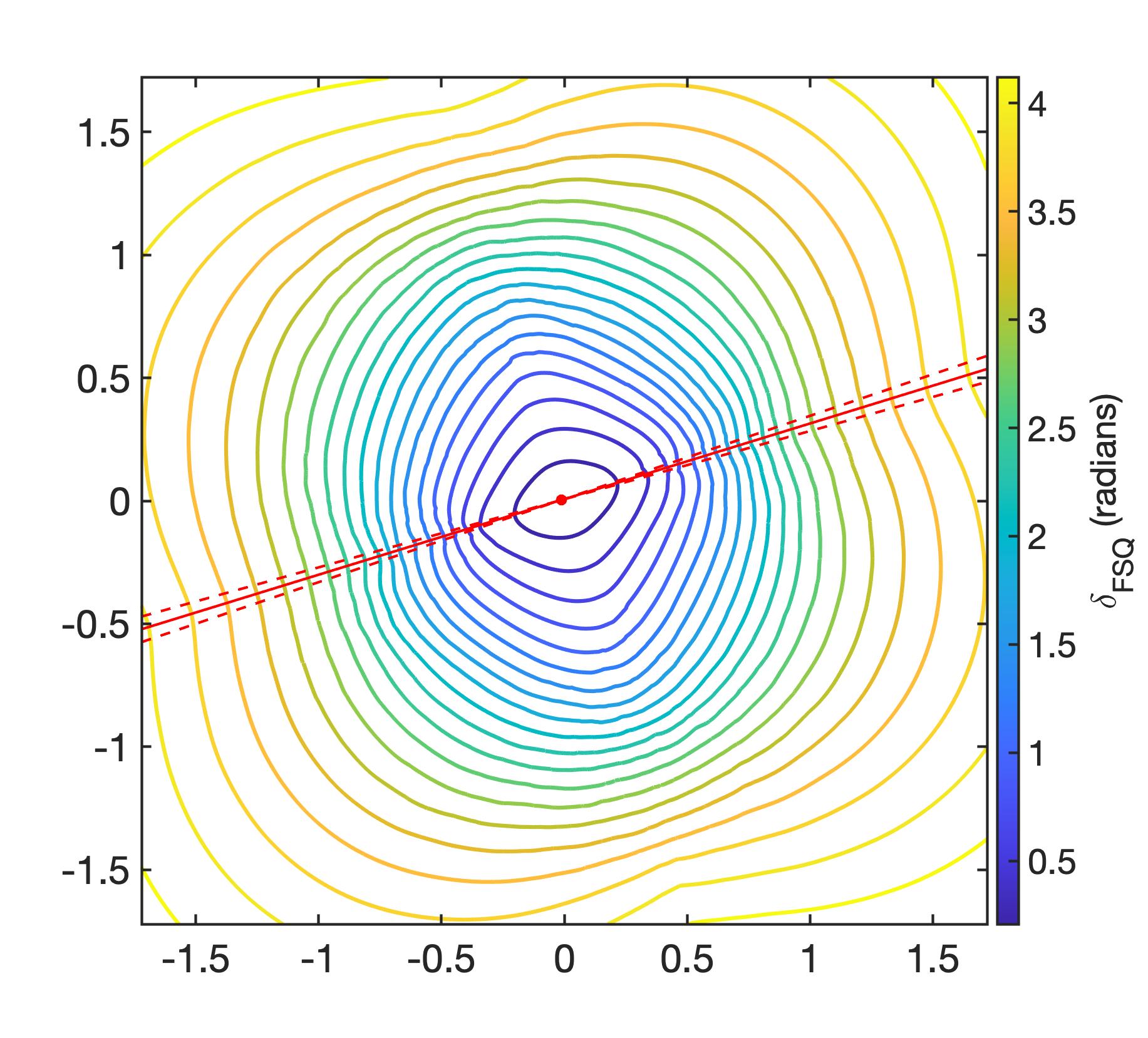}
    \caption{Four examples of mock clusters without substructure. Spatial coordinates measured in units of Einstein radii from the cluster center. See Section~\ref{sec:mock} for the definition of the Einstein radius. {\it Left:} Quad images produced by 300 sources, colored by arrival order: blue$\to$red$\to$green$\to$magenta. {\it Middle:} \llrw{linearly spaced} contours show the mass distribution. The red line is the true position angle of ellipticity, $\theta_T$; see Section \ref{sec:smooth}. The red dot is the true center, $(x_T,y_T)$.  {\it Right:} Visualization of the three estimated properties of the cluster. The contours are those of $\delta_{\rm FSQ}$, and the location of the minima corresponds to the estimated cluster center ($x_Q,y_Q$), represented by the red dot. The red solid line shows estimated $\theta_Q$. \lasko{Black dots show the quad-estimated center for one of $30$ subsets of $10$ quads each, which is used to give an estimate on the error for $\theta_Q$, shown by the red dashed lines (Note that black dots are nearly coincident with the red dot)}. \llrw{The contours demonstrate that the FSQ-based method is a robust way to locate the center.}}
    \label{fig:smooth}
\end{figure*}

\begin{figure*}
    \centering
    \includegraphics[height=0.31\textwidth]{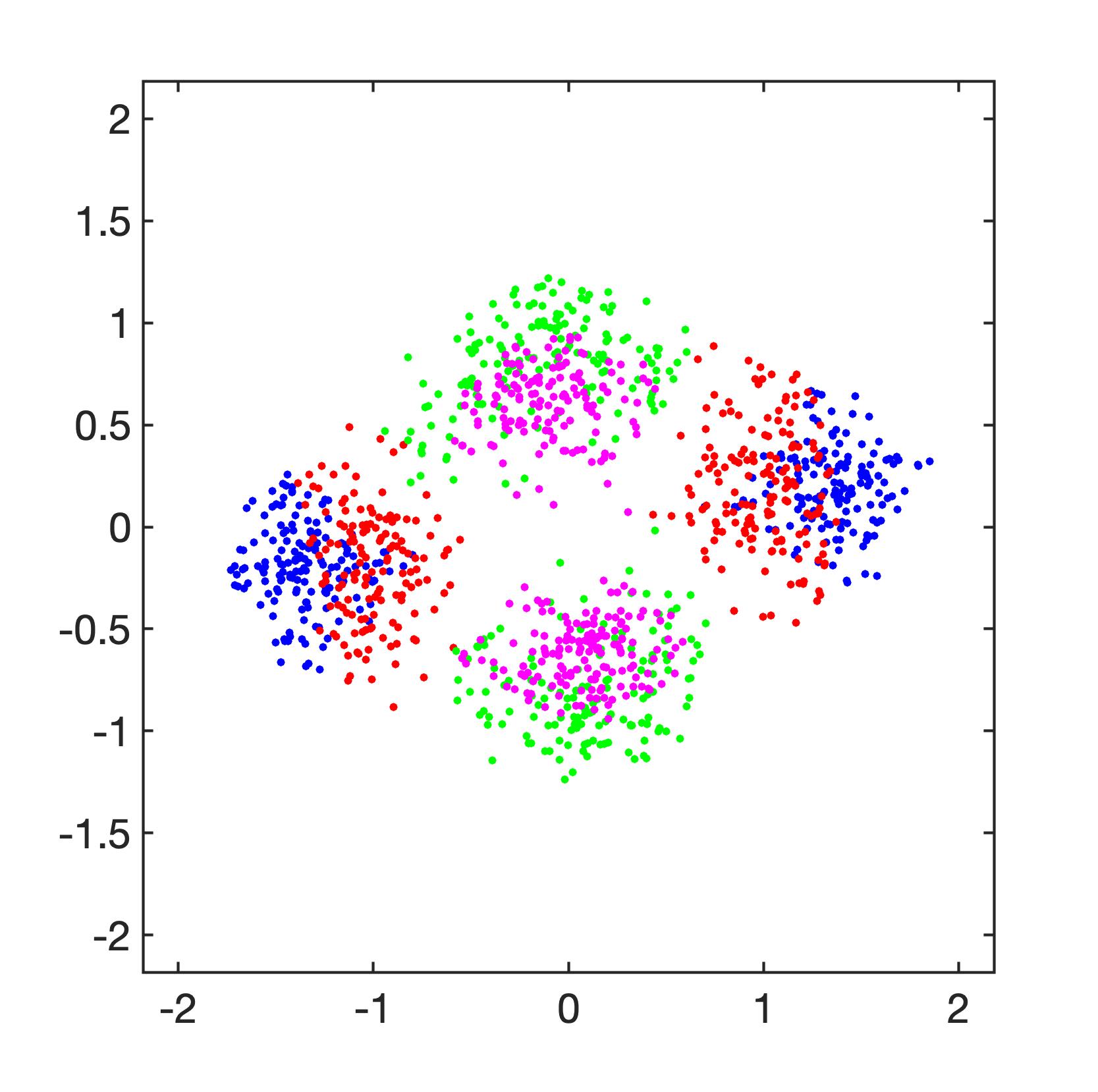}%
    \includegraphics[height=0.31\textwidth]{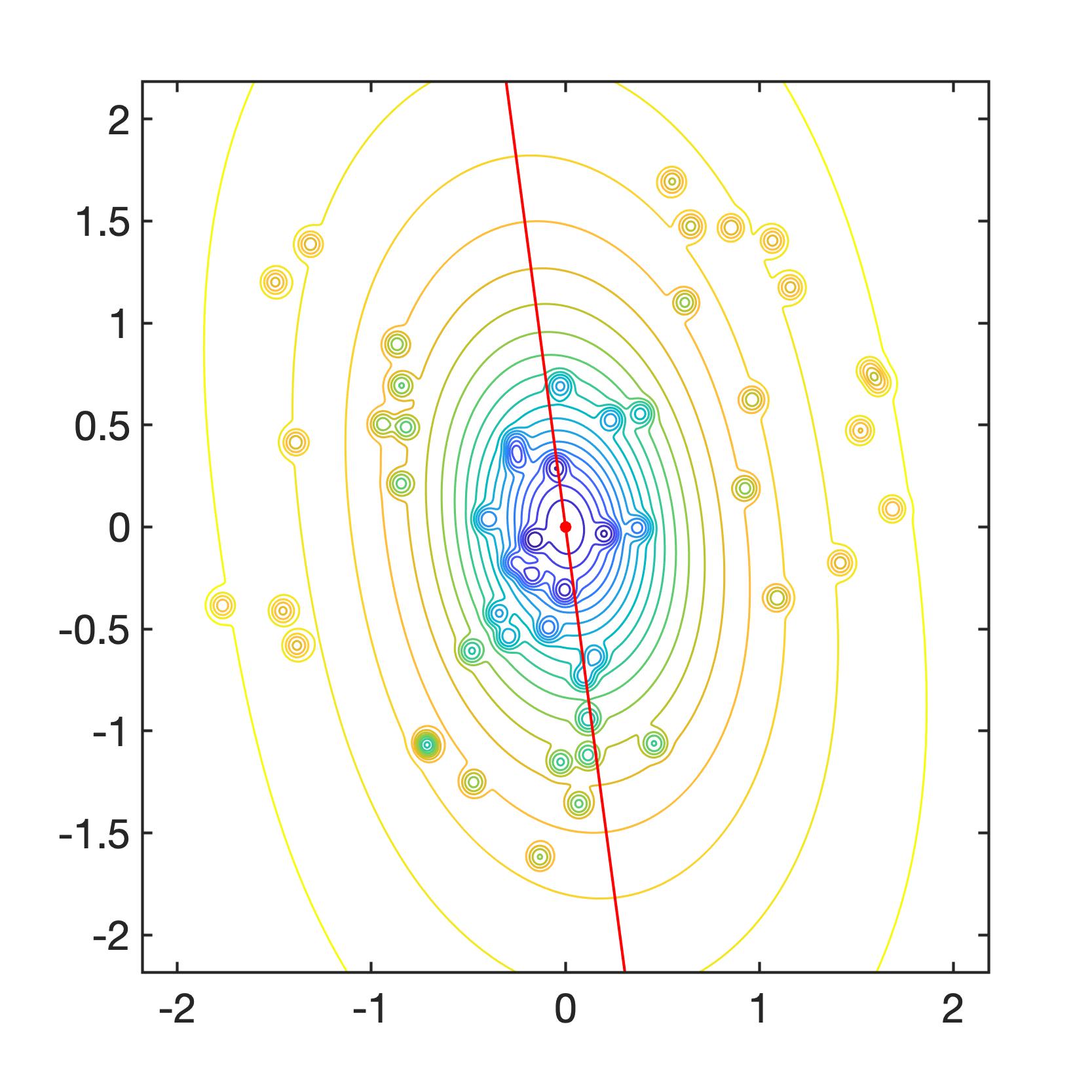}%
    \includegraphics[height=0.31\textwidth]{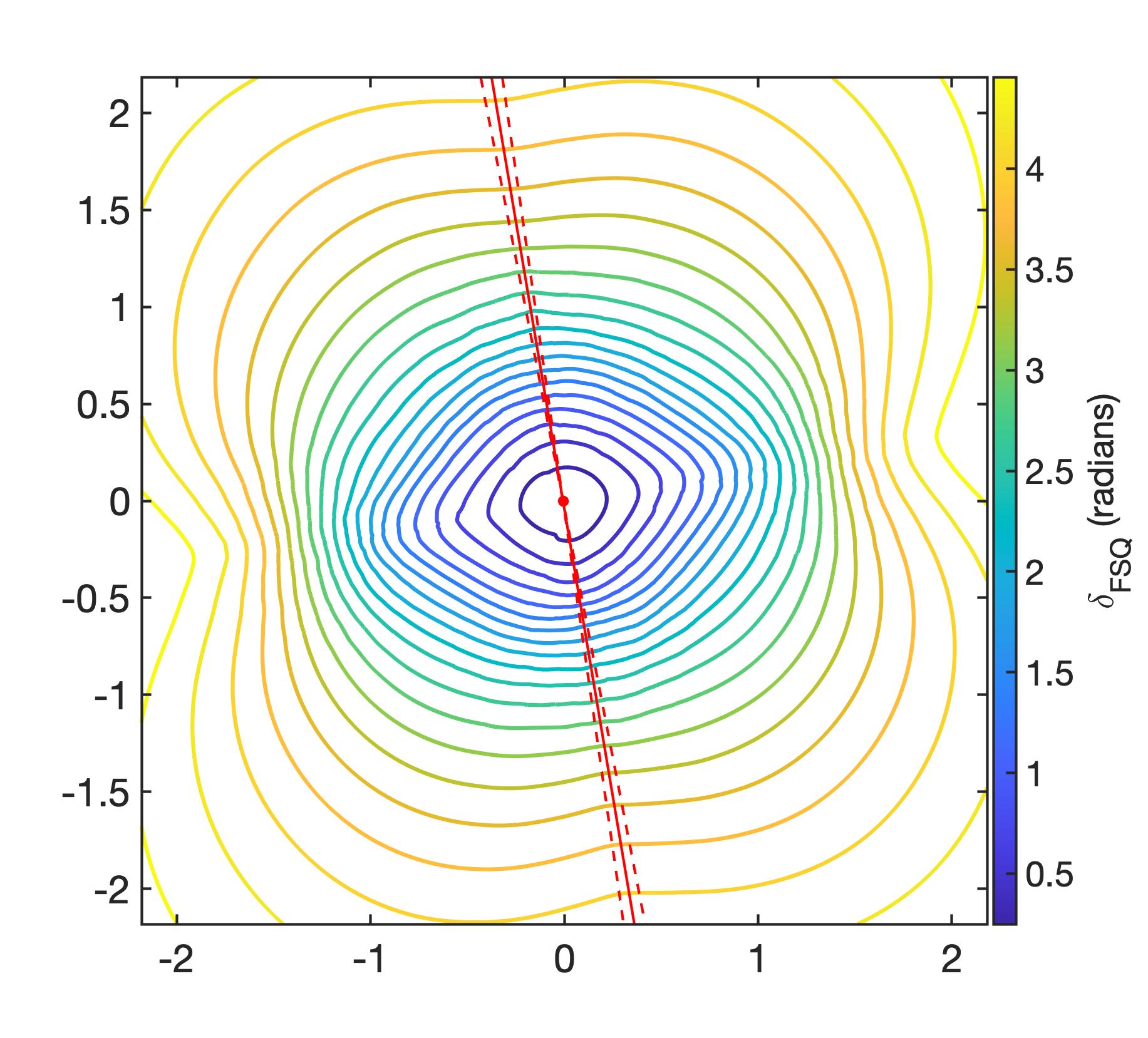}\\
    \includegraphics[height=0.31\textwidth]{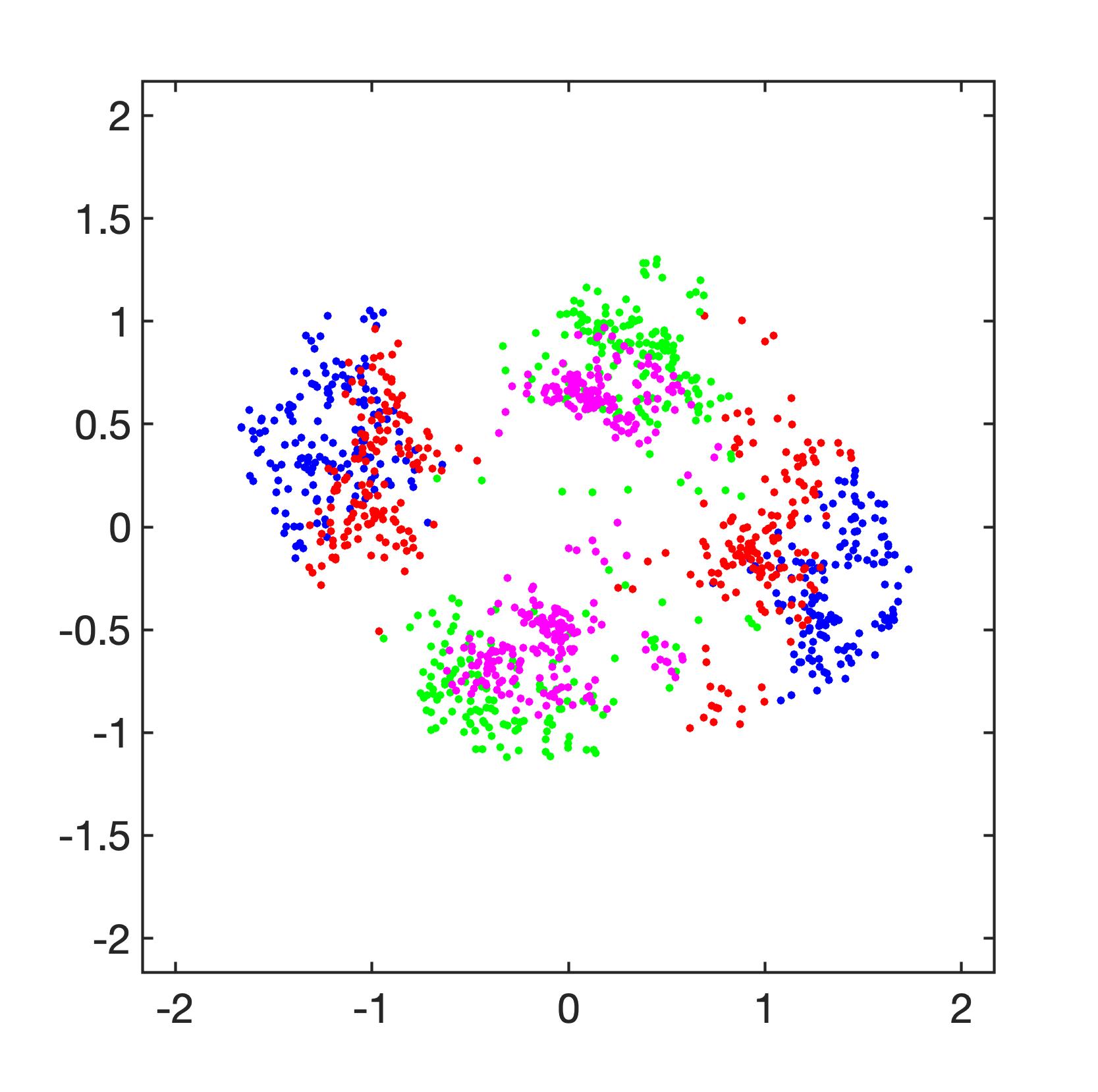}%
    \includegraphics[height=0.31\textwidth]{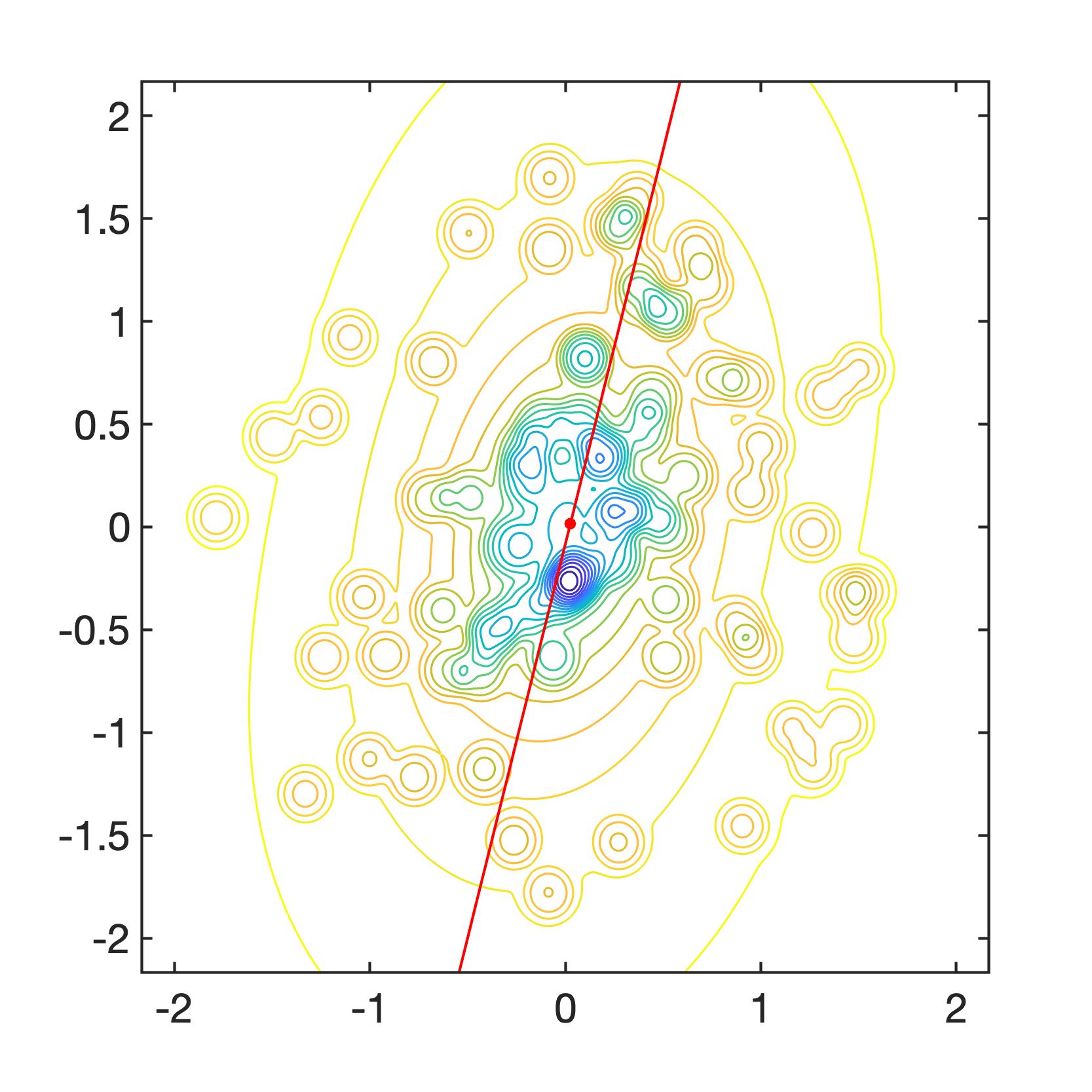}%
    \includegraphics[height=0.31\textwidth]{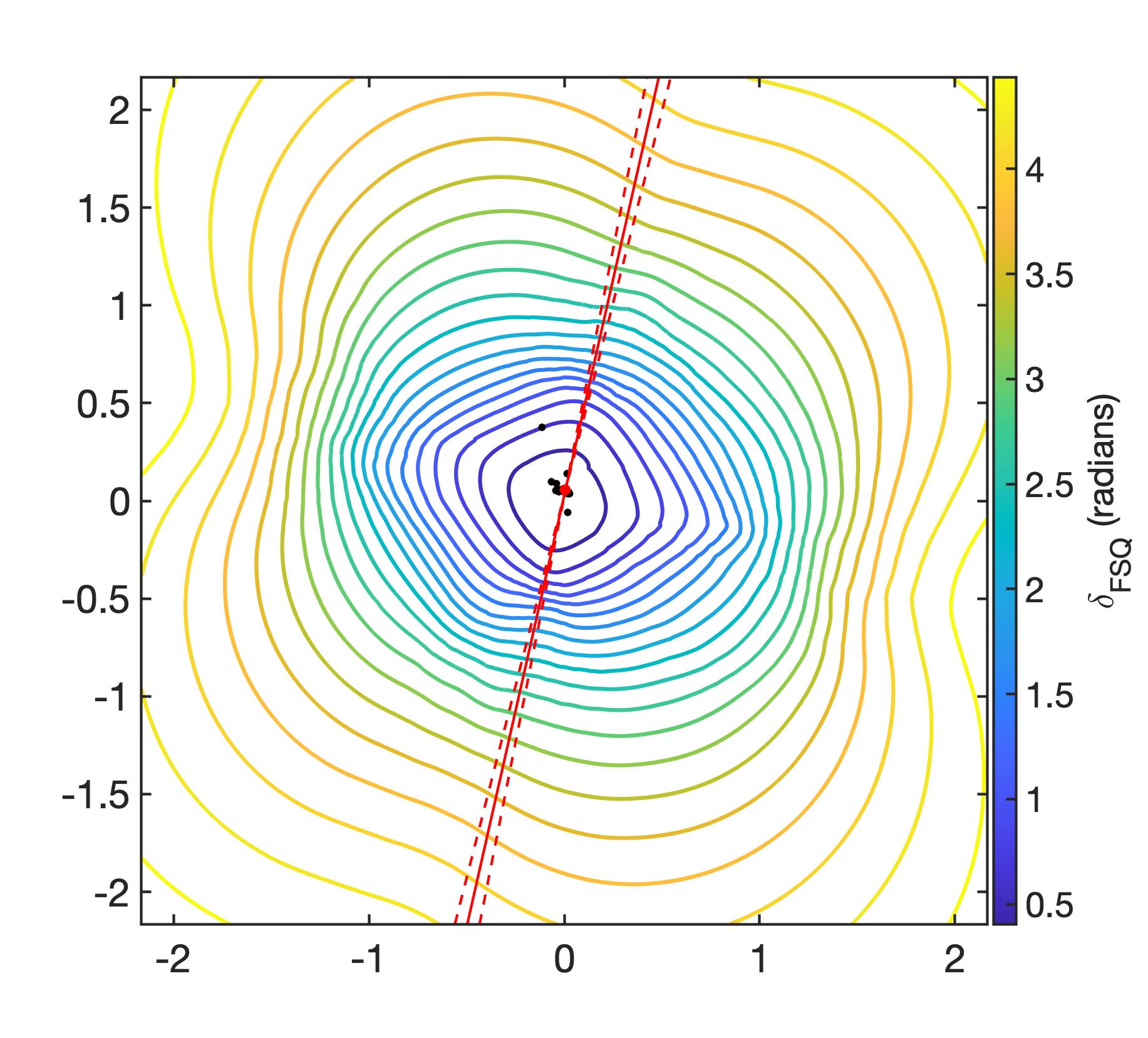}\\
    \includegraphics[height=0.31\textwidth]{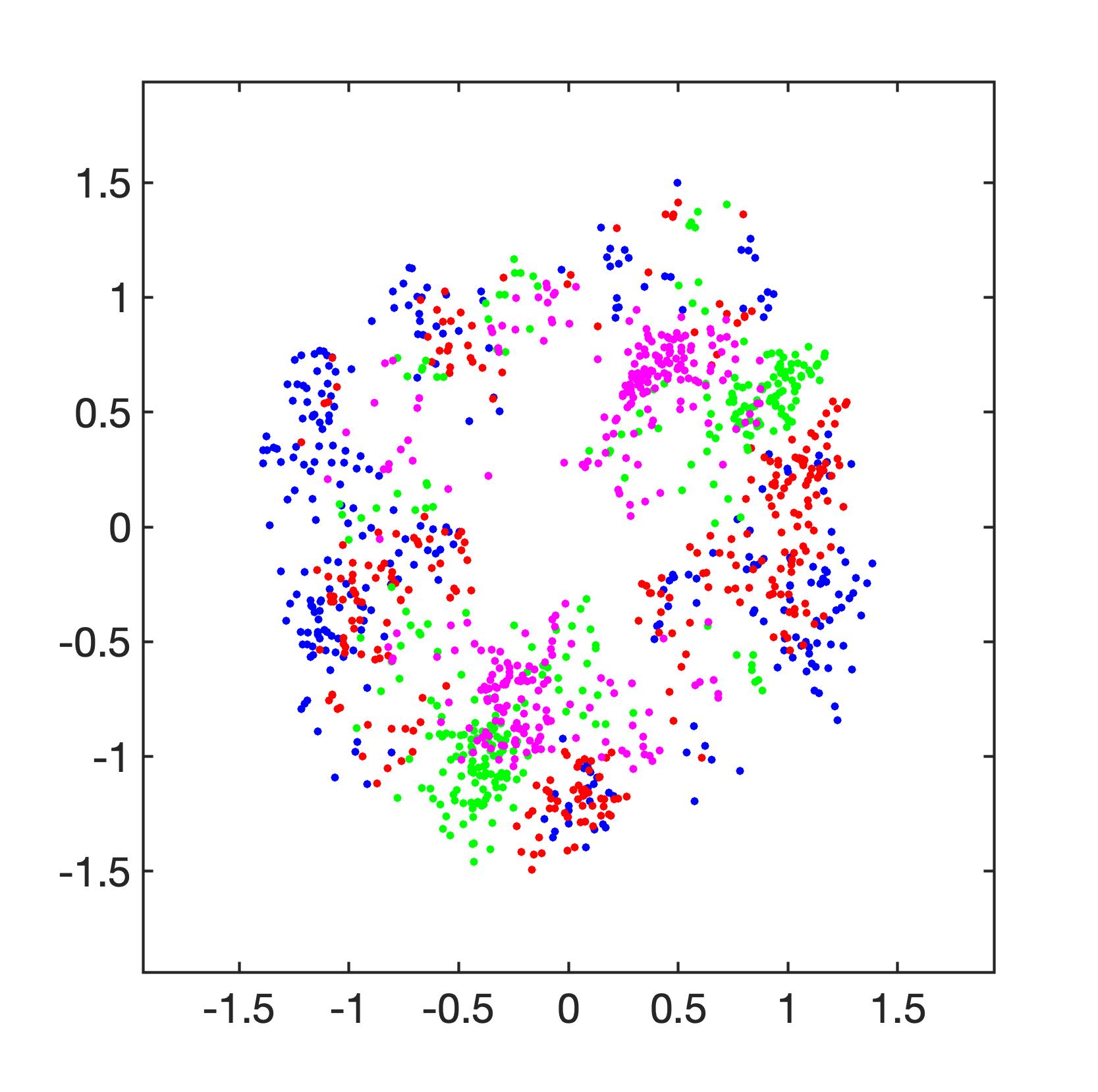}%
    \includegraphics[height=0.31\textwidth]{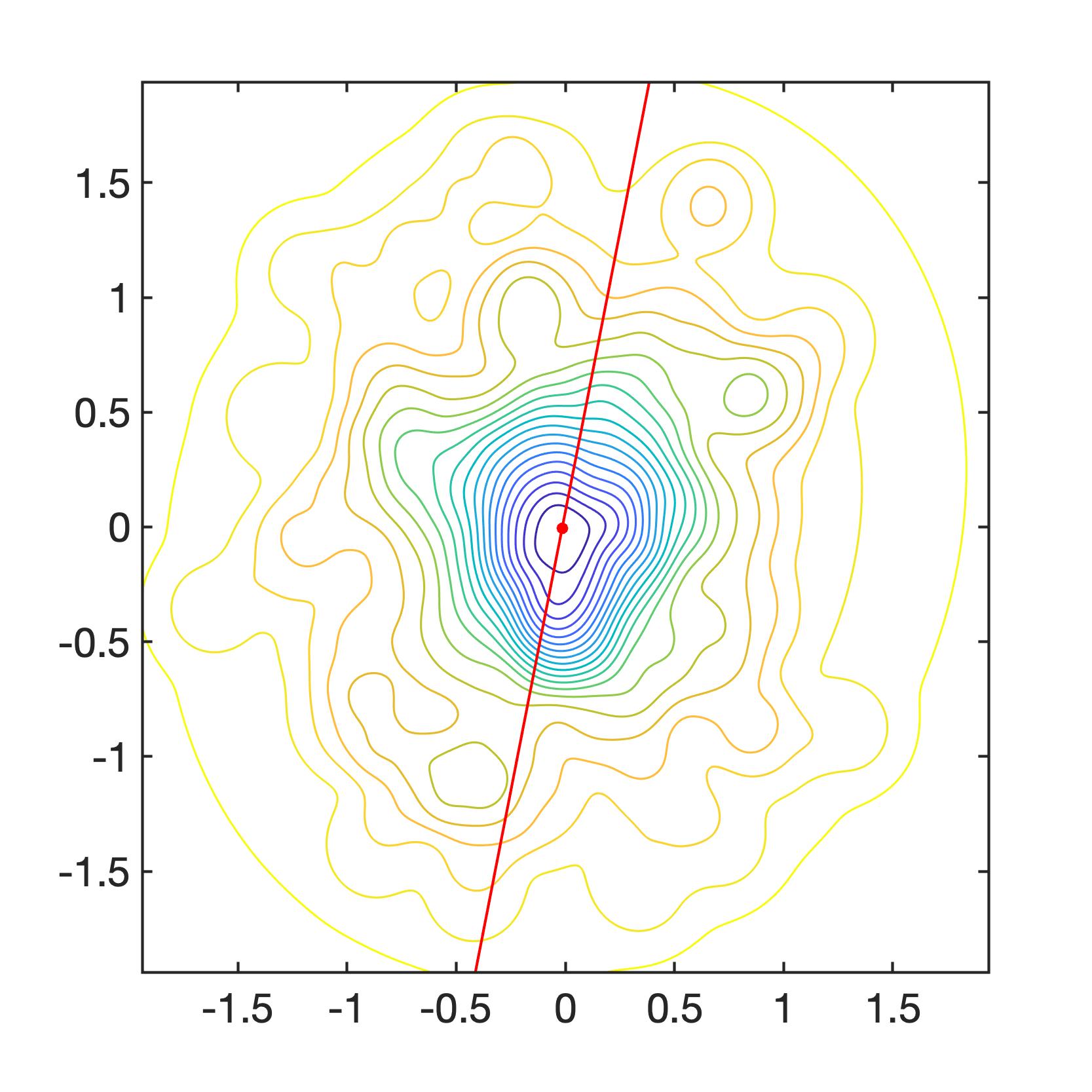}%
    \includegraphics[height=0.31\textwidth]{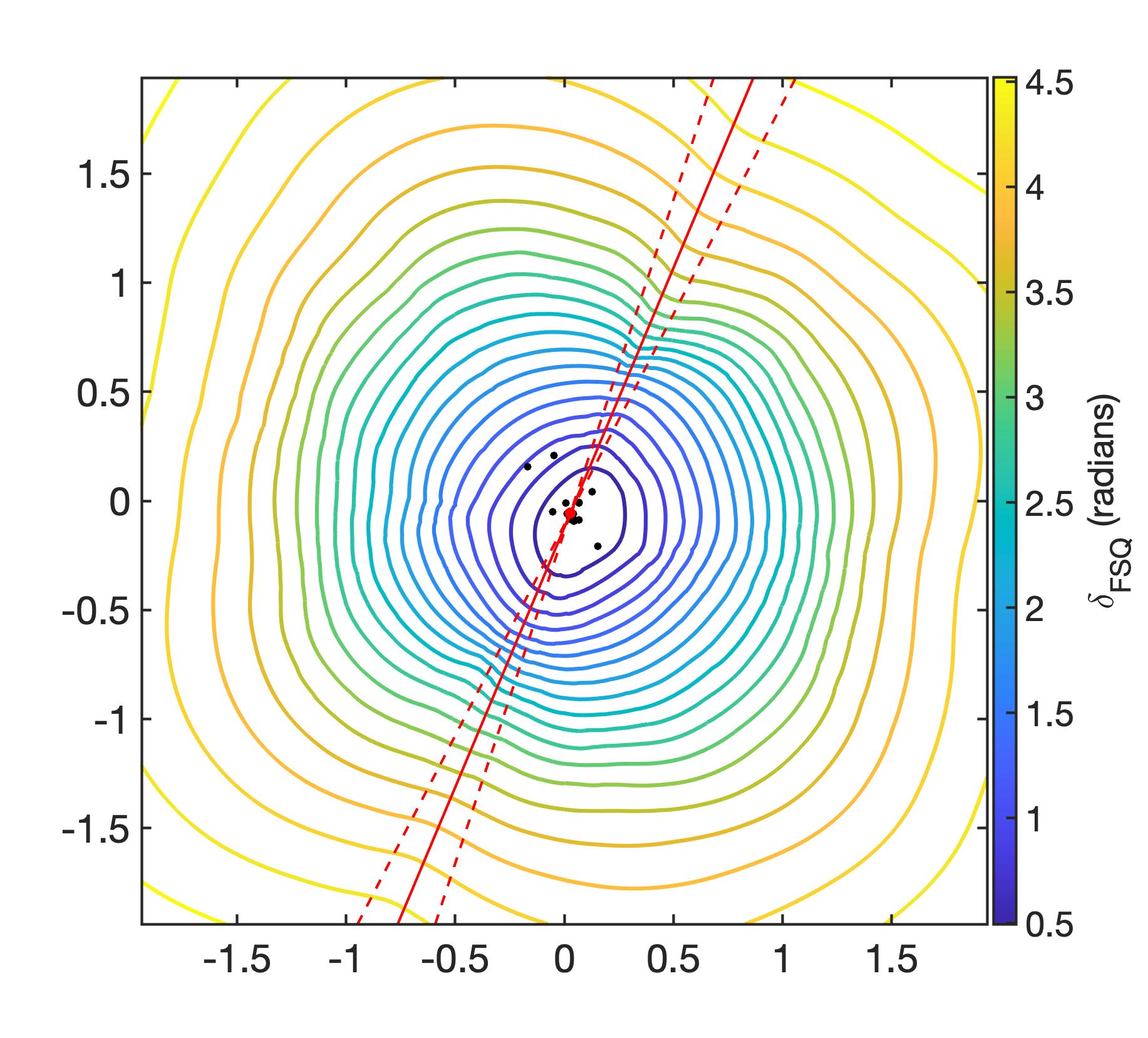}\\
    \includegraphics[height=0.31\textwidth]{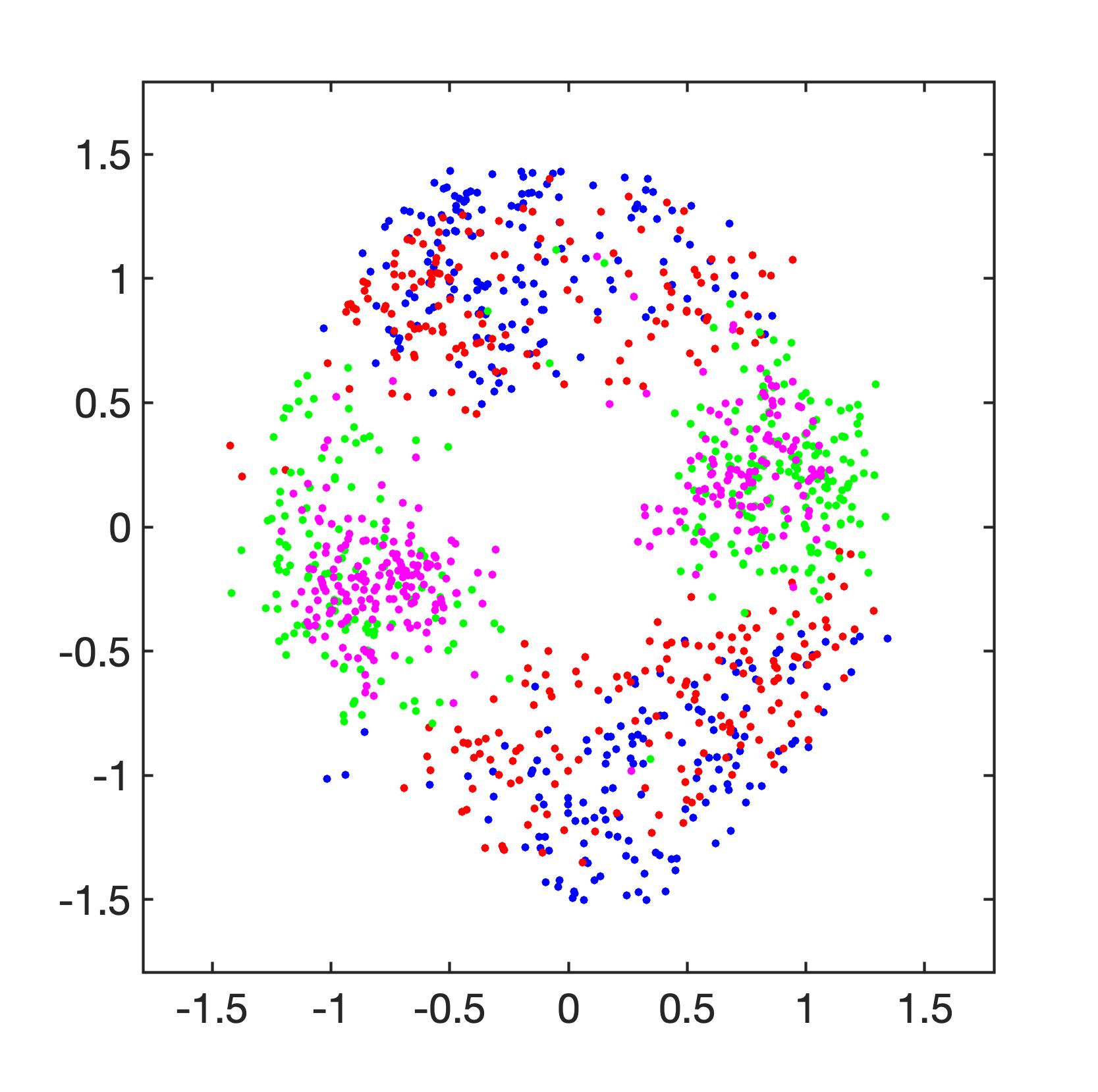}%
    \includegraphics[height=0.31\textwidth]{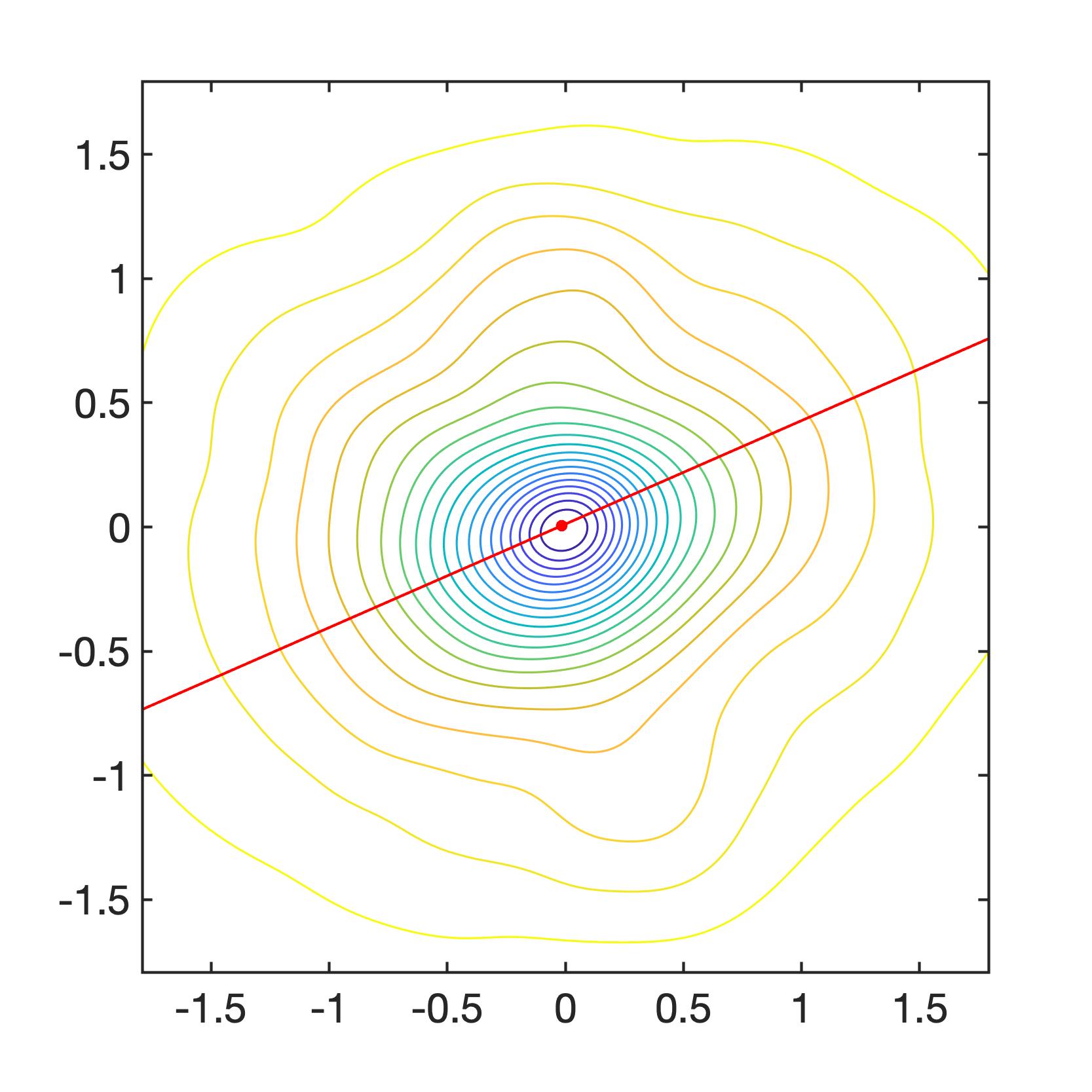}%
    \includegraphics[height=0.31\textwidth]{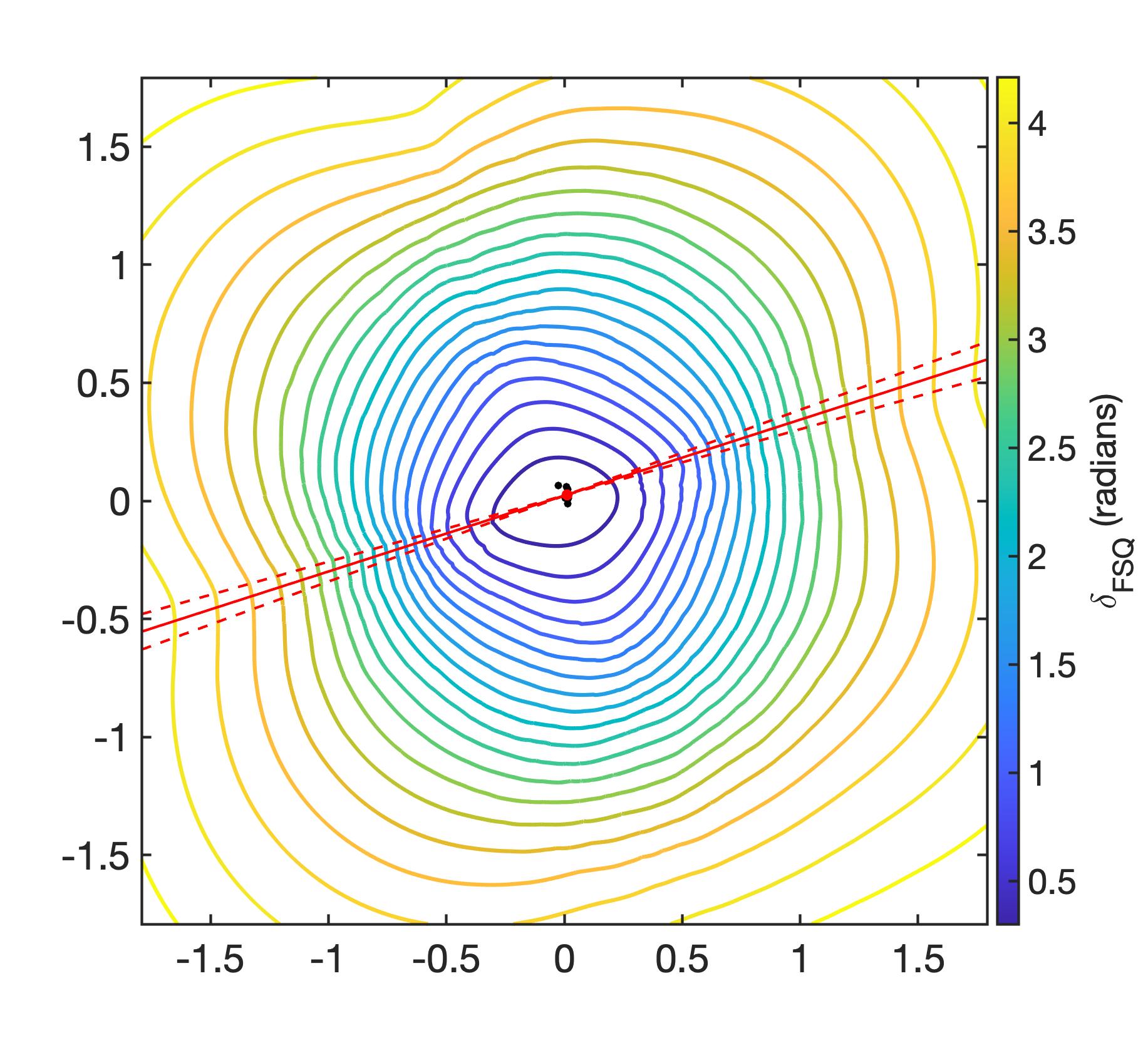}
    \caption{Similar to Fig.~\ref{fig:smooth}, but clusters have imposed substructure over their smooth counterparts. See Section~\ref{sec:substruc} for details.}
    \label{fig:structured}
\end{figure*}

\section{Comparing true and quad-estimated properties of clusters}

\subsection{Smooth elliptical clusters}\label{sec:smooth}

We first carry out a test of our proposed method: we apply it to smooth elliptical clusters, with no substructure. We use 100 mock clusters; Figure \ref{fig:smooth} shows a representative subset of four. The contours in the left and middle columns are contours of equal surface mass density. The colored points in the left panel show images from 300 quads: blue, red, green and magenta points represent 1st, 2nd, 3rd, and 4th arriving images, respectively. The wide radial distribution of images is the consequence of a wide range of source redshifts used. The red solid dot and red line in the middle panels shows the true center and position angle of the cluster.

In the right panels of Figure \ref{fig:smooth} the red solid line shows our quad-estimated position angle using all 300 quads per cluster, while the dashed red lines show its rms dispersion, which was calculated from 30 subsets of 10 quads, with every quad appearing in exactly one subset. The black points in the right panel show centers estimated from these 30 independent subsets, as described in Section~\ref{sec:est_cen}. In each of the 4 clusters shown in \ref{fig:smooth}, the black points and the red point nearly coincide, implying that even 10 quads are enough to locate the center. 

The contours in the right panels are contours of the rms deviation of quads from the FSQ. The minima of these contours indicates the quad-estimated center, $(x_Q,y_Q)$, and the rms of deviations from the FSQ at this location is referred to as $\delta_{\rm FSQ}$. \llrw{We show these contours in the lens plane to demonstrate that the FSQ-based center finding method has only one minimum, and is therefore robust.}

\begin{figure*}[!bh]
    \centering
    \includegraphics[width=.48\textwidth]{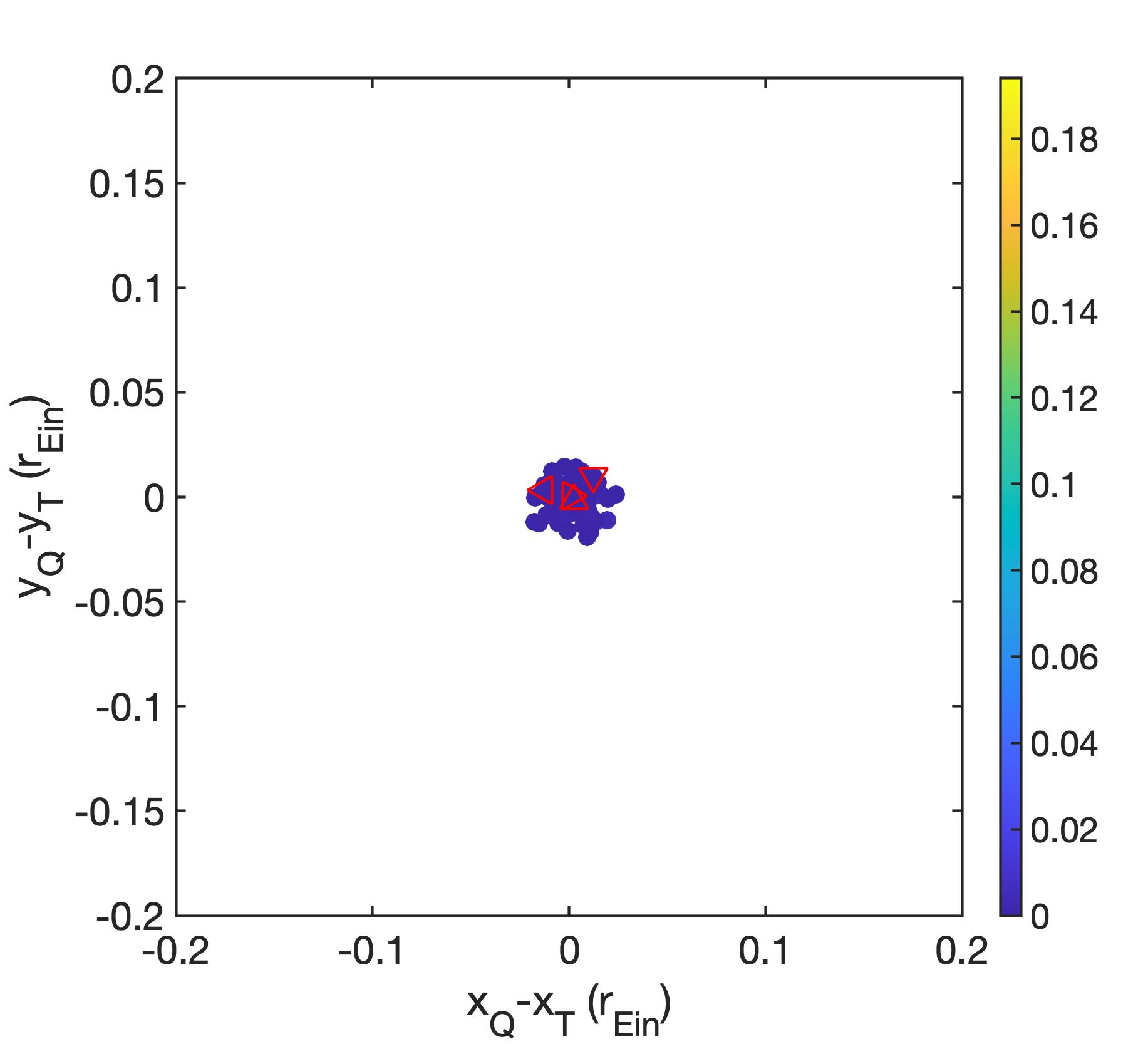}
    \includegraphics[width=.48\textwidth]{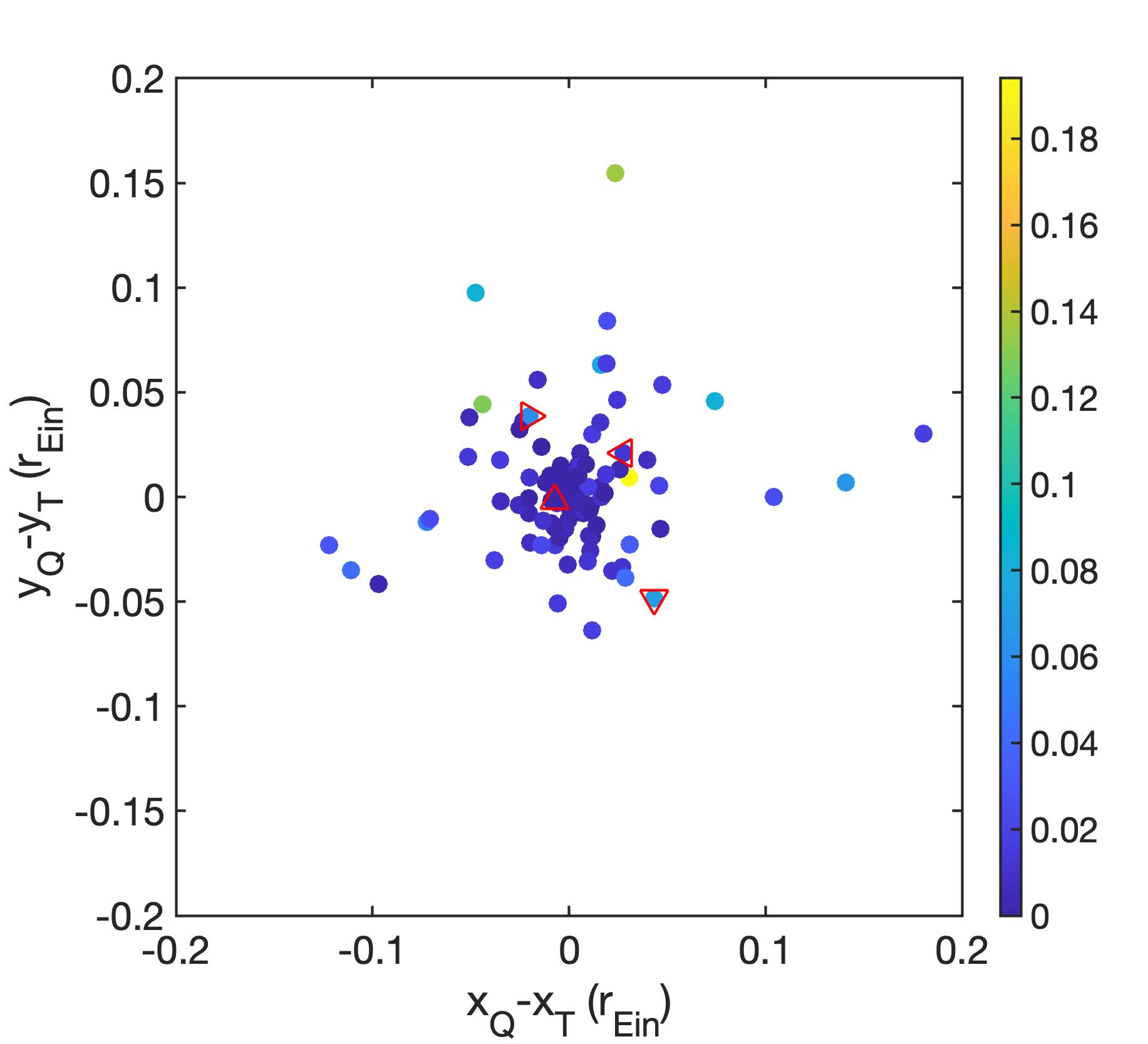}
    \caption{Comparison of the true, $(x_T,y_T)$, vs. estimated, $(x_Q,y_Q)$, cluster centers for all $100$ smooth purely elliptical mock clusters ({\it left}), and $100$ substructured mock clusters ({\it right}). 
    (See Sections~\ref{sec:true_cen} and \ref{sec:est_cen}.)
    Color indicates the uncertainty in the quad-estimated center, in units of cluster Einstein radius, obtained using 30 random subsets of 10 quads. The points marked by red triangles correspond to the clusters shown in Figures \ref{fig:smooth} and \ref{fig:structured}: triangle pointing up is the first row, pointing right is the second row, pointing down is the third row, and pointing left is the fourth row. \llrw{To facilitate comparison, the scale of the two panels is the same.}
    %Notice that the estimation of the centers and their uncertainties is much more accurate with smooth clusters than clusters with substructure.
    }
    \label{fig:deltaCenters}
\end{figure*}

\begin{figure*}
    \centering
    \includegraphics[width=.48\textwidth]{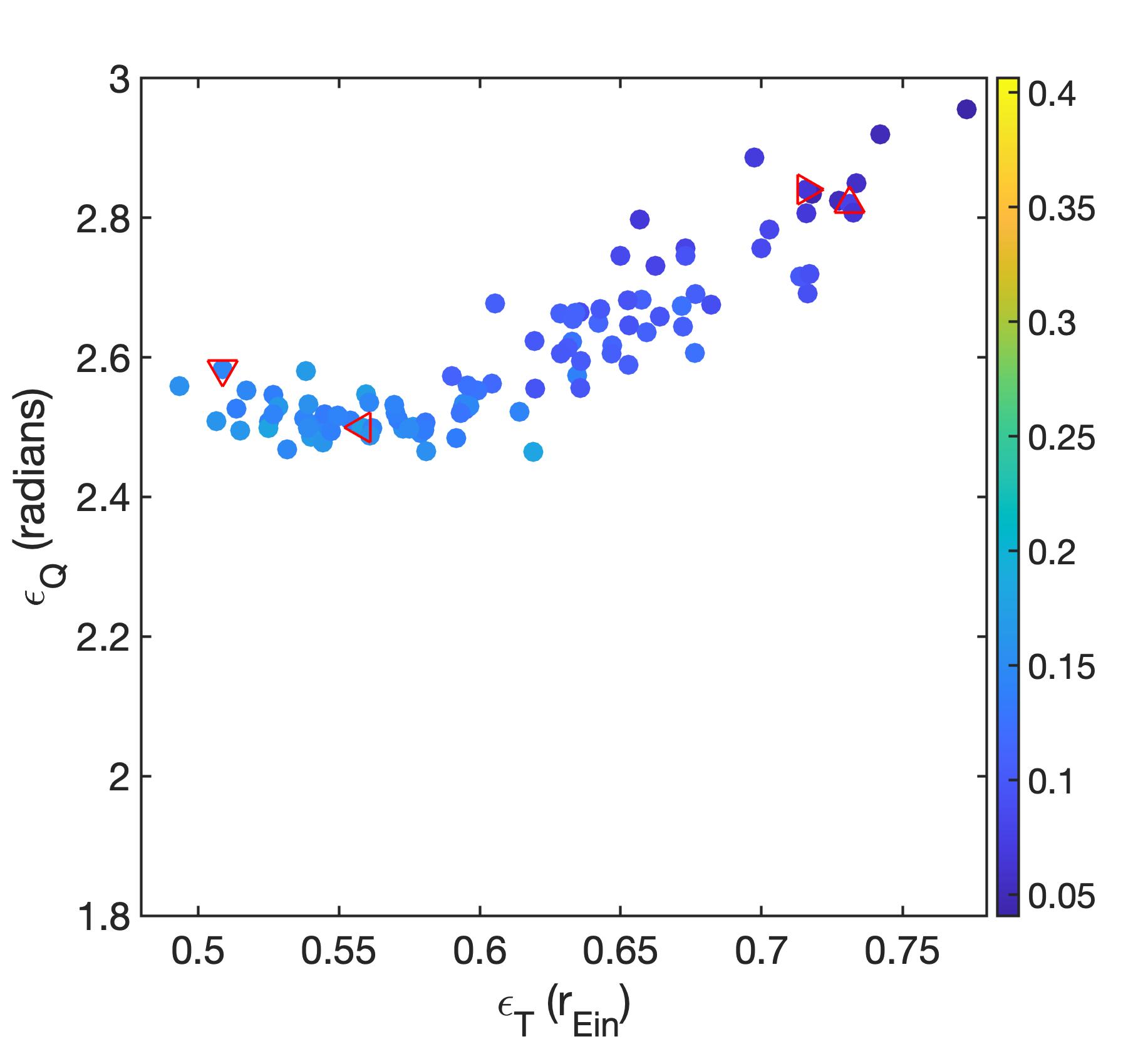}
    \includegraphics[width=.48\textwidth]{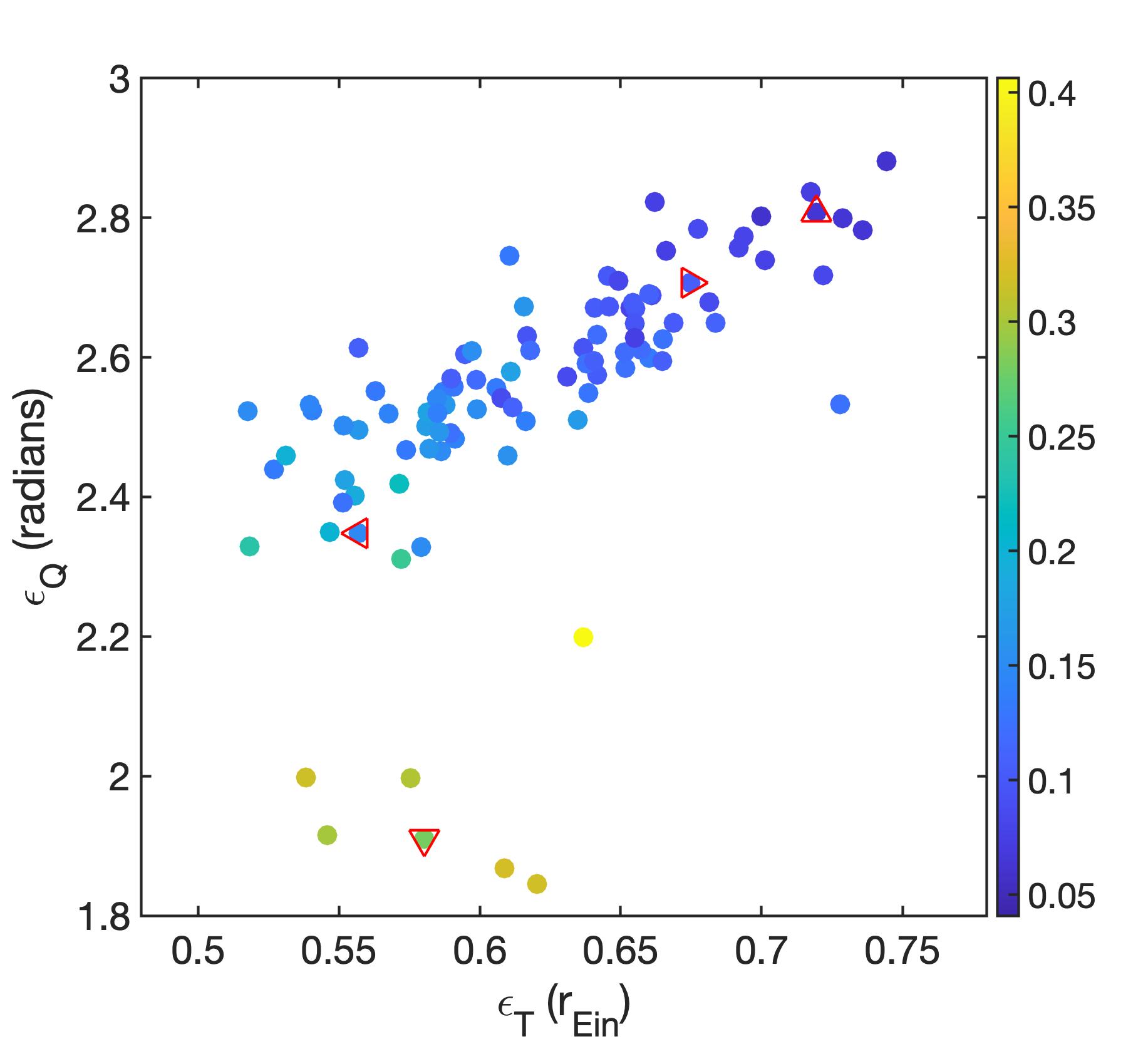}
    \caption{Comparison of true ellipticity, $\epsilon_T$, vs. quad-estimated elongation, $\epsilon_Q$, for all $100$ smooth mock clusters ({\it left}), and $100$ substructured mock clusters ({\it right}). True ellipticity is measured using Eq.~\ref{eq:ellipticity} and normalized by the respective cluster's Einstein radius. Estimated elongation is $\epsilon_Q=\langle\theta_{12}\rangle$.  Though these two measures, $\epsilon_T$ and $\epsilon_Q$, are different, the tight relation in the left panel can be used to translate between the two measures. Most importantly, the two measures of ellipticity correlate, showing that quads contain ellipticity information. \lasko{Color indicates the uncertainty on $\epsilon_Q$ for each cluster}, measured in radians. The points marked by red triangles correspond to the clusters shown in Figures \ref{fig:smooth} and \ref{fig:structured}: triangle pointing up is the first row, pointing right is the second row, pointing down is the third row, and pointing left is the fourth row. See Sections~\ref{sec:smooth} and \ref{sec:substruc} for details.
    }
    \label{fig:ellipticity}
\end{figure*}

\begin{figure*}
    \centering
    %\vskip-2cm
    \includegraphics[trim={0cm 1cm 0cm 4cm},clip,width=.48\textwidth]{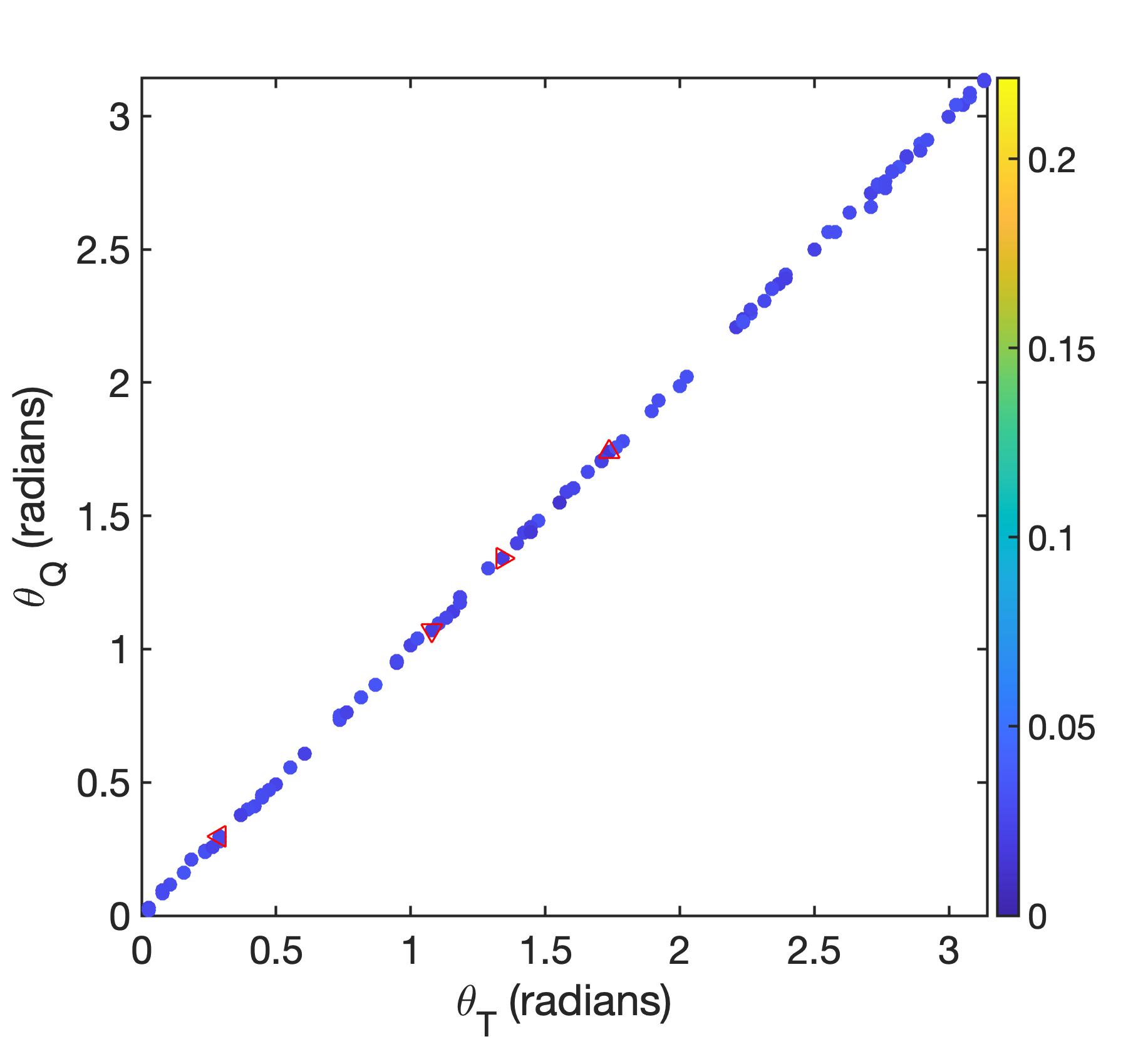}
    \includegraphics[trim={0cm 1cm 0cm 4cm},clip,width=.48\textwidth]{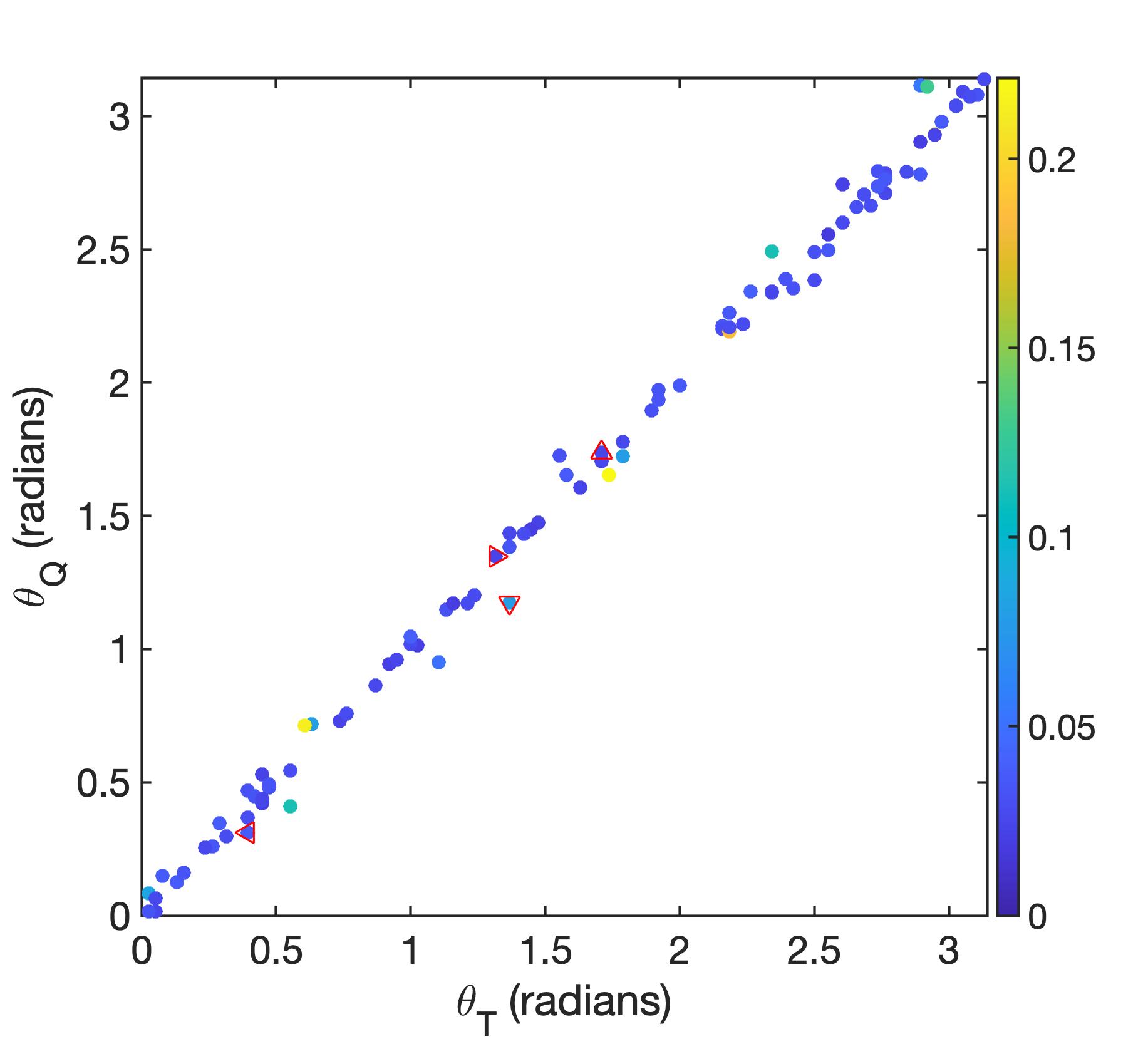}
    \caption{Comparison of true position angle, $\theta_T$, vs. estimated position angle, $\theta_Q$, for all $100$ smooth mock clusters ({\it left}), and $100$ substructured mock clusters ({\it right}). The true position angle is calculated using Eq.~\ref{eq:ellipticity}. The estimated PA is the average of angles that bisect $\theta_{12}$ of each of the $300$ quads in a mock cluster. Color indicates the uncertainty associated with the estimated $\theta_Q$ in radians, calculated by taking the standard deviation on $\theta_Q$ using $30$ subsets with $10$ quads each. The points marked by red triangles correspond to the four clusters shown in Figures~\ref{fig:smooth} and \ref{fig:structured}: triangle pointing up is the first row, pointing right is the second row, pointing down is the third row, and pointing left is the fourth row. For substructure-less clusters the quad estimation of PA is excellent: the points follow a diagonal in the left panel.
    }
    \label{fig:positionAngles}
    %\vskip-2cm
\end{figure*}

A summary of quad-estimated cluster properties for all 100 clusters is presented in the left panels of Figures~\ref{fig:deltaCenters},  \ref{fig:ellipticity}, and \ref{fig:positionAngles}. In each one of these figures the 4 clusters of Figure~\ref{fig:smooth} are highlighted as red triangles. The average and rms dispersion in the displacement between the true and quad-estimated center and PA are $9.0\times10^{-3}$ $\pm$ $5.4\times10^{-3}$ in units of cluster Einstein radius, and $2.64\times10^{-2}$ $\pm$ $7.2\times10^{-3}$ radians, respectively. 

The quad-estimated, $\epsilon_Q$, and true, $\epsilon_T$ ellipticities are tightly correlated. The quad-estimated elongation, $\epsilon_Q$ was measured from lensed image angles and is in units of radians, while the true ellipticity, $\epsilon_T$ was measured from the true mass distribution and is in units of length, normalized by $r_{\rm Ein}$. The actual values and units of $\epsilon$'s are not relevant, but what is important for us here is that the two are correlated, implying that quads alone contain the information about the ellipticity. The points in Fig.~\ref{fig:ellipticity} are colored by their $\delta_{FSQ}$ values. Less elliptical mass distributions follow the FSQ better than more elliptical ones, though $\delta_{FSQ}$ is small for all clusters.

Summing up our results from smooth substructure-less clusters, the estimation of all three properties is very good. This establishes that our quad-based method works as intended. We can now apply it to substructured clusters.

\subsection{Substructured clusters}\label{sec:substruc}

Figure~\ref{fig:structured} is similar to Fig.~\ref{fig:smooth}, but for 4 out of our 100 substructured mock clusters, chosen to span the range of cluster morphologies. \llrw{The right panels show that even for substructured clusters our FSQ-based method finds the global minimum.}  The 30 black points in the right panels represent the cluster center locations estimated using 10 quads each. Typically, these points, which are calculated without any reference to the true cluster center, sit at a distance of \lasko{$(1.64\times10^{-2} \pm 3.07\times10^{-2})\,r_{\rm Ein}$} from the center found using all 300 quads.

The right panel of Figure~\ref{fig:deltaCenters} shows the comparison between the true and the quad-estimated cluster center locations, using 300 quads per cluster. The color of the points represents the uncertainty on the cluster's center estimation found through the 30 subsets of 10 quads each. Though not as accurate as for the smooth elliptical clusters (left panel), the average and rms dispersions of these are still small, \lasko{$(3.59\times10^{-2}\pm 3.53\times10^{-2})\, r_{\rm Ein}$}.

The right panel in Figure~\ref{fig:ellipticity} presents the quad-estimated elongation vs. true ellipticity. Just as for the smooth clusters (left panel), $\epsilon_Q$ was measured from lensed image angles and is in units of radians, while $\epsilon_T$ was measured from the true mass distribution and is in units of length, normalized by $r_{\rm Ein}$. As stated in Section~\ref{sec:smooth}, the actual values of $\epsilon$'s are not important; what matters here is that the two are correlated. As expected, the correlation is not as tight as for clusters without substructure, but it is well defined nonetheless. Most of the scatter is in the lower left portion of the panel, which is the region of least elongation. Points in that region of the plot tend to have high values of $\delta_{\rm FSQ}$, which indicates a high amount of substructure (see the color scale). An example is shown on the third row of Figure~\ref{fig:structured}, where we can see a clear influence of substructure on the mass profile of the cluster. For clusters with low elongation, considerable substructure, or both, elongation is hard to define. Thus it is not surprising that our estimation method struggles to assign these clusters an accurate value.

The right panel in Figure~\ref{fig:positionAngles} shows the true position angle of 100 clusters vs. the quad-estimated position angle. The average and rms of the deviation between true and estimated PAs are \lasko{$4.13\times10^{-2}$ $\pm$ $4.78\times10^{-2}$} radians, respectively.

We conclude that the center, ellipticity and PA are well estimated just from quads alone, with no model priors. That means these properties of clusters will have similar values when recovered by lens inversions methods with different methodologies. The uncertainties we quote above should be representative of their spread. 

Figure~\ref{fig:substructure} shows the relation between $\delta_{\rm FSQ}$ and $s_T$, the true amount of substructure, in units of $r_{\rm Ein}$. (We note that we tried using reduced shear in place of $\kappa$ to measure true amount of substructure, but the correlation with $\delta_{FSQ}$ turned out to be worse.) There is a clear positive relation between the two, though with considerable scatter. The width of scatter can be as large as an order of magnitude, indicating that quads alone do not provide good constraints on $s_T$. %One may wonder if differently defined quad-based estimator would have produced a tighter correlation with $s_T$. It is possible, but our choice is very reasonable: $\delta_{\rm FSQ}$ is the lowest order, single-value indicator of deviation from ellipticity.

Given the large scatter, we conclude that quads alone do not constrain substructure nearly as well as they constrain the center, ellipticity and PA. We continue this discussion in Section~\ref{sec:last}, after examining three other clusters in the next Section.

\section{Three example clusters from the literature} \label{sec:examples}

We apply our quad-based estimators to a simulated galaxy cluster, Ares, and two observed clusters, Abell 1689 (A1689) and RXJ1347.5-1145 (RXJ1347). All three are reasonably centrally concentrated, so that their lensed sources have enough quads to apply our method. Coincidentally, all three clusters have 13 quads each. 

The quad images of Ares \citep[$z=0.5$,][]{Meneghetti_2017} are shown in the left panel of Figure~\ref{fig:ares}. Even though it is a merging cluster, it has one dominant center located at $(x_T,y_T)=(-19.8", -31.5")$; see the middle panel of Figure~\ref{fig:ares}. Ares' quad-estimated center lies $1.94"$, or $5.91\times10^{-2}$ $r_{\rm Ein}$ of the dominant center in the mass profile. This corresponds to $<2$ of the rms uncertainty we found in Section~\ref{sec:substruc}
%{\color{dgreen} If we are talking about rms uncertainty on single centers, then yes. There's also the rms uncertainty for all 100 substructured clusters, which is $5.54\times10^{-2} r_{\rm Ein}$}, or $2\sigma$ in the center determination obtained for the mock clusters in Section~\ref{sec:substruc}, 
and the right panel of Figure~\ref{fig:deltaCenters}.

The contours of $\delta_{FSQ}$ are shown in the right panel. Unlike the right panels of Figures~\ref{fig:smooth} and \ref{fig:structured} which were based on 300 quads per cluster, the small number of quads in Ares results in sharp changes in the contours:
%This is because a smaller number of quads gives the images located at points a stronger influence on the total deviation. 
when the trial center is very near the location of an image, there is a rapid change in the value of two of the angles $\theta_{12}, \theta_{23}$, and $\theta_{34}$, resulting in a steep change in deviation from the FSQ. However, these sharp features are far from the estimated center, so do not effect the results.

Ares' $\delta_{\rm FSQ}=4.08\times10^{-2}$ radians, which places roughly in the middle of the range of $\delta_{\rm FSQ}$'s of our mock clusters; see Figure~\ref{fig:substructure}. Visual inspection shows that the position angle for Ares, indicated by red lines in all three panels of Figure~\ref{fig:ares},
aligns very well with the elongation of its true isodensity contours.

A1689, at $z_l=0.183$, shown in the three panels of Figure~\ref{fig:abell}, could be a line of sight merger, but on the plane of the sky it is very roughly circularly symmetric \citep[e.g.,][]{bro05,coe10,die15,bin16,gho22}. Its quad-estimated center is about $2.7"$, or $5.86\times10^{-2}\,r_{\rm Ein}$ from the BCG, towards the next brightest galaxy, G1. This offset between true and estimated center is $<2$ times the (rms) uncertainty estimated using out substructured clusters. The true mass distribution is obviously not known for this cluster, and the location of the BCG is not necessarily the mass center of the cluster \citep{lau14}, so we cannot directly compare our estimated values to the true ones. In the middle panel of Figure~\ref{fig:abell} we plot the mass distribution recovered by free-form \grale\ \citep{gho22}, but we do not use it to measure true cluster properties. Using the method of minimizing deviations from the FSQ, we obtain a value of $\delta_{\rm FSQ} = 6.74\times10^{-2}$ radians, which is in the upper half of our mock clusters (see Figure~\ref{fig:substructure}), so probably has more substructure than Ares.

RXJ1347, at $z=0.451$ is a massive, X-ray luminous merging galaxy cluster \citep[e.g.,][]{hal08,koh14,ued18,ric21}. 
%We use the mass distribution recovered by \cite{ric21} in lieu of the true mass. They fit the cluster using 3 cluster-scale dark matter halos. We calculate the cluster center as the center of mass of these 3 halos, weighted by the square of their velocity dispersion, which yields $(-5.11",-0.60")$. Our quad-estimated center, $(-6.18",-3.83")$, is displaced from it by $3.4"$, or $XXX r_{\rm Ein}$. 
Our quad-estimated center, $(x_Q,y_Q)=(-6.18",-3.83")$, is roughly between the two brightest galaxies of the cluster, located at $(0",0")$ and $(-17.8",-2.1")$ \citep{ric21}, while the Einstein radius of the cluster, as obtained from its 13 observed quads is $35.5"$. The middle panel of Figure~\ref{fig:rxj} shows the mass distribution recovered by simply parametrized {\textsc{Lenstool}} \citep{ric21}, for reference. As with A1689 we do not use it to estimate true cluster properties. The quad-estimated position angle appears to align well with the elongation of the cluster's mass; see Figure~\ref{fig:rxj}. Its $\delta_{\rm FSQ} = 9.34\times10^{-2}$ radians, which places it above the average $\delta_{\rm FSQ}$ of our mock clusters, and suggests that it is the most substructured of the 3 clusters examined in this Section.

\section{Conclusions and Discussion}\label{sec:last}

The paper's goal is to better understand the relative roles of multiple image data vs. model assumptions in lensing mass reconstructions, not to carry out such reconstructions. \llrw{Our analysis, and the methods we use are not meant to complement or substitute for the existing lens inversion methods. We use our analysis tools with a different goal in mind.}
We wanted to know how much information comes from the lensing images themselves, and how much is determined by the modeling priors. To that end, we asked, how well can we estimate global cluster properties from images alone, without any lens models? To do that we estimated global cluster properties---cluster center, ellipticity, its position angle and amount of substructure---using only the angular distribution of images of quadruply imaged sources around the cluster center, in a model-free way, \llrw{i.e., without doing a lens-inversion to get a mass model}. \llrw{(While our mock cluster construction is model-based, our estimation of cluster properties is model-free.)} Our rationale is that if a given cluster property is robustly estimated based on images alone, without a mass model, at least for the range of mass distributions used in this paper, then model priors play a minor role in their determination, while images play a dominant role, and all lens inversion methods should agree. If a lens property is estimated based on quads only very approximately, then modeling priors will have a stronger influence on it, and various lens inversion methods will yield a range of results.

A caveat to our method is that it does not use all the lensing information: it cannot use sources that do not result in quads, and for quads, it cannot use the image distances from the lens center. We are not aware of any existing method that would use all the lensing observables to estimate global cluster properties in a model-free way.

\subsection{Estimation of cluster centers, ellipticity and position angle}

For smooth elliptical mock clusters we show that one can estimate the center, ellipticity and its position angle accurately and precisely using our quad-based method (see the left panels of Figures~\ref{fig:deltaCenters}, \ref{fig:ellipticity}, and \ref{fig:positionAngles}).  

For mock clusters with substructure, quad-based estimation works reasonably well (see the right panels of Figures~\ref{fig:deltaCenters}, \ref{fig:ellipticity}, and \ref{fig:positionAngles}): \lasko{92\%} of estimated centers lie within \lasko{$0.1 r_{\rm Ein}$} of the true center, and \lasko{93\%} of cluster position angles are within \lasko{$0.1$} radians of the true position angle. Estimation of ellipticity depends on the amount of substructure and cluster ellipticity, with less substructured and more elliptical clusters yielding better precision.  

In the case of Ares, where the true mass distribution is known, we use its 13 quads to estimate the center, which is $<2"$, or $0.059\, r_{\rm Ein}$ ($<2$ of the rms uncertainty estimated from our synthetic clusters) away from the true center. As Figure~\ref{fig:ares} shows our estimated position angle lines up very well with the elongation of the mass density contours. 

The true mass distributions of A1689 and RXJ1347 are not known, but quad-based estimation recovers their centers near the central dominant galaxies, and the ellipticity position angle aligns well with the visual appearance as well as the lens inversion reconstructions.

We conclude that cluster centers, ellipticity and position angle are estimated robustly with our quad-based method with no model priors, and should therefore also be recovered well and by all lensing inversion methods, with a good degree of consistency between methods. This is largely in agreement with the findings of \citep{Meneghetti_2017}. While they do not discuss how well the reconstructions recover cluster center, they do compare the recovery of ellipticity, position angle and substructure in their Figure~27. Of these 3 properties the best recovered one is the position angle, especially for the more realistic cluster Hera. The next best estimated property is ellipticity, which is recovered by lens reconstructions considerably better than substructure: of the 20 lens reconstructions, 16 estimate ellipticity at the same level or better than substructure. This order of how well properties are recovered agrees with our findings.

\subsection{Estimation of amount of substructure, \llrw{or deviations from ellipticity}}\label{sec:conc_substruc}

\begin{figure}
    \centering
    \includegraphics[width=.48\textwidth]{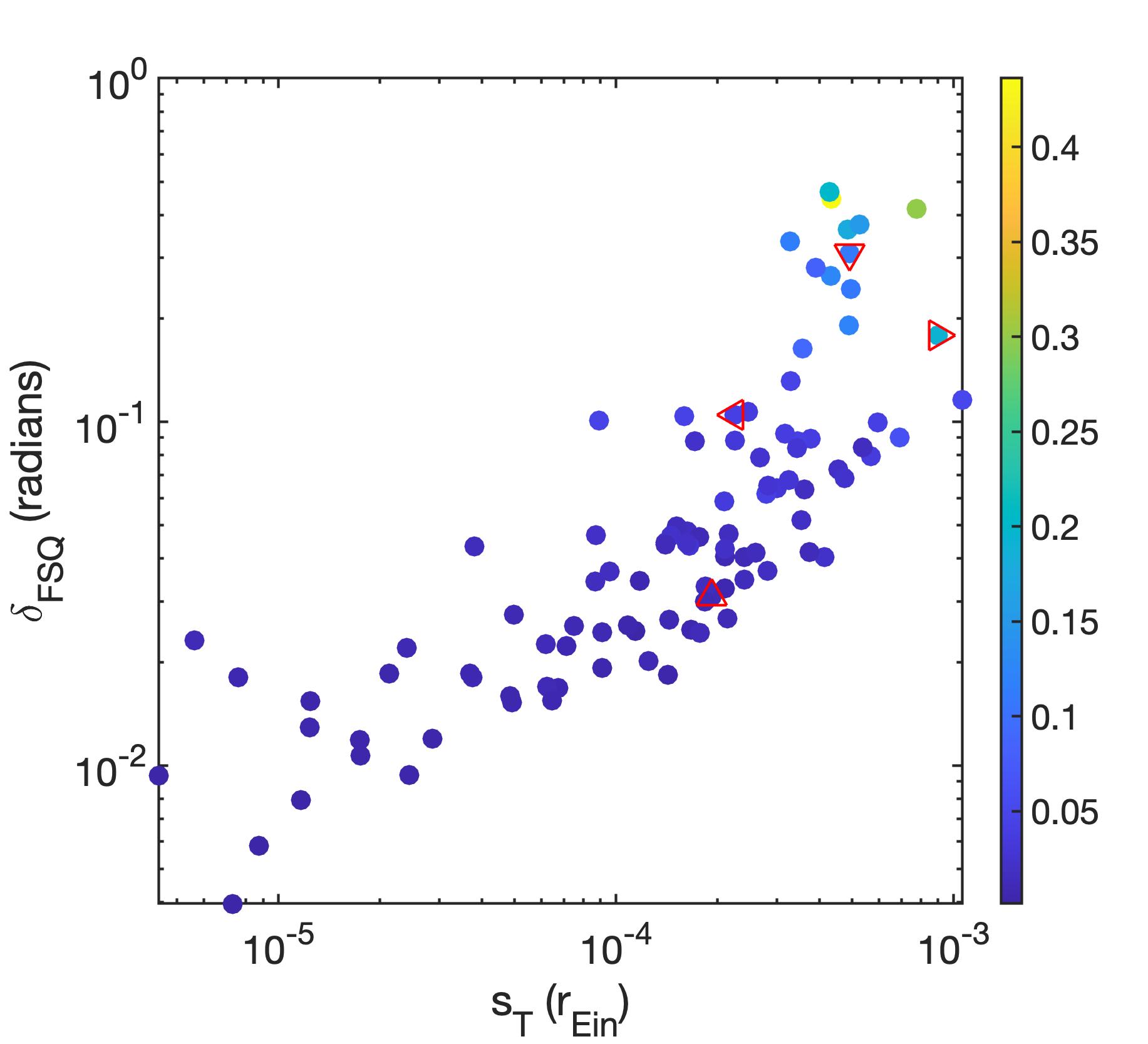}
    \caption{Comparison of quad-estimated amount of substructure, $\delta_{\rm FSQ}$, and the true amount of substructure, $s_T$. (Both quantities are dimensionless.)  It is apparent that there is a correlation between these two quantities, but the scatter is large. Color indicates the cluster's ellipticity, $\epsilon_T$.
    See Sections~\ref{sec:substruc} and \ref{sec:conc_substruc} for details.
    %, and it becomes more defined if we look at only the clusters with $\epsilon_Q \lesssim 2.6$. The points marked by triangles correspond to the clusters seen in Figures \ref{fig:smooth} and \ref{fig:structured}, the orientation of the triangle indicating which cluster is shown.
    }
    \label{fig:substructure}
\end{figure}

\begin{figure*}
    \centering
    \includegraphics[width=.31\textwidth]{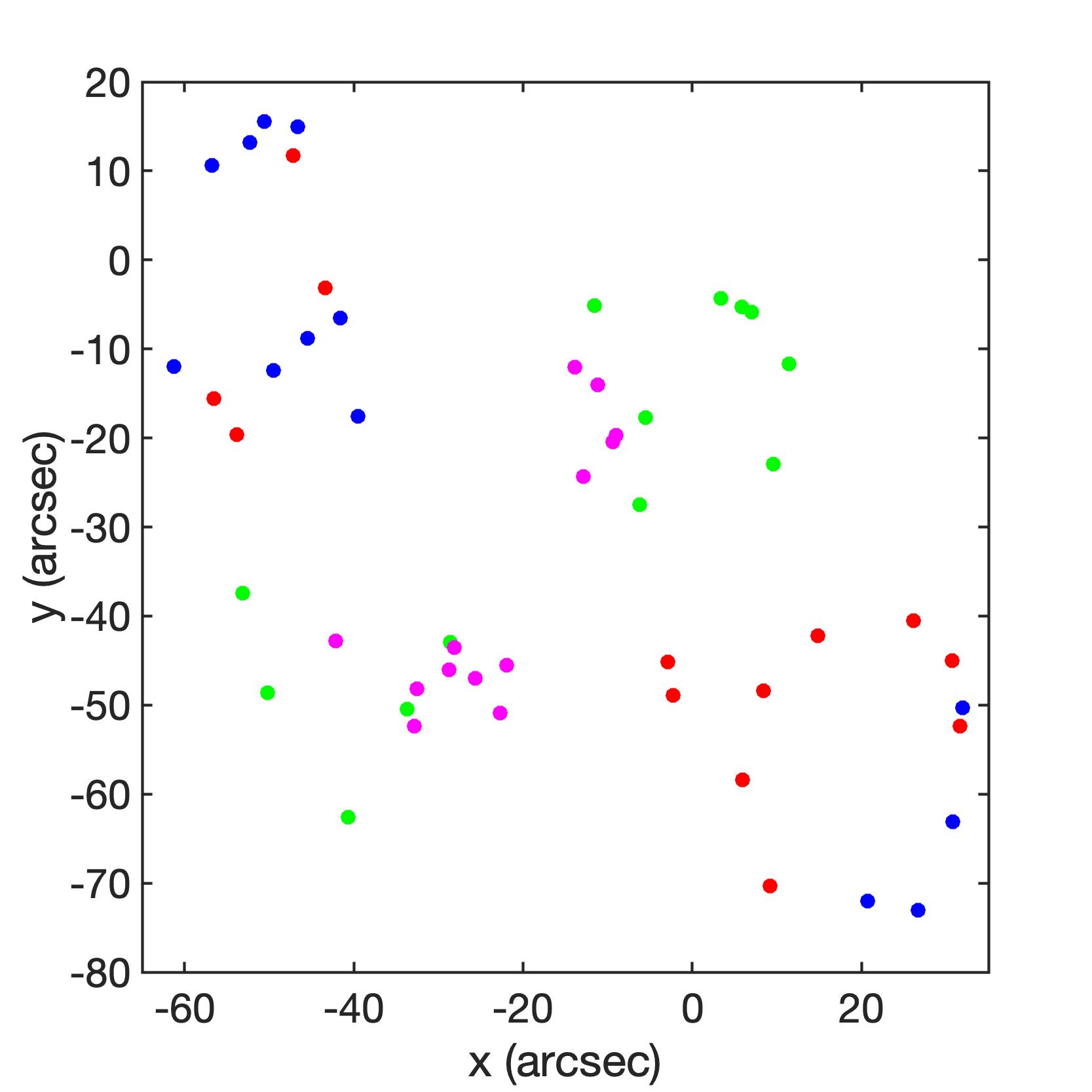}%
    \includegraphics[trim={0cm 5cm 0cm 4cm},clip,width=.31\textwidth]{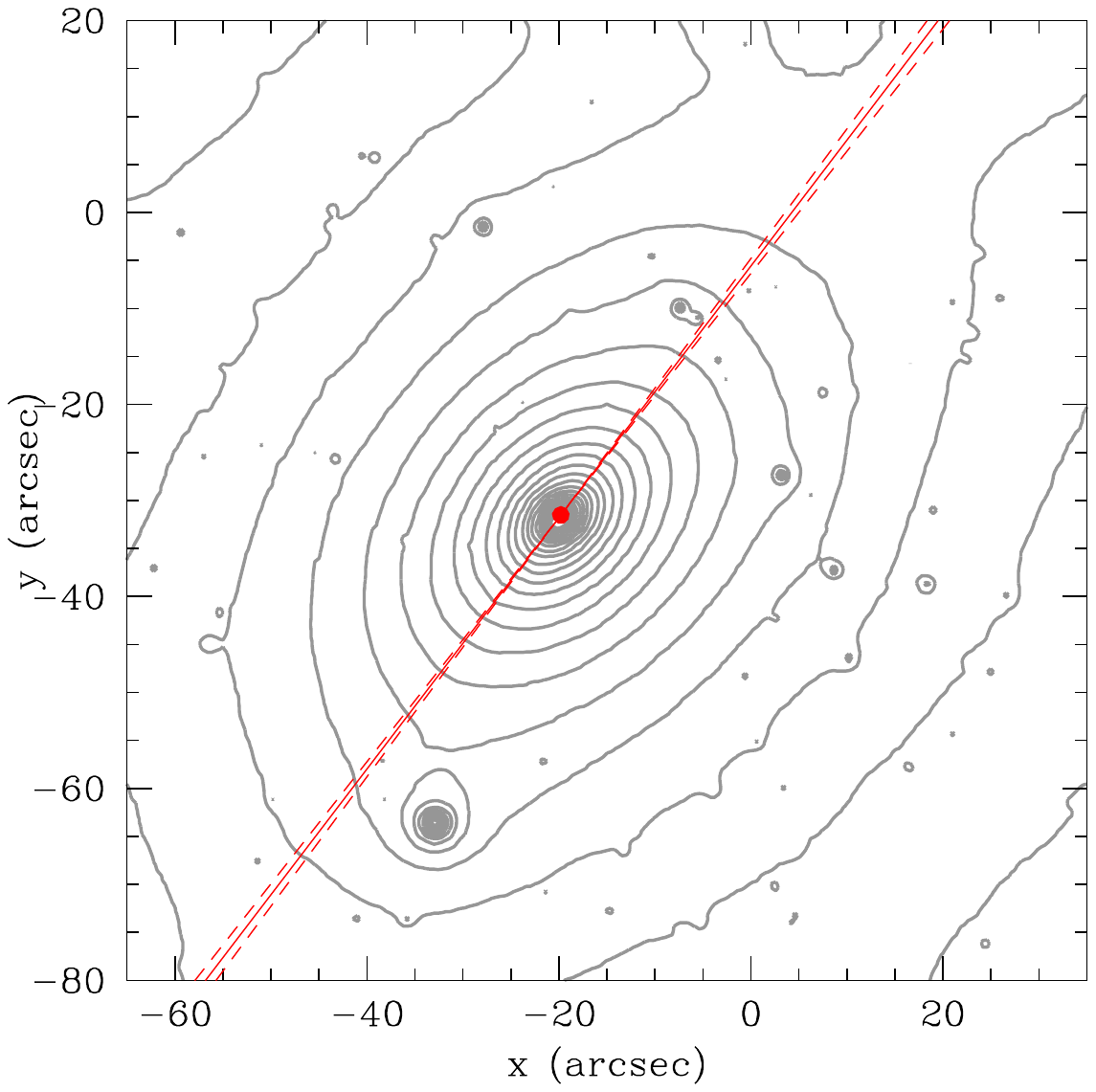}%
    \includegraphics[width=.34\textwidth]{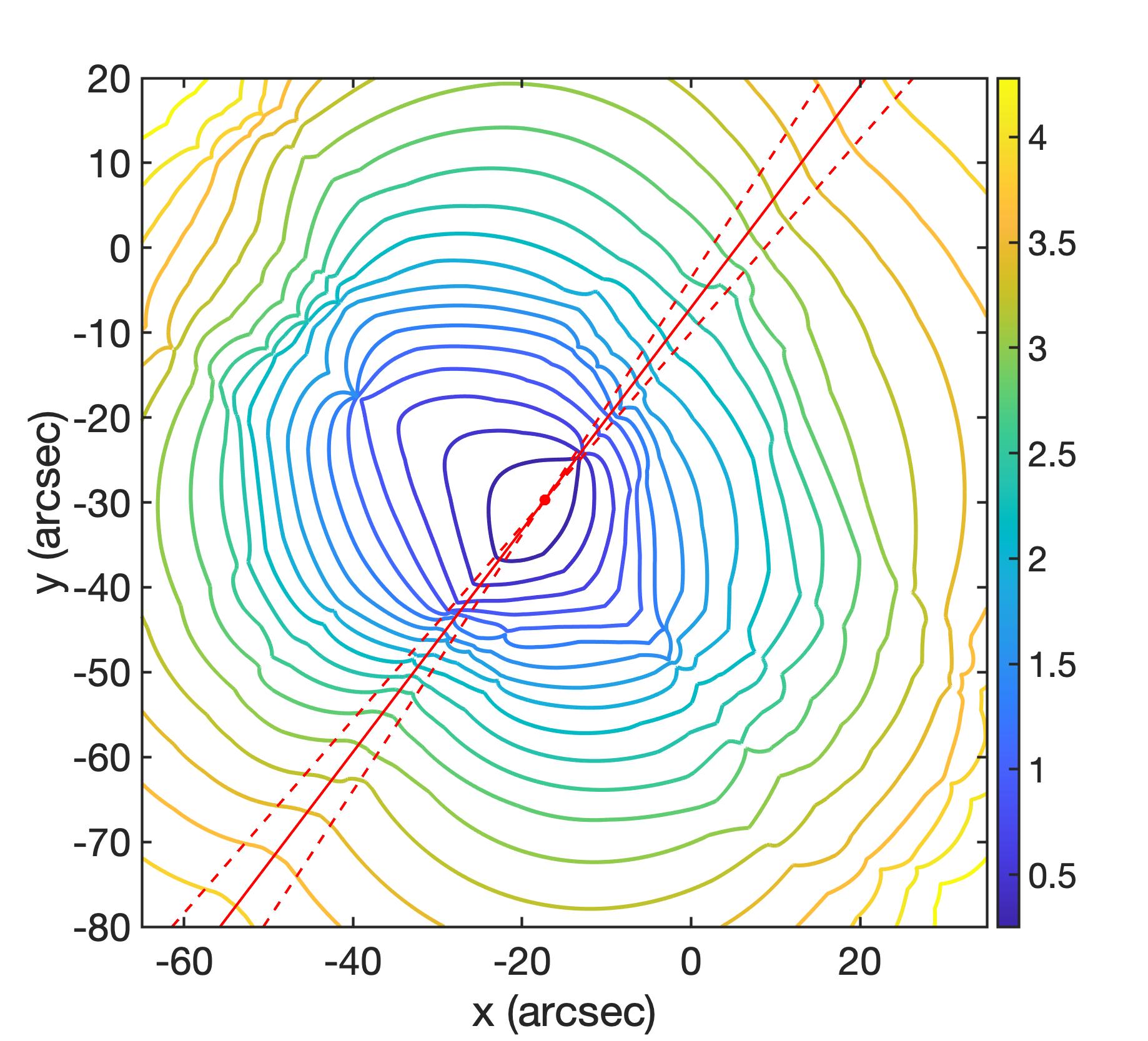}
    \caption{Ares simulated cluster. {\it Left:} Quad images are colored by arrival sequence: blue$\to$red$\to$green$\to$magenta. {\it Middle:} The true mass distribution \citep{Meneghetti_2017}, with the center at $(x_T,y_T)=(-19.8",-31.5")$. {\it Right:} Contours of $\delta_{\rm FSQ}$ generated using $13$ quads. The minimum $\delta_{\rm FSQ}$ is the quad-estimated center, and is at $(x_Q,y_Q)=(-18.07",-30.63")$. The slanted line in all 3 panels is the position angle derived by our method using all $13$ quads. The dashed lines in the left and right panels represent the standard deviation on $\theta_Q$ found by taking $30$ random subsets containing $7$ quads each. Both the center and the position angle fit those obtained visually from the mass profile.}
    \label{fig:ares}
\end{figure*}

\begin{figure*}
    %\vskip-2cm
    \centering
    \includegraphics[width=.31\textwidth]{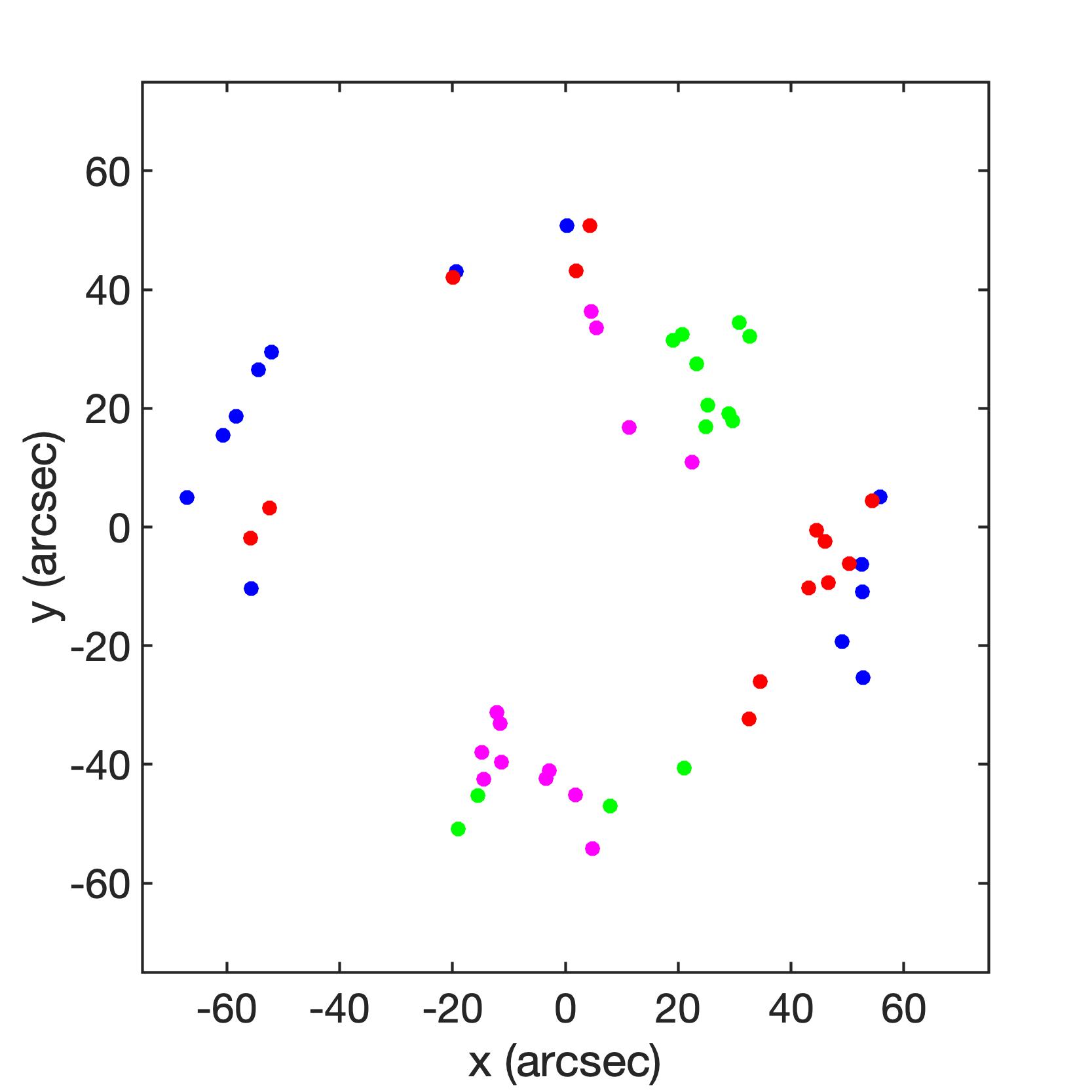}%
    \includegraphics[width=.36\textwidth]{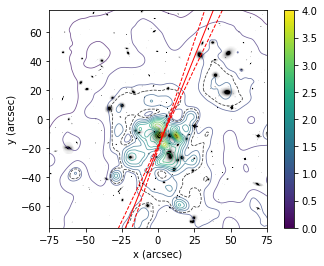}%
    \includegraphics[width=.34\textwidth]{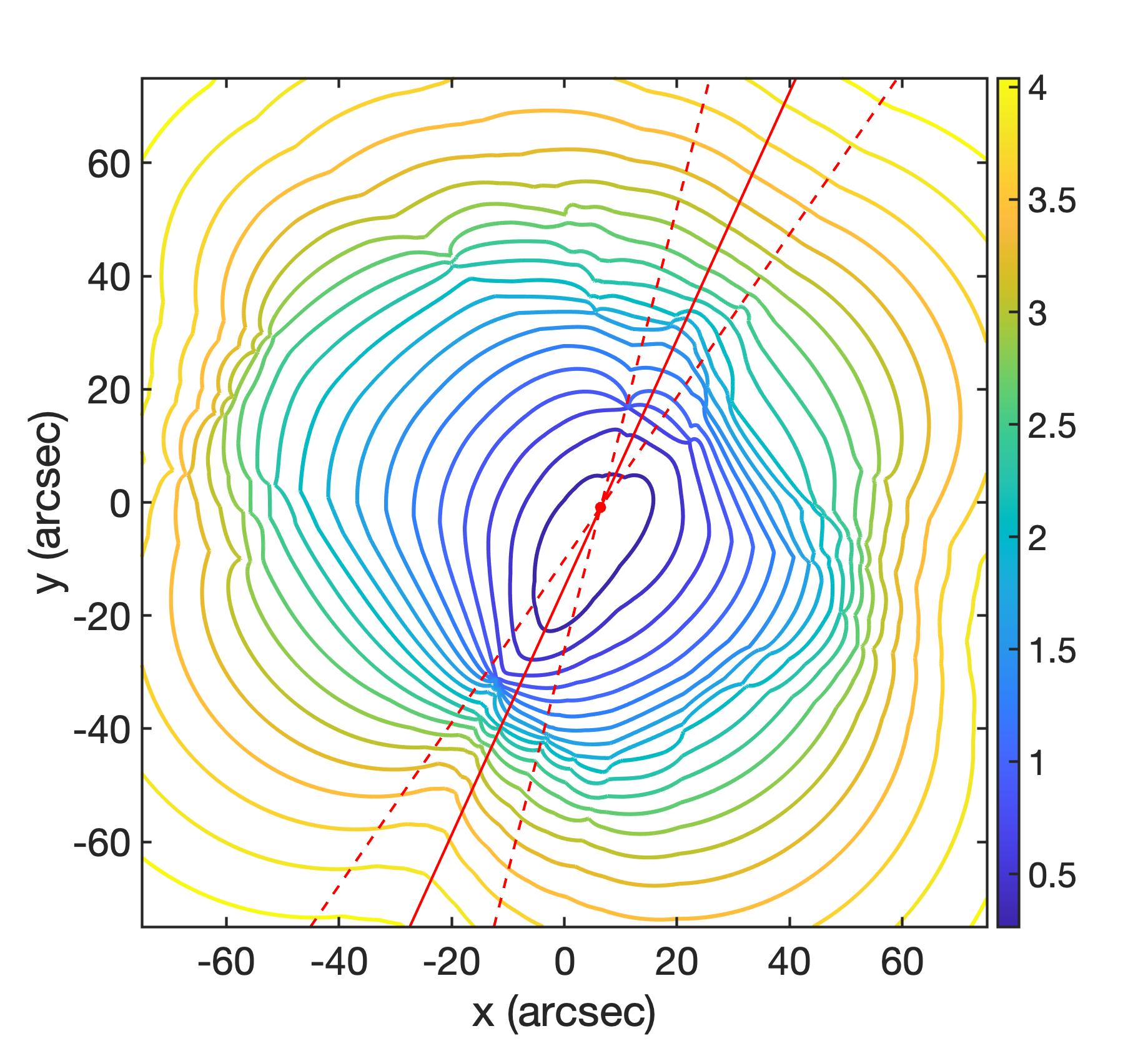}
    \caption{Similar to Figure~\ref{fig:ares}, but for A1689. {\it Left:} Quad images of A1689, colored by arrival sequence: blue$\to$red$\to$green$\to$magenta. {\it Middle:} Reconstructed mass distribution as presented in \cite{gho22}. {\it Right:} Contours of $\delta_{\rm FSQ}$ generated using $13$ quads. The minimum $\delta_{\rm FSQ}$ is the quad-estimated center, and is at $(x_Q,y_Q)=(3.57",-10.46")$, about $2.7"$ from the BCG, towards the next brightest galaxy, G1. The slanted line in all 3 panels is the position angle derived by our method using all $13$ quads, with the dashed lines representing the standard deviation on $\theta_Q$ found by taking $30$ random subsets containing $7$ quads each. Both the center and the position angle fit those obtained visually from the mass profile.}
    \label{fig:abell}
\end{figure*}

\begin{figure*}
    \centering
    \includegraphics[width=.31\textwidth]{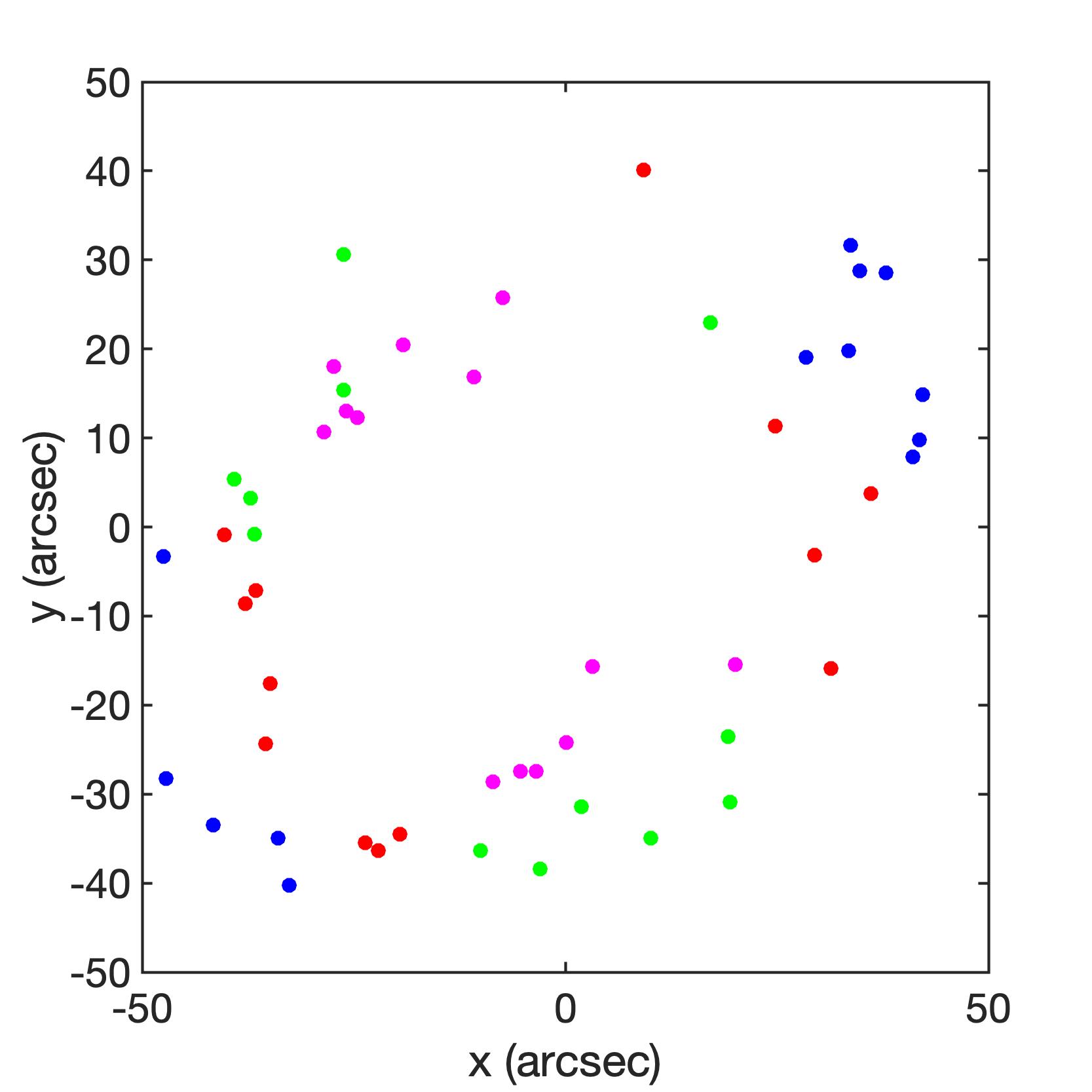}%
    \includegraphics[width=.31\textwidth]{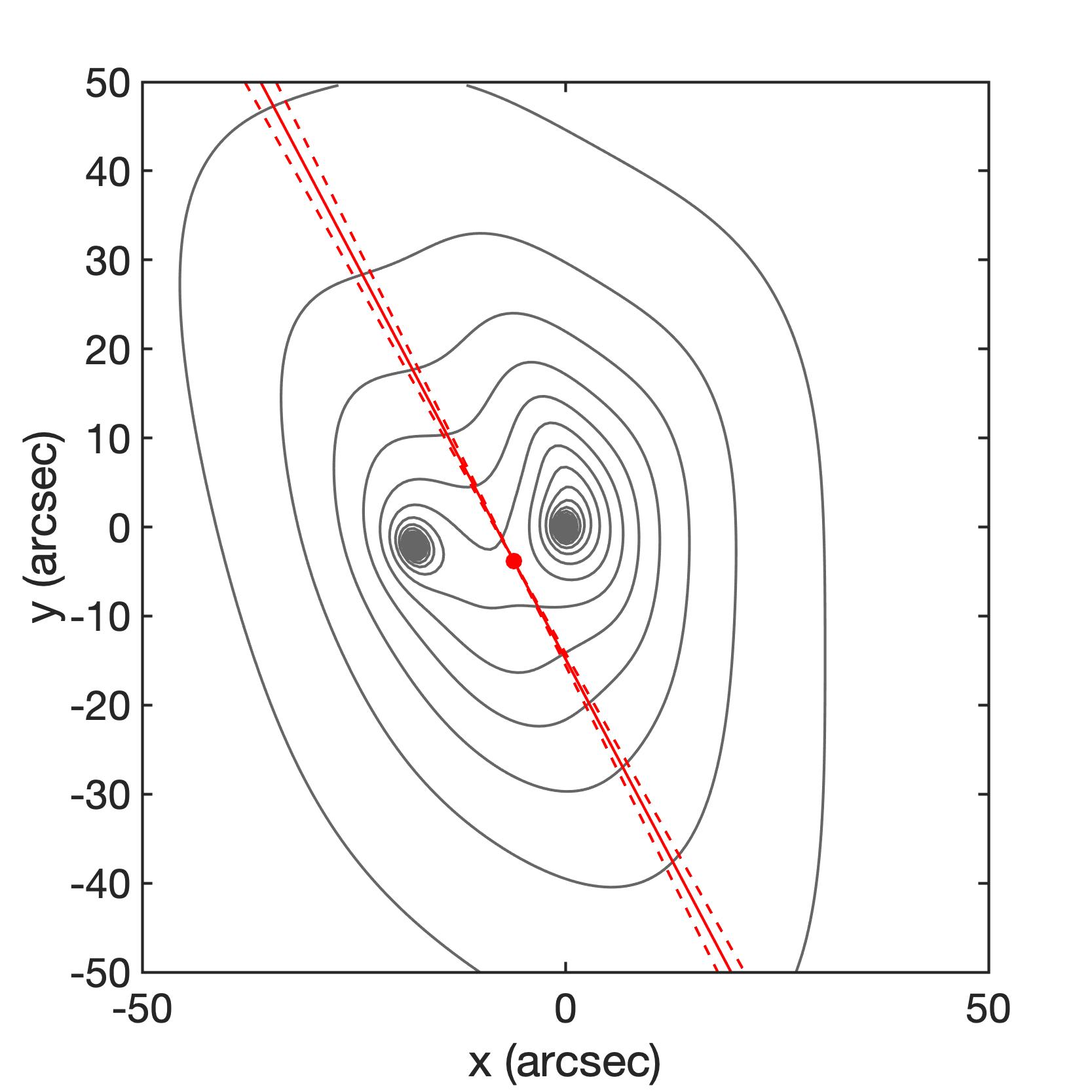}%
    \includegraphics[width=.34\textwidth]{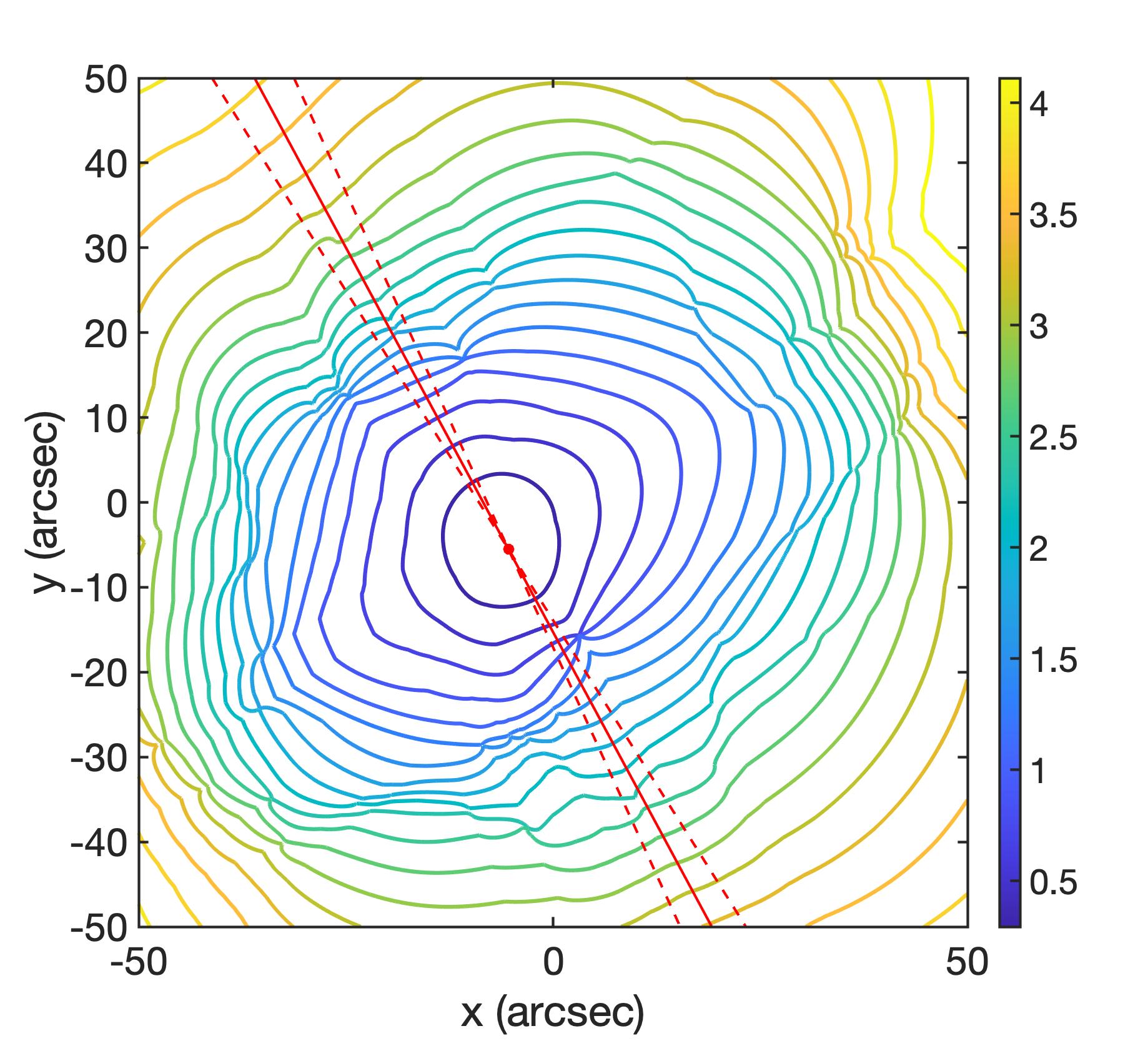}
    \caption{Similar to Figure~\ref{fig:ares}, but for RXJ1347. {\it Left:} Quad images colored by arrival sequence:  blue$\to$red$\to$green$\to$magenta. {\it Middle:} Reconstructed mass model produced using dPIE model parameters \citep{ric21}. Red line shows the quad-estimated position angle, $\theta_Q$, which matches well with $\theta_T$. {\it Right:} Contours of $\delta_{\rm FSQ}$ generated using $13$ quads. The minimum $\delta_{\rm FSQ}$ is the quad-estimated center, and is at $(x_Q,y_Q)=(-6.18",-3.83")$. The slanted line in both panels is the position angle derived by our method using all $13$ quads, with the dashed lines representing the standard deviation on $\theta_Q$ found by taking $30$ random subsets containing $7$ quads each.}
    \label{fig:rxj}
\end{figure*}

The estimation of the amount of substructure yields interesting and instructive conclusions. Fig.~\ref{fig:substructure} shows that for our mock clusters there is a positive relation between the quad-estimated amount of substructure and the true amount of substructure measured from the mass distribution. Before we discuss the scatter in the next paragraph, we note that the correlation also holds for Ares, A1689 and RXJ1347: the quad-estimated amount of substructure increases progressively from Ares ($\delta_{\rm FSQ}=4.08\times 10^{-2}$), to A1689 ($\delta_{\rm FSQ}=6.74\times 10^{-2}$), to RXJ1347 ($\delta_{\rm FSQ}=9.34\times 10^{-2}$). This sequence of increasing $\delta_{\rm FSQ}$ values is consistent with what we know about these 3 clusters: Ares, though bimodal, has a dominant central mass concentration that hosts most of the images, and a small amount of substructure (see the middle panel of Fig.~\ref{fig:ares}). A1689 is a cluster near equilibrium, but with known substructure \citep{gho22}, while RXJ1347 is a merging cluster with complex structure \citep{ric21}. 

The correlation between $\delta_{\rm FSQ}$ and the true amount of substructure for our mock clusters (Fig.~\ref{fig:substructure}) has a lot of scatter. For example, in Fig.~\ref{fig:structured} the clusters in the 1st and 4th rows have very similar $s_T$ values, within a factor of $1.17$ of each other, yet their values of $\delta_{FSQ}$ differ by a factor of $3.35$. We can see even larger disparities in clusters with roughly equal $\delta_{\rm FSQ}$--many with differences in $s_T$ spanning more than an order of magnitude. One reason why $\delta_{\rm FSQ}$ cannot predict $s_T$ well is that we are not able to use all the multiple image information in our analysis: we do not use image distances from the cluster center as input (see Section~\ref{sec:estimated}), and we cannot use doubles. The other reason is that lensed images alone cannot uniquely determine the mass distribution.
%, resulting in lensing degeneracies.

\begin{figure*}
    \centering
    \includegraphics[width=0.33\textwidth]{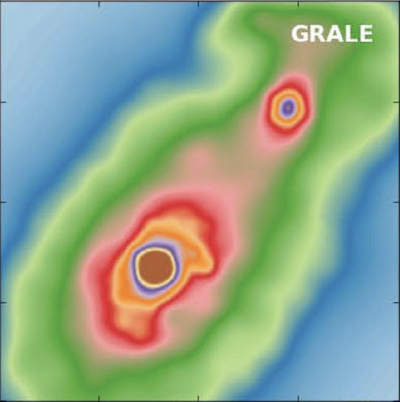}%
    \includegraphics[width=0.33\textwidth]{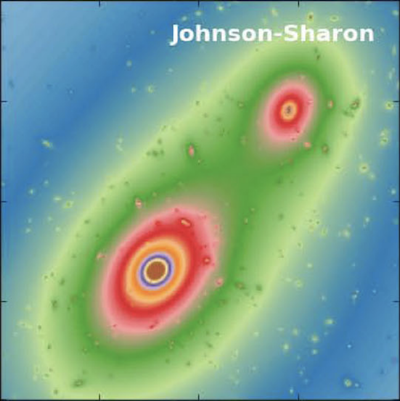}%
    \includegraphics[width=0.33\textwidth]{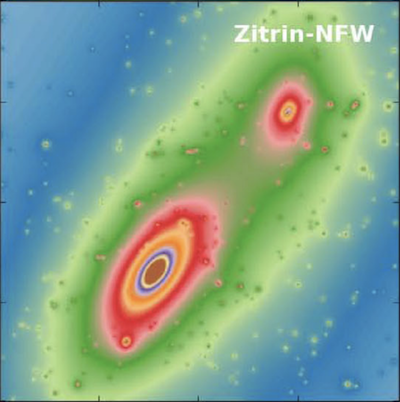}\\
    \includegraphics[width=0.33\textwidth]{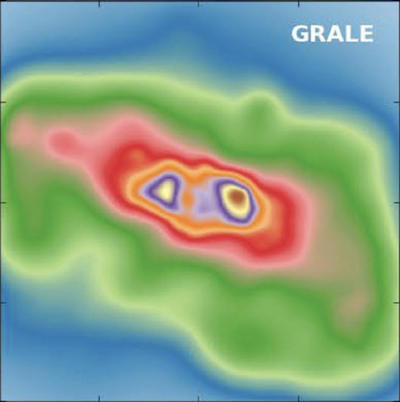}%
    \includegraphics[width=0.33\textwidth]{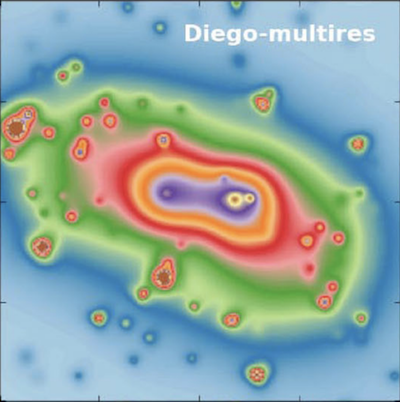}%
    \includegraphics[width=0.33\textwidth]{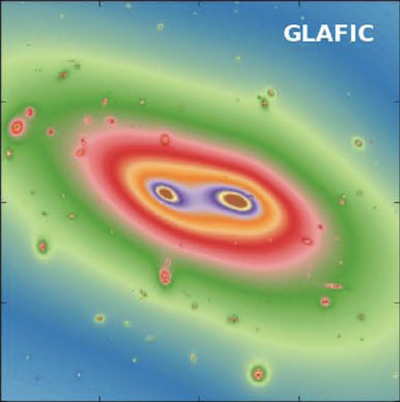}
    \caption{Three different lens mass reconstructions of the Ares simulated cluster ({\it top}) discussed in Section~\ref{sec:examples}, and the Hera simulated cluster ({\it bottom}), which has been omitted from our analysis due to a lack of quads \citep{Meneghetti_2017}. All reconstruction in each row used very similar multiple image sets. This illustrates that images alone cannot fully constrain the substructure, and apparently, leave room for priors to fill in the information gaps (see also Figure~\ref{fig:substructure}).}
    \label{fig:massProfiles}
\end{figure*}

The latter helps explain why given the same multiple image data set, different reconstruction methodologies---parametric vs. LTM vs. free-form vs. hybrid---can reproduce observed multiple images with qualitatively different lens mass distributions: Figure~\ref{fig:massProfiles} shows 3 examples of mass distributions of simulated clusters Ares and Hera, reconstructed by different lens inversion methods: free-form \grale, simply parametrized Johnson-Sharon, Glafic, Light-Traces-Mass (LTM) Zitrin-NFW, and hybrid Diego-multires \citep[for references, see][]{Meneghetti_2017}. The mass distribution of simply-parametrized and LTM models is dominated by smooth dark matter and very localized mass due to individual member galaxies, while that of the free-form (and to a lesser degree hybrid) methods has more diffuse substructure with less very small scale structure. 
%\llrw{In other words, same lens data leads to different substructure characteristics, implying that model priors have an important impact on the recovered mass distributions.}
%(similar to the overlay in the middle panel of Fig.~\ref{fig:abell}, for A1689). 

This visual appearance of mass distribution can also be cast in terms of mass power spectrum. \cite{moh16} showed that reconstructions by simply-parametrized {\textsc {Lenstool}} are dominated by power on large and small scales, with a somewhat of a deficit at intermediate scales. On the other hand, free-form \grale\ includes intermediate length scales, which result in diffuse substructures, but misses power on the smallest scales.

We conclude that multiple images by themselves, i.e., without model priors, allow for a range of substructure characteristics and hence differently shaped mass distributions, corresponding to different shape degeneracies \citep{sah06}. Lens model priors `select' what scale substructure to use to fit the observed images. In the future, if the number of multiple images in clusters increases, for example with data from the {\it James Webb Space Telescope}, it will be possible to let lensed images, not modeling priors, be the dominant factor in determining the mass distribution on a wide range of scales within galaxy clusters.

\acknowledgements
The authors would like to thank Johan Richard (Observatoire de Lyon, France) for helping us with the analysis of the lensing cluster RXJ 1347.

\section*{Data Availability}
The data and and scripts used for the purposes of this paper are publicly available from the GitHub repository: \href{https://github.com/kekoalasko/FSQ_Modeling}{https://github.com/kekoalasko/FSQ\_Modeling}. Any requests for supplementary data or code will be shared upon reasonable request to the corresponding author.

\bibliography{references.bib}

\end{document}